\documentclass[preprint,noshowpacs,noshowkeys,aps]{revtex4}%
\usepackage{amsfonts}
\usepackage{amsmath}
\usepackage{amssymb}
\usepackage{graphicx}%
\providecommand{\U}[1]{\protect\rule{.1in}{.1in}}
\begin{document}
\preprint{ }
\title[Nonlocal Phases of Wavefunctions]{Beyond the Dirac phase factor$\boldsymbol{:}$ Dynamical Quantum
Phase-Nonlocalities in the Schr\"{o}dinger Picture}
\author{Konstantinos Moulopoulos}
\email{cos@ucy.ac.cy}
\affiliation{University of Cyprus, Department of Physics, 1678 Nicosia, Cyprus}

\begin{abstract}
Generalized solutions of the standard gauge transformation equations are
presented and discussed in physical terms. They go beyond the usual Dirac
phase factors and they exhibit nonlocal quantal behavior, with the well-known
Relativistic Causality of classical fields affecting directly the
\textit{phases} of wavefunctions in the Schr\"{o}dinger Picture. These
nonlocal phase behaviors, apparently overlooked in path-integral approaches,
give a natural account of the dynamical nonlocality character of the various
(even static) Aharonov-Bohm phenomena, while at the same time they seem to
respect Causality. Indeed, for particles passing through nonvanishing magnetic
or electric fields they lead to cancellations of Aharonov-Bohm phases at the
observation point, generalizing earlier semiclassical experimental
observations (of Werner \& Brill) to delocalized (spread-out) quantum states.
This leads to a correction of previously unnoticed sign-errors in the
literature, and to a natural explanation of the deeper reason why certain
time-dependent semiclassical arguments are consistent with static results in
purely quantal Aharonov-Bohm configurations. These nonlocalities also provide
a remedy for misleading results propagating in the literature (concerning an
uncritical use of Dirac phase factors, that persists since the time of
Feynman's work on path integrals). They are shown to conspire in such a way as
to exactly cancel the instantaneous Aharonov-Bohm phase and recover
Relativistic Causality in earlier \textquotedblleft
paradoxes\textquotedblright\ (such as the van Kampen
thought-experiment),$\boldsymbol{\ }$and to also complete Peshkin's discussion
of the electric Aharonov-Bohm effect in a \textit{causal} manner. The present
formulation offers a direct way to address time-dependent single- \textit{vs}
double-slit experiments and the associated causal issues -- issues that have
recently attracted attention, with respect to the inability of current
theories to address them.

\end{abstract}
\volumeyear{2011}
\volumenumber{number}
\issuenumber{number}
\eid{identifier}
\date[May 10, 2011]{}
\maketitle

\section{\bigskip Introduction}

\bigskip The Dirac phase factor $-$ with a phase containing spatial or
temporal integrals of potentials (of the general form $%
%TCIMACRO{\dint \limits^{\boldsymbol{r}}}%
%BeginExpansion
{\displaystyle\int\limits^{\boldsymbol{r}}}
%EndExpansion
\boldsymbol{A}\cdot d\boldsymbol{x}^{\prime}-c%
%TCIMACRO{\dint \limits^{t}}%
%BeginExpansion
{\displaystyle\int\limits^{t}}
%EndExpansion
\phi dt^{\prime}$) $-$ is the standard and widely used solution of the gauge
transformation equations of Electrodynamics (with $\boldsymbol{A}$ and $\phi$
vector and scalar potentials respectively). In a quantum mechanical context,
it connects wavefunctions of two systems (with different potentials) that
experience the same classical fields at the observation point $(\boldsymbol{r}%
,t)$, the two more frequently discussed cases being: either systems that are
completely gauge-equivalent (a trivial case with no physical consequences), or
systems that exhibit phenomena of the Aharonov-Bohm type (magnetic or
electric)\cite{AB} $-$ and then this Dirac phase has nontrivial observable
consequences (mathematically, this being due to the fact that the
corresponding \textquotedblleft gauge function\textquotedblright\ is now
multiple-valued). In the above two cases, the classical fields experienced by
the two (mapped)\ systems are equal \textit{at every point }of the accessible
spacetime region. However, it has not been widely realized that the gauge
transformation equations, viewed in a more general context, can have
\textit{more general solutions} than simple Dirac phases, and these lead to
wavefunction-\textit{phase-nonlocalities} that have been widely overlooked and
that seem to have important physical consequences. These nonlocal solutions
are applicable to cases where the two systems are allowed to experience
\textit{different fields} at spacetime points (or regions) that are
\textit{remote} to (and do \textit{not} contain) the observation point
$(\boldsymbol{r},t)$. In this article we rigorously show the existence of
these generalized solutions, demonstrate them in simple physical examples, and
fully explore their structure, presenting cases (and closed analytical results
for the wavefunction-phases) that actually connect (or map) two quantal
systems that are \textbf{neither physically equivalent nor of the usual
Aharonov-Bohm type}. We also fully investigate the consequences of these
generalized (\textit{nonlocal}) influences (on wavefunction-phases) and find
them to be numerous and important; we actually find them to be of a different
type in static and in time-dependent field-configurations (and in the latter
cases we show that they lead to Relativistically \textit{causal} behaviors,
that apparently resolve earlier \textquotedblleft paradoxes\textquotedblright%
\ arising in the literature from the use of standard Dirac phase factors). The
nonlocal phase behaviors discussed in the present work may be viewed as a
justification for the (recently emphasized\cite{Popescu}) terminology of
\textquotedblleft dynamical nonlocalities\textquotedblright\ associated with
all Aharonov-Bohm effects (even static ones), although in our approach these
nonlocalities seem to also respect Causality (without the need to
independently invoke the Uncertainty Principle) -- and, to the best of our
knowledge, this is the first theoretical picture with such characteristics.

In order to introduce some background and further motivation for this article
let us first remind the reader of a very basic property that will be central
to everything that follows, which however is usually taken to be valid only in
a restricted context (but is actually more general than often realized). This
property is a simple (U(1)) phase-mapping between quantum systems, and is
usually taken in the context of gauge transformations, ordinary or
singular$\boldsymbol{;}$ here, however, it will appear in a more general
framework, hence the importance of reminding of its independent, basic and
more general origin. We begin by recalling that, if $\Psi_{1}(\mathbf{r},t)$
and $\Psi_{2}(\mathbf{r},t)$ are solutions of the time-dependent
Schr\"{o}dinger (or Dirac) equation for a quantum particle of charge $q$ that
moves (as a test particle) in two distinct sets of (predetermined and
classical) vector and scalar potentials ($\mathbf{A}_{1},\phi_{1}$) and
($\mathbf{A}_{2},\phi_{2}$), that are generally spatially- and
temporally-dependent [and such that, at the spacetime point of observation
$(\mathbf{r},t)$, the magnetic and electric fields are the same in the two
systems], then we have the following formal connection between the solutions
(wavefunctions) of the two systems%

\begin{equation}
\Psi_{2}(\mathbf{r},t)=e^{i\frac{q}{\hbar c}\Lambda(\mathbf{r},t)}\Psi
_{1}(\mathbf{r},t), \label{Basic1}%
\end{equation}
with the function $\Lambda(\mathbf{r},t)$ required to satisfy%

\begin{equation}
\nabla\Lambda(\mathbf{r,t})=\mathbf{A}_{2}(\mathbf{r},t)-\mathbf{A}%
_{1}(\mathbf{r},t)\qquad and\qquad-\frac{1}{c}\frac{\partial\Lambda
(\mathbf{r},t)}{\partial t}=\phi_{2}\left(  \mathbf{r},t\right)  -\phi
_{1}(\mathbf{r},t). \label{Basic11}%
\end{equation}
The above property can be immediately proven by substituting each $\Psi_{i}$
into its corresponding ($i_{th}$) time-dependent Schr\"{o}dinger equation
(namely with the set of potentials ($\mathbf{A}_{i}(\mathbf{r},t),\phi
_{i}(\mathbf{r},t)$))$\boldsymbol{:}$ one can then easily see that
(\ref{Basic1}) and (\ref{Basic11}) guarantee that both Schr\"{o}dinger
equations are indeed satisfied together (after cancellation of a global phase
factor in system 2, see Appendix A for a detailed proof). [In addition, the
equality of all classical fields at the observation point, namely
$\mathbf{B}_{2}(\mathbf{r},t)=\nabla\times\mathbf{A}_{2}(\mathbf{r}%
,t)=\nabla\times\mathbf{A}_{1}(\mathbf{r},t)=$ $\mathbf{B}_{1}(\mathbf{r},t)$
for the magnetic fields and $\mathbf{E}_{2}(\mathbf{r},t)=-\nabla\phi
_{2}\left(  \mathbf{r},t\right)  -\frac{1}{c}\frac{\partial\mathbf{A}%
_{2}(\mathbf{r},t)}{\partial t}=-\nabla\phi_{1}\left(  \mathbf{r},t\right)
-\frac{1}{c}\frac{\partial\mathbf{A}_{1}(\mathbf{r},t)}{\partial t}%
=\mathbf{E}_{1}(\mathbf{r},t)$ for the electric fields, is obviously
consistent with all equations (\ref{Basic11}) (as is easy to see if we take
the \textit{curl} of the 1st and the \textit{grad} of the 2nd) $-$ provided,
at least, that $\Lambda(\mathbf{r},t)$ is such that interchanges of partial
derivatives with respect to all spatial and temporal variables (at the point
$\left(  \mathbf{r},t\right)  $) are allowed].

As already mentioned, the above fact is of course well-known within the
framework of the theory of quantum mechanical gauge transformations (the usual
case being for $\mathbf{A}_{1}=\phi_{1}=0$, hence for a mapping from a system
with no potentials)$\boldsymbol{;}$ but in that framework, these
transformations are supposed to connect (or map) two \textit{physically
equivalent systems} (more rigorously, this being true for ordinary gauge
transformations, in which case the function $\Lambda(\mathbf{r},t)$, the
so-called gauge function, is unique (single-valued) in spacetime coordinates).
In a formally similar manner, the above argument is also often used in the
context of the so-called \textquotedblleft singular gauge
transformations\textquotedblright, where $\Lambda$ is multiple-valued, but the
above equality of classical fields is still imposed (at the observation point,
which always lies in a physically accessible region)$\boldsymbol{;}$ then the
above simple phase mapping (at all points of the physically accessible
spacetime region, that always and everywhere experience equal fields) leads to
the standard phenomena of the Aharonov-Bohm type, reviewed later below, where
\textit{unequal fields in physically-inaccessible regions} have observable
consequences. However, we should keep in mind that that above property
((\ref{Basic1}) and (\ref{Basic11}) taken together) can be \textit{more
generally valid} $-$ and in this article we will present cases (and closed
analytical results for the appropriate phase connection $\Lambda
(\mathbf{r},t)$) that actually connect (or map) two systems (in the sense of
(\ref{Basic1})) that are \textit{neither physically equivalent nor of the
usual Aharonov-Bohm type}. And naturally, because of the above provision of
field equalities at the observation point, it will turn out that any
nonequivalence of the two systems will involve \textit{remote} (although
\textit{physically accessible}) regions of spacetime, namely regions that do
\textit{not} contain the observation point $(\mathbf{r},t)$ (and in which
regions, as we shall see, the classical fields experienced by the particle may
be \textit{different} in the two systems).

\section{Motivation}

One may wonder on the actual reasons why one should be looking for more
general cases of a simple phase mapping of the type (\ref{Basic1}) between
\textit{nonequivalent} systems. To answer this, let us take a step back and
first recall some simple and well-known results that originate from the above
phase mapping. It is standard knowledge, for example, that, if we want to find
solutions $\Psi(x,t)$ of the $t$-dependent Schr\"{o}dinger (or Dirac) equation
for a quantum particle (of charge $q)$ that moves along a (generally curved)
one-dimensional (1-D) path, and in the presence (somewhere in the embedding
3-dimensional (3-D) space) of a fairly localized (and time-independent)
classical magnetic flux $\Phi$ that \textit{does \textbf{not} pass through any
point of the path,} then we formally have%

\begin{equation}
\Psi(x,t)^{(\Phi)}\sim e^{i\frac{q}{\hbar c}\int_{x_{0}}^{x}\mathbf{A(r}%
^{\prime})\cdot d\mathbf{r}^{\prime}}\Psi(x,t)^{(0)} \label{standard}%
\end{equation}
(the dummy variable $\mathbf{r}^{\prime}$ describing points along the 1-D
path, and the term \textquotedblleft formally\textquotedblright\ meaning that
the above is valid \textit{before imposition of any boundary conditions }(that
are meant to be imposed only on the system with the flux $\Phi$\cite{ring})).
In (\ref{standard}), $\Psi(x,t)^{(0)}$ is a formal solution of the same system
in the case of absence of any potentials (hence $\Phi=0$ everywhere in the 3-D
space). The above holds because, for \textit{all points} $\mathbf{r}^{\prime}$
of the 1-D path, the particle experiences a vector potential
$\mathbf{A(r\prime)}$ of the form $\mathbf{A(r\prime)=\nabla}^{\prime
}\mathbf{\Lambda(r}^{\prime}\mathbf{)}$ (since the magnetic field is
$\mathbf{\nabla}^{\prime}\mathbf{\times A(r}^{\prime})=0$\textbf{ }for
\textit{all} $\mathbf{r}^{\prime}$, by assumption), in combination with the
above phase-mapping (with a phase $\frac{q}{\hbar c}\mathbf{\Lambda(r)}$)
between two quantum systems, one in the presence and one in the absence of a
vector potential (i.e. the potentials in (\ref{Basic11}) being $\mathbf{A}%
_{1}=0$ and $\mathbf{A}_{2}=$ $\mathbf{A}$, together with $\phi_{2}=\phi
_{1}=0$ if we decide to attribute everything to vector potentials only). In
this particular system, the obvious $\mathbf{\Lambda(r)}$ that solves the
above $\mathbf{A(\mathbf{r})=\nabla\Lambda(r)}$ (for \textbf{all} points of
the 1-D space available to the particle) is indeed $\mathbf{\Lambda
(r)=\Lambda(r}_{0}\mathbf{)+}\int_{\mathbf{r}_{0}}^{\mathbf{r}}\mathbf{A(r}%
^{\prime})\cdot d\mathbf{r}^{\prime}$, and this gives (\ref{standard}) (if
$\mathbf{r}$ denotes the above point $x$ of observation and $\mathbf{r}_{0}$
the arbitrary initial point $x_{0}$ (both lying on the physical path), and if
the constant $\mathbf{\Lambda(r}_{0}\mathbf{)}$ is taken to be zero).

What if, however, some parts of the magnetic field that comprise the magnetic
flux $\Phi$ actually \textit{pass through} some points or a whole region
(interval) of the path available to the particle? In such a case, the above
standard argument is not valid (as $\mathbf{A}$ cannot be written as a grad at
any point of the interval where the magnetic field $\mathbf{\nabla\times
A}\neq0$). Are there however general results that we can still write for
$\Psi^{(\Phi)}(x,t)$, if the spatial point of observation $x$ is again outside
the interval with the nonvanishing magnetic field? Or, what if in the previous
problems, the magnetic flux (either remote, or partly passing through the
path) is time-dependent $\Phi(t)$? (In that case then, there exists in general
an additional electric field $E$ induced by Faraday's law of Induction on
points of the path, and the usual gauge transformation argument is once again
not valid).

Returning to another standard (solvable) case (which is actually the
\textquotedblleft dual\textquotedblright\ or the \textquotedblleft electric
analog\textquotedblright\ of the above), if along the 1-D physical path the
particle experiences only a spatially-uniform (but generally time-dependent)
classical scalar potential $\phi(t)$, we can again formally map $\Psi
(x,t)^{(\phi)}$ to a potential-free solution $\Psi(x,t)^{(0)}$, through a
$\Lambda(t)$ that now solves $-\frac{1}{c}\frac{\partial\Lambda(t)}{\partial
t}=\phi(t)$, and this gives $\Lambda(t)=\Lambda(t_{0})-c\int_{t_{0}}^{t}%
\phi(t^{\prime})dt^{\prime}$, leading to the \textquotedblleft electric
analog\textquotedblright\ of (\ref{standard}), namely%

\begin{equation}
\Psi(x,t)^{(\phi)}\sim e^{-i\frac{q}{\hbar}\int_{t_{0}}^{t}\phi(t^{\prime
})dt^{\prime}}\Psi(x,t)^{(0)} \label{standard2}%
\end{equation}
with obvious notation. (Notice that, for either of the two mapped systems in
this problem, the electric field is zero at all points of the path). What if,
however, the scalar potential has also some $x$-dependence along the path
(that leads to an electric field (in a certain interval) that the particle
passes through)? In such a case, the above standard argument is again not
valid. Are there however general results that we can still write for
$\Psi(x,t)^{(\phi)}$, if the spatial point of observation $x$ is again outside
the interval with the nonvanishing electric field?

We state here directly that this article will provide affirmative answers to
questions of the type posed above, by actually giving the corresponding
general results in closed analytical forms.

At this point it is also useful to briefly reconsider the earlier mentioned
case, namely of a time-dependent $\Phi(t)$ that is remote to the 1-D physical
path, because in this manner we can immediately provide another motivation for
the present work$\boldsymbol{:}$ this time-dependent problem is surrounded
with a number of important misconceptions in the literature (the same being
true about its electric analog, as we shall see)$\boldsymbol{:}$ the formal
solution that is usually written down for a $\Phi(t)$ is again (\ref{standard}%
), namely the above spatial line integral of $\mathbf{A}$, in spite of the
fact that $\mathbf{A}$ is now $t$-dependent$\boldsymbol{;}$ the problem then
is that, because of the first of (\ref{Basic11}), $\Lambda$ must now have a
$t$-dependence and, from the second of (\ref{Basic11}), there must necessarily
be scalar potentials involved in the problem (which have been by force set to
zero, in our pre-determined mapping between vector potentials only). Having
decided to use systems that experience only vector (and not scalar)
potentials, the correct solution cannot be simply a trivial $t$-dependent
extension of (\ref{standard}). A corresponding error is usually made in the
electric dual of the above, namely in cases that involve $\mathbf{r}%
$-dependent scalar potentials, where (\ref{standard2}) is still erroneously
used, giving an $\mathbf{r}$-dependent $\Lambda$, although this would
necessarily lead to the involvement of vector potentials (through the first of
(\ref{Basic11}) and the $\mathbf{r}$-dependence of $\Lambda$) that have been
neglected from the beginning $-$ a situation (and an error) that appears, in
exactly this form, in the description of the so-called electric Aharonov-Bohm
effect, as we shall see.

Speaking of errors in the literature, it might here be the perfect place to
also point to the reader the most common misleading statement often made in
the literature (and again, for notational simplicity, we restrict our
attention to a one-dimensional system, with spatial variable $x$, although the
statement is obviously generalizable to (and often made for systems of) higher
dimensionality by properly using line integrals over arbitrary curves in
space)$\boldsymbol{:}$ It is usually stated [e.g. in Brown \&
Holland\cite{BrownHolland}, see i.e. their eq. (57) applied for vanishing
boost velocity $\mathbf{v}=0$] that the general gauge function that connects
(through a phase factor $e^{i\frac{q}{\hbar c}\Lambda(x,t)}$) the
wavefunctions of a quantum system with no potentials (i.e. with a set of
potentials $(\mathbf{0},0)$) to the wavefunctions of a quantum system that
moves in vector potential $\boldsymbol{A}(x,t)$ and scalar potential
$\phi(x,t)$ (i.e. in a set of potentials $(\boldsymbol{A},\phi)$) is the
obvious combination (and a natural extension) of (\ref{standard}) and
(\ref{standard2}), namely%

\begin{equation}
\Lambda(x,t)=\Lambda(x_{0},t_{0})+%
%TCIMACRO{\dint \limits_{x_{0}}^{x}}%
%BeginExpansion
{\displaystyle\int\limits_{x_{0}}^{x}}
%EndExpansion
\boldsymbol{A}(x^{\prime},t)dx^{\prime}-c%
%TCIMACRO{\dint \limits_{t_{0}}^{t}}%
%BeginExpansion
{\displaystyle\int\limits_{t_{0}}^{t}}
%EndExpansion
\phi(x,t^{\prime})dt^{\prime}, \label{wrong}%
\end{equation}
which, however, is \textbf{incorrect } for $x$ and $t$ uncorrelated
variables$\boldsymbol{:}$ it does \textbf{not }satisfy the standard
system\textbf{\ }of gauge transformation equations%

\begin{equation}
\nabla\Lambda(x,t)=\mathbf{A(}x,t)\mathbf{\qquad and\qquad-}\frac{1}{c}%
\frac{\partial\Lambda(x,t)}{\partial t}=\phi(x,t). \label{BasicPDE}%
\end{equation}
The reader can easily see why: (i) when the $\nabla$ operator acts on
eq.(\ref{wrong}), it gives the correct $A(x,t)$ from the 1st term, but it also
gives some annoying additional nonzero quantity from the 2nd term (that
survives because of the $x$-dependence of $\phi$); hence it invalidates the
first of the basic system (\ref{BasicPDE}). (ii) Similarly, when the
$\mathbf{-}\frac{1}{c}\frac{\partial}{\partial t}$ operator acts on
eq.(\ref{wrong}), it gives the correct $\phi(x,t)$ from the 2nd term, but it
also gives some annoying additional nonzero quantity from the 1st term (that
survives because of the $t$-dependence of $\mathbf{A}$); hence it invalidates
the second of the basic system (\ref{BasicPDE}). It is only when $\mathbf{A}$
is $t$-independent, and $\phi$ is spatially-independent, that eq.(\ref{wrong})
is correct (as the above annoying terms do not appear and the basic system is
satisfied). [An alternative form that is also given in the literature is again
eq.(\ref{wrong}), but with the variables that are not integrated over
implicitly assumed to belong to the initial point (hence a $t_{0}$ replaces
$t$ in $\boldsymbol{A}$, and simultaneously an $x_{0}$ replaces $x$ in $\phi
$). However, one can see again that the system (\ref{BasicPDE}) is not
satisfied (the above differential operators, when acted on $\Lambda$, give
$\boldsymbol{A}(x,t_{0})$ and $\phi(x_{0},t)$, hence not the values of the
potentials at the point of observation $(x,t)$ as they should), this not being
an acceptable solution either].

What is the problem here? Or, better put, what is the deeper reason for the
above inconsistency? The short answer is the uncritical use of Dirac phase
factors that come from path-integral treatments. It is indeed obvious that the
form (\ref{wrong}) that is often used in the literature (in canonical
(non-path-integral) formulations where $x$ and $t$ are \textbf{uncorrelated}
variables (and not correlated to produce a path $x(t)$)) \textit{is not
generally correct}, and that is one of the main points that has motivated this
work. We will find \textit{generalized results} that actually \textit{correct}
eq.(\ref{wrong}) through extra nonlocal terms, and through the proper
appearance of $x_{0}$ and $t_{0}$ (as in eq.(\ref{LambdaStatic1}) and
eq.(\ref{LambdaStatic2}) to be found later in Section VII), and these are the
\textbf{exact} ones (namely the exact $\Lambda(x,t),$ that at the end, upon
action of $\nabla$ and $\mathbf{-}\frac{1}{c}\frac{\partial}{\partial t}$
satisfies exactly the basic system (\ref{BasicPDE})). And the formulation that
gives these results is generalized later in the article, for $\Lambda(x,y)$
(in the 2-D static case) and also for $\Lambda(x,y,t)$ (in the full dynamical
2-D case), and leads to the exact (nontrivial) forms of the phase function
$\Lambda$ that satisfy (in all cases) the system (\ref{BasicPDE}) $-$ with
\textit{the direct verification\ (i.e. proof, by \textquotedblleft going
backwards\textquotedblright, that these forms are indeed the exact solutions)
being given in the main text, }and with the rigorous mathematical derivations
being presented in Appendices\textit{.}

This article gives a full exploration of issues related to the above
motivating discussion, by pointing to a \textquotedblleft
practical\textquotedblright\ (and generalized) use of gauge transformation
mapping techniques, that at the end lead to these generalized (and, at first
sight, unexpected) solutions for the general form of $\Lambda$. For cases such
as the ones discussed above, or even more involved ones, there still appears
to exist a simple phase mapping (between two inequivalent systems), but the
phase connection $\Lambda$ seems to contain not only integrals of potentials,
but also \textquotedblleft fluxes\textquotedblright\ of the classical fields
from \textit{remote} spacetime regions. The above mentioned systems are the
simplest ones where these new results can be applied, but apart from this, the
present investigation seems to lead to a number of nontrivial corrections of
misleading (or even incorrect) reports of the above type in the literature,
that are not at all marginal (and are due to an incorrect use of a
path-integral viewpoint in an otherwise canonical framework). The generalized
$\Lambda$-forms also lead to an honest resolution of earlier \textquotedblleft
paradoxes\textquotedblright\ (involving Relativistic Causality), and in some
cases to a new interpretation of known semiclassical experimental
observations, corrections of certain sign-errors in the literature, and
nontrivial extensions of earlier semiclassical results to general (even
completely delocalized) states. Most importantly, however, the new formulation
seems capable of treating issues of Causality in time-dependent slit
experiments as we shall see\cite{jphysa}.

\section{The standard background}

Having clearly stated that the above phase-mapping can be a more general
property than is usually realized, let us briefly recall the history of the
standard framework that such mappings have appeared. It was essentially from
Weyl's work (1929), but also from independent proposals by Schr\"{o}dinger
(1922), Fock (1927) and London (1927)\cite{Weyl}, that it was firmly
established that a simple unitary (U(1)) phase factor certainly connects two
quantum systems, when these are gauge-equivalent (and then the phase that
connects their wavefunctions is basically the gauge function of an ordinary
gauge transformation). A simple unitary phase-connection of this type is also
reserved for quantum systems moving in multiple-connected spacetimes (with
enclosed appropriately defined \textquotedblleft fluxes\textquotedblright\ in
the physically inaccessible regions) the corresponding \textquotedblleft gauge
transformation\textquotedblright\ termed singular, and the corresponding
\textquotedblleft gauge function\textquotedblright\ now being multiple-valued
(although the wavefunctions of the \textquotedblleft final\textquotedblright%
\ (mapped) system are still single-valued) leading to phenomena of the
Aharonov-Bohm type\cite{Kleinert}. As we already stressed, in the present work
we report on a phase-connection between systems that are \textit{not}
\textquotedblleft equivalent\textquotedblright\ (in the sense of the above
two), since they can go through \textit{different} classical fields in remote
regions of space and/or time, and we give explicit forms of the appropriate
\textquotedblleft gauge functions\textquotedblright. The results are exact, in
analytical form, and they generalize the standard Dirac phase factors derived
from path integral treatments (that are very often used in an incorrect way as
we shall demonstrate); apart from a discussion of such misconceptions
propagating in the literature, we also give actual applications of the new
results in static and time-dependent experiments involving quantum charged
particles inside electromagnetic potentials, both of the Aharonov-Bohm type
(i.e. with inaccessible fields and their fluxes) but also with the particles
actually passing through classical magnetic and electric fields, and even
being in completely general (spead-out) quantum states (and not necessarily
narrow wavepackets in semiclassical motion as typically done in the literature).

In Section I we recalled the mapping (\ref{Basic1}) and (\ref{Basic11}),
together with an outline of its proof. Let us here briefly restate it for
completeness and for a better flow of the arguments that will
follow$\boldsymbol{:}$ if $\Psi_{1}(\mathbf{r},t)$ and $\Psi_{2}%
(\mathbf{r},t)$ are solutions of the time-dependent Schr\"{o}dinger (or Dirac)
equation for a quantum particle of charge $q$ that moves (as a test particle)
in two distinct sets of (predetermined and classical) vector and scalar
potentials ($\mathbf{A}_{1},\phi_{1}$) and ($\mathbf{A}_{2},\phi_{2}$) [and
such that, at the spacetime point of observation $(\mathbf{r},t)$, the
magnetic and electric fields are the same in the two systems], then we have
the following formal connection between the solutions (wavefunctions) of the
two systems%

\begin{equation}
\Psi_{2}(\mathbf{r},t)=e^{i\frac{q}{\hbar c}\Lambda(\mathbf{r},t)}\Psi
_{1}(\mathbf{r},t), \label{UsualPsi}%
\end{equation}
\ (by \textquotedblleft formal\textquotedblright\ connection meaning again
that this is valid \textit{before imposition of any boundary conditions}, and
these will have to be applied only on \textit{our} system, i.e. system 2 (not
necessarily on system 1, see \cite{ring})). In (\ref{UsualPsi}) $\Lambda
(\mathbf{r},t)$ is a general function that must satisfy (\ref{Basic11}), which
we here want to treat as a system of Partial Differential Equations (PDEs)%

\begin{equation}
\nabla\Lambda(\mathbf{r,t})=\mathbf{A}_{2}(\mathbf{r},t)-\mathbf{A}%
_{1}(\mathbf{r},t)\qquad and\qquad-\frac{1}{c}\frac{\partial\Lambda
(\mathbf{r},t)}{\partial t}=\phi_{2}\left(  \mathbf{r},t\right)  -\phi
_{1}(\mathbf{r},t). \label{gaugetransf}%
\end{equation}
\ \qquad In the static case, and if, for simplicity, we start from system 1
being completely free of potentials ($\mathbf{A}_{1}=\phi_{1}=0$), the
wavefunctions of the particle in system 2 (moving only in a static vector
potential $A(\mathbf{r})$) will acquire an extra phase with an appropriate
\textquotedblleft gauge function\textquotedblright\ $\Lambda(\mathbf{r})$ that
must satisfy$\ $%

\begin{equation}
\nabla\Lambda(\mathbf{r})=\mathbf{A}(\mathbf{r}). \label{usualgrad}%
\end{equation}
\ 

As mentioned in Section II, the standard (and widely-used) solution of this is
the line integral%

\begin{equation}
\Lambda(\mathbf{r})=\Lambda(\mathbf{r}_{\mathbf{0}})+\int_{\mathbf{r}_{0}%
}^{\mathbf{r}}\mathbf{A}(\mathbf{r}^{\prime})\boldsymbol{.}d\mathbf{r}%
^{\prime} \label{usualABLambda}%
\end{equation}
(which, by considering two paths encircling an enclosed inaccessible magnetic
flux, formally leads to the well-known magnetic Aharonov-Bohm effect\cite{AB}%
). It should however be stressed again that the above is only true if
(\ref{usualgrad}) is valid for \textbf{all} points $\mathbf{r}$ of the region
where the particle moves, i.e. if the particle in system 2 moves (as a narrow
wavepacket) always outside magnetic fields ($\nabla\times\mathbf{A}=0$
everywhere) as already emphasized in Section II. Similarly, if the particle in
system 2 moves only in a spatially homogeneous scalar potential $\phi(t)$, the
appropriate $\Lambda$ must satisfy\ \ \ \
\begin{equation}
-\frac{1}{c}\frac{\partial\Lambda(t)}{\partial t}=\phi(t), \label{usualdt}%
\end{equation}
the standard solution being%

\begin{equation}
\Lambda(t)=\Lambda(t_{0})-c\int_{t_{0}}^{t}\phi(t^{\prime})dt^{\prime}
\label{usualelectricABLambda}%
\end{equation}
that gives the extra phase acquired by system 2 (this result formally leading
to the electric Aharonov-Bohm effect\cite{AB,Peshkin} by applying it to two
equipotential regions, such as two metallic cages held in distinct
time-dependent scalar potentials). Once again, it should be stressed that the
above is only true if (\ref{usualdt}) \ (and the assumed spatial homogeneity
of the scalar potential $\phi$ and of $\Lambda$) is valid at \textbf{all}
times $t$ of interest, i.e. if the particle in system 2 moves (as a narrow
wavepacket) always outside electric fields ($\mathbf{E}=-\nabla\phi-\frac
{1}{c}\frac{\partial\mathbf{A}}{\partial t}=0$ \ at all times). (In the
electric Aharonov-Bohm setup, the above is ensured by the fact that $t$ lies
in an interval of a finite duration $T$\ for which the potentials are turned
on, in combination with the narrowness of the wavepacket$\boldsymbol{;}$ this
guarantees that, during $T$, the particle has vanishing probability of being
at the edges of the cage where the potential starts having a spatial
dependence. The reader is referred to Appendix B of Peshkin\cite{Peshkin} that
demonstrates the intricasies of the electric Aharonov-Bohm effect, to which we
return with an important comment at the end of Section XII).

\bigskip

In the present work, we relax the above assumptions and present more general
solutions of the system of PDEs (\ref{gaugetransf}), covering cases where the
particle is \textit{not} necessarily a narrow wavepacket (it can actually be
in completely delocalized states) and is \textit{not} excluded from
\textit{remote} regions (in spacetime) of nonvanishing (or, more generally, of
unequal) fields (magnetic or electric), regions therefore that are actually
accessible to the particle (hence non-Aharonov-Bohm cases $-$ or even
combinations of spatial multiple-connectivity of the magnetic Aharonov-Bohm
type, but simultaneous simple-connectivity in spacetime (i.e. in the
$(x,t)-$plane)). We find analytically nonlocal influences of these remote
fields on $\Lambda(\mathbf{r},t)$ (with $(\mathbf{r},t)$ always denoting the
observation point in spacetime), and therefore on the phases of wavefunctions
at $(\mathbf{r},t)$, that seem to have a number of important
consequences$\boldsymbol{:}$ they provide (i) a natural justification of
earlier or more recent experimental observations for semiclassical behavior in
simple-connected space (when the particles pass through nonvanishing (or, more
generally, unequal) magnetic or electric fields), and also extensions to more
general cases of delocalized (spread-out) quantum states, (ii) a nontrivial
correction to misleading or even incorrect results that appear often in the
literature (errors that are of a different type for static and for
time-dependent situations), and (iii) a natural remedy for Causality
\textquotedblleft paradoxes\textquotedblright\ in time-dependent Aharonov-Bohm
configurations. These dynamical nonlocalities of quantum mechanical phases
seem to have escaped from all path-integral treatments. An extension of the
method applied to Maxwell's equations governing the behavior of the fields
(rather than to the equations that give the \textquotedblleft gauge
function\textquotedblright\ $\Lambda$) indicates that these \textit{phase
nonlocalities} demonstrate in part a causal propagation of phases of quantum
mechanical wavefunctions in the Schr\"{o}dinger Picture (and these can address
causal issues in time-dependent single- \textit{vs} double-slit experiments,
an area that seems to have recently attracted
attention\cite{Tollaksen,Popescu,He}).

\section{Example of Generalized Solutions in Static Cases}

\bigskip By way of an example we immediately provide a simple result that will
be found later (in Section X) for a static $(x,y)$-case (and for
simple-connected space) that generalizes the standard Dirac phase
(\ref{usualABLambda}), namely
\begin{equation}
\Lambda(x,y)=\Lambda(x_{0},y_{0})+\int_{x_{0}}^{x}A_{x}(x^{\prime
},y)dx^{\prime}+\int_{y_{0}}^{y}A_{y}(\mathbf{x}_{\mathbf{0}},y^{\prime
})dy^{\prime}+\left\{
%TCIMACRO{\dint \limits_{y_{0}}^{y}}%
%BeginExpansion
{\displaystyle\int\limits_{y_{0}}^{y}}
%EndExpansion
dy^{\prime}%
%TCIMACRO{\dint \limits_{x_{0}}^{x}}%
%BeginExpansion
{\displaystyle\int\limits_{x_{0}}^{x}}
%EndExpansion
dx^{\prime}B_{z}(x^{\prime},y^{\prime})+g(x)\right\}  \label{static1}%
\end{equation}

\[
with\text{ \ }g(x)\text{ \ }chosen\text{ \ }so\text{ \ }that\ \ \ \left\{
%TCIMACRO{\dint \limits_{y_{0}}^{y}}%
%BeginExpansion
{\displaystyle\int\limits_{y_{0}}^{y}}
%EndExpansion
dy^{\prime}%
%TCIMACRO{\dint \limits_{x_{0}}^{x}}%
%BeginExpansion
{\displaystyle\int\limits_{x_{0}}^{x}}
%EndExpansion
dx^{\prime}B_{z}(x^{\prime},y^{\prime})+g(x)\right\}  \boldsymbol{:}\text{ is
}\mathsf{independent\ of\ }\ x.
\]

In the above $B_{z}={\huge (}\boldsymbol{B}_{2}-\boldsymbol{B}_{1}%
{\huge )}_{z}$ is the difference of perpendicular magnetic fields in the two
systems, which can be nonvanishing at regions remote to the observation point
$(x,y)$ (see below). (It is reminded that at the point of observation
$B(x,y)=0$, signifying the essential fact that the fields in the two systems
are identical (recall that $B_{z}=B_{2z}-B_{1z}$) at the point of observation
$(x,y)$). The reader should note that the first 3 terms of (\ref{static1}) are
the Dirac phase (\ref{usualABLambda}) along two perpendicular segments that
continuously connect the initial point $(x_{0},y_{0})$ to the point of
observation $(x,y)$, \textit{in a clockwise sense }(see for example the
red-arrow paths in Fig.1(b)). But apart from this Dirac phase, we also have
nonlocal contributions from $B_{z}$ and its flux within the \textquotedblleft
observation rectangle\textquotedblright\ (see i.e. the rectangle being formed
by the red- and green-arrow paths in Fig.1(b)). Below we will directly verify
that (\ref{static1}) is indeed a solution of (\ref{usualgrad}) (even for
$B_{z}(x^{\prime},y^{\prime})\neq0$ for $(x^{\prime},y^{\prime})\neq(x,y)$),
i.e. of the system \bigskip of PDEs%

\begin{equation}
\frac{\partial\Lambda(x,y)}{\partial x}=A_{x}(x,y)\qquad and\qquad
\frac{\partial\Lambda(x,y)}{\partial y}=A_{y}(x,y). \label{usualgradcomps}%
\end{equation}
(Although the former is trivially satisfied (at least for cases where
interchanges of integrals with derivatives are legitimate), for the latter to
be verified one needs to simply substitute $\frac{\partial A_{x}(x^{\prime
},y)}{\partial y}$ with \ $\frac{\partial A_{y}(x^{\prime},y)}{\partial
x^{\prime}}-B_{z}(x^{\prime},y)$ \ and then carry out the integration with
respect to $x^{\prime}$ -- the reader should note the crucial appearance (and
proper placement) of $\mathbf{x}_{\mathbf{0}}\boldsymbol{\ }$ in
(\ref{static1}) for the verification of both (\ref{usualgradcomps})). It
should be noted again that (\ref{static1}) satisfies (\ref{usualgradcomps})
even for nonzero $B_{z}$ (i.e. when the particle passes through unequal
magnetic fields in remote regions), in contradistinction to the standard
result (\ref{usualABLambda}). (For the benefit of the reader we clearly
provide in the next Section all the steps for the direct verification of
(\ref{static1})).

\bigskip

Equivalently, we will later obtain the result%

\begin{equation}
\Lambda(x,y)=\Lambda(x_{0},y_{0})+\int_{x_{0}}^{x}A_{x}(x^{\prime}%
,\mathbf{y}_{\mathbf{0}})dx^{\prime}+\int_{y_{0}}^{y}A_{y}(x,y^{\prime
})dy^{\prime}+\left\{  {\Huge -}%
%TCIMACRO{\dint \limits_{x_{0}}^{x}}%
%BeginExpansion
{\displaystyle\int\limits_{x_{0}}^{x}}
%EndExpansion
dx^{\prime}%
%TCIMACRO{\dint \limits_{y_{0}}^{y}}%
%BeginExpansion
{\displaystyle\int\limits_{y_{0}}^{y}}
%EndExpansion
dy^{\prime}B_{z}(x^{\prime},y^{\prime})+h(y)\right\}  \label{static2}%
\end{equation}%
\[
with\text{ \ }h(y)\text{ \ }chosen\text{ \ }so\text{ \ }that\text{
\ \ }\left\{  {\Huge -}%
%TCIMACRO{\dint \limits_{x_{0}}^{x}}%
%BeginExpansion
{\displaystyle\int\limits_{x_{0}}^{x}}
%EndExpansion
dx^{\prime}%
%TCIMACRO{\dint \limits_{y_{0}}^{y}}%
%BeginExpansion
{\displaystyle\int\limits_{y_{0}}^{y}}
%EndExpansion
dy^{\prime}B_{z}(x^{\prime},y^{\prime})+h(y)\right\}  \boldsymbol{:}\text{ is
}\mathsf{independent\ of\ }\ y,
\]
and again the reader should note that, apart from the first 3 terms (the Dirac
phase (\ref{usualABLambda}) along the two other (alternative) perpendicular
segments (connecting $(x_{0},y_{0})$ to $(x,y)$), now \textit{in a
counterclockwise sense }(the green-arrow paths in Fig.1(b))), we also have
nonlocal contributions from the flux of $B_{z}$ that is enclosed within the
same \textquotedblleft observation rectangle\textquotedblright\ (that is
naturally defined by the four segments of the two solutions (Fig.1(b))). It
can also be easily verified that (\ref{static2}) also satisfies the system
(\ref{usualgradcomps}) (for this $\frac{\partial A_{y}(x,y^{\prime})}{\partial
x}$ \ needs to be substituted with $\frac{\partial A_{x}(x,y^{\prime}%
)}{\partial y^{\prime}}+B_{z}(x,y^{\prime})$ \ and then integration with
respect to $y^{\prime}$ needs to be carried out, with the proper appearance
(and placement)\ of \ $\mathbf{y}_{\mathbf{0}}$ in (\ref{static2}) now being
the crucial element $-$ see direct verification in the next Section).

In all the above, $A_{x}$ and $A_{y}$ are the Cartesian components of
$\ \boldsymbol{A}\mathbf{(r)}=\boldsymbol{A}(x,y)=\boldsymbol{A}%
_{\boldsymbol{2}}(\mathbf{r})-\boldsymbol{A}_{\boldsymbol{1}}(\mathbf{r})$,
and, as already mentioned, \ $B_{z}$ \ is the difference between
(perpendicular) magnetic fields that the two systems may experience in regions
that \textit{do not contain} the observation point $(x,y)$ (i.e.
$B_{z}(x^{\prime},y^{\prime})={\huge (}\boldsymbol{B}_{\boldsymbol{2}%
}(x^{\prime},y^{\prime})-\boldsymbol{B}_{\boldsymbol{1}}(x^{\prime},y^{\prime
}){\huge )}_{z}=\frac{\partial A_{y}(x^{\prime},y^{\prime})}{\partial
x^{\prime}}-$ $\frac{\partial A_{x}(x^{\prime},y^{\prime})}{\partial
y^{\prime}}$, \ and, although at the point of observation $(x,y)$ we have
$B_{z}(x,y)=0$ (already emphasized in the Introductory Sections), this
$B_{z}(x^{\prime},y^{\prime})$ can be nonzero for \ $(x^{\prime},y^{\prime
})\neq(x,y)$). It should be noted that it is because of $B_{z}(x,y)=0$ that
the functions $g(x)$ and $h(y)$ of (\ref{static1}) and (\ref{static2}) can be
found, and the new solutions therefore exist (and are nontrivial). For the
impatient reader, simple physical examples with the associated analytical
forms of $g(x)$ and $h(y)$ are given in detail in Section VI.

In the present and following Section we place the emphasis in pointing out
(and proving) the new solutions (that apparently have been widely overlooked
in the literature). In later Sections, we will see that these results actually
demonstrate that the passage of particles through magnetic fields has the
effect of cancelling Aharonov-Bohm types of phases. And in the special case of
narrow wavepacket states in semiclassical motion we will provide an
understanding of this cancellation in terms of the experimentally observed
compatibility (or consistency) between the Aharonov-Bohm fringe-displacement
and the trajectory-deflection due to the Lorentz force. (The corresponding
\textquotedblleft electric analog\textquotedblright\ of this consistency of
semiclassical trajectory-behavior will also be pointed out $-$ through an
elementary physical picture that is here given for the first time, to the best
of our knowledge). However, the above cancellations are true even for
completely delocalized states (and the deeper reason for this will be obvious
from the derivation of the above two solutions (presented in detail in
Appendix D) $-$ the origin of the cancellations being essentially the
single-valuedness of phases for simple-connected space). Therefore, these
generalized results go beyond the usual Aharonov-Bohm behaviors reviewed in
the Introductory Sections, and give an extended description of physical
systems in more involved physical arrangements (where the particle also passes
through remote fields). [It is also simply added here that cancellations of
the above type will be extended and generalized further to cases that also
involve the time variable $t$; these will be presented in later Sections, with
a detailed mathematical derivation given in Appendix G. Interpreted in a
different way, such cancellations $-$ through the new nonlocal terms $-$ will
take away the \textquotedblleft mystery\textquotedblright\ of why certain
classical arguments (based on past history and the Faraday's law of Induction)
seem to \textquotedblleft work\textquotedblright\ (give the correct
Aharonov-Bohm phases in static arrangements, by invoking the \textit{history}
of how the experimental set up was built at earlier times). Although we will
give very general methods of deriving even more generalized results,
applicable to a large number of physical cases, we will mostly restrict
attention to detailed applications that provide a natural remedy for earlier
discussed \textquotedblleft paradoxes\textquotedblright\ in time-dependent
Aharonov-Bohm configurations, and are indicative of an even more general
causal propagation of wavefunction phases in the Schr\"{o}dinger Picture].

\section{\textbf{Elementary\ Verification\ Of\ Above Solutions (even for cases
with }$\boldsymbol{B}_{z}\boldsymbol{\neq0}$ in remote regions\textbf{)}}

In static cases, and simple-connected space, let us call our solution
(\ref{static1}) $\Lambda_{1}$, namely%

\[
\Lambda_{1}(x,y)=\Lambda_{1}(x_{0},y_{0})+%
%TCIMACRO{\dint \limits_{x_{0}}^{x}}%
%BeginExpansion
{\displaystyle\int\limits_{x_{0}}^{x}}
%EndExpansion
A_{x}(x^{\prime},y)dx^{\prime}+%
%TCIMACRO{\dint \limits_{y_{0}}^{y}}%
%BeginExpansion
{\displaystyle\int\limits_{y_{0}}^{y}}
%EndExpansion
A_{y}(x_{0},y^{\prime})dy^{\prime}+\left\{
%TCIMACRO{\dint \limits_{y_{0}}^{y}}%
%BeginExpansion
{\displaystyle\int\limits_{y_{0}}^{y}}
%EndExpansion
dy^{\prime}%
%TCIMACRO{\dint \limits_{x_{0}}^{x}}%
%BeginExpansion
{\displaystyle\int\limits_{x_{0}}^{x}}
%EndExpansion
dx^{\prime}B_{z}(x^{\prime},y^{\prime})+g(x)\right\}
\]
with\ $g(x)$ chosen so that\ $\ \left\{
%TCIMACRO{\dint \limits^{y}}%
%BeginExpansion
{\displaystyle\int\limits^{y}}
%EndExpansion%
%TCIMACRO{\dint \limits^{x}}%
%BeginExpansion
{\displaystyle\int\limits^{x}}
%EndExpansion
B_{z}+g(x)\right\}  \boldsymbol{\ }$is independent of $x.$

Verification that it solves the system of PDEs (\ref{usualgradcomps}) (even
for $B_{z}(x^{\prime},y^{\prime})\neq0$ for $(x^{\prime},y^{\prime})\neq
(x,y)$)$\boldsymbol{:}$

\textbf{A) \ }$\frac{\partial\Lambda_{1}(x,y)}{\partial x}=A_{x}(x,y)\qquad
$satisfied trivially\qquad$\checkmark$

(because $\left\{  ...\right\}  $ is independent of $x$).

\textbf{B) \ }$\frac{\partial\Lambda_{1}(x,y)}{\partial y}=%
%TCIMACRO{\dint \limits_{x_{0}}^{x}}%
%BeginExpansion
{\displaystyle\int\limits_{x_{0}}^{x}}
%EndExpansion
\frac{\partial A_{x}(x^{\prime},y)}{\partial y}dx^{\prime}+A_{y}(x_{0},y)+%
%TCIMACRO{\dint \limits_{x_{0}}^{x}}%
%BeginExpansion
{\displaystyle\int\limits_{x_{0}}^{x}}
%EndExpansion
B_{z}(x^{\prime},y)dx^{\prime}+\frac{\partial g(x)}{\partial y},$

(the last term being trivially zero, $\frac{\partial g(x)}{\partial y}=0$),
and then with the substitution

$\frac{\partial A_{x}(x^{\prime},y)}{\partial y}=\frac{\partial A_{y}%
(x^{\prime},y)}{\partial x^{\prime}}-B_{z}(x^{\prime},y)$

we obtain

$\frac{\partial\Lambda_{1}(x,y)}{\partial y}=%
%TCIMACRO{\dint \limits_{x_{0}}^{x}}%
%BeginExpansion
{\displaystyle\int\limits_{x_{0}}^{x}}
%EndExpansion
\frac{\partial A_{y}(x^{\prime},y)}{\partial x^{\prime}}dx^{\prime}-%
%TCIMACRO{\dint \limits_{x_{0}}^{x}}%
%BeginExpansion
{\displaystyle\int\limits_{x_{0}}^{x}}
%EndExpansion
B_{z}(x^{\prime},y)dx^{\prime}+A_{y}(x_{0},y)+%
%TCIMACRO{\dint \limits_{x_{0}}^{x}}%
%BeginExpansion
{\displaystyle\int\limits_{x_{0}}^{x}}
%EndExpansion
B_{z}(x^{\prime},y)dx^{\prime}.$

(i) We see that the 2nd and 4th terms of the right-hand-side (rhs)
\textit{cancel each other}, and

(ii) the 1st term of the rhs is $%
%TCIMACRO{\dint \limits_{x_{0}}^{x}}%
%BeginExpansion
{\displaystyle\int\limits_{x_{0}}^{x}}
%EndExpansion
\frac{\partial A_{y}(x^{\prime},y)}{\partial x^{\prime}}dx^{\prime}%
=A_{y}(x,y)-A_{y}(x_{0},y).$

Hence finally

$\frac{\partial\Lambda_{1}(x,y)}{\partial y}=A_{y}(x,y).\qquad\checkmark$

\bigskip

We have directly shown therefore (by \textquotedblleft going
backwards\textquotedblright) that the basic system of PDEs
(\ref{usualgradcomps}) is indeed satisfied by our \textbf{generalized}
solution $\Lambda_{1}(x,y),$ \textbf{even for any nonzero} $B_{z}(x^{\prime
},y^{\prime})$\ (in regions $(x^{\prime},y^{\prime})\neq(x,y)\boldsymbol{;}$
recall that always $B_{z}(x,y)=0$). To fully appreciate the above simple
proof, the reader is urged to look at the cases of \textquotedblleft
striped\textquotedblright\ $B_{z}$-distributions in the next Section, the
point of observation $(x,y)$ always lying outside the strips, so that the
above function $g(x)$ can easily be determined, and the new solutions really
\textit{exist }- and they are nontrivial. (As already noted, a formal
derivation of the above solution (\ref{static1}) - rather than its above
\textquotedblleft backwards\textquotedblright\ verification - is given in
Appendix D).

In a completely analogous way, one can easily see that our alternative
solution (eq.(\ref{static2})) also satisfies the basic system of PDEs above.
Indeed, if we call our second static solution (eq.(\ref{static2}))
$\Lambda_{2}$, namely%

\[
\Lambda_{2}(x,y)=\Lambda_{2}(x_{0},y_{0})+\int_{x_{0}}^{x}A_{x}(x^{\prime
},y_{0})dx^{\prime}+\int_{y_{0}}^{y}A_{y}(x,y^{\prime})dy^{\prime}+\left\{
{\Huge -}%
%TCIMACRO{\dint \limits_{x_{0}}^{x}}%
%BeginExpansion
{\displaystyle\int\limits_{x_{0}}^{x}}
%EndExpansion
dx^{\prime}%
%TCIMACRO{\dint \limits_{y_{0}}^{y}}%
%BeginExpansion
{\displaystyle\int\limits_{y_{0}}^{y}}
%EndExpansion
dy^{\prime}B_{z}(x^{\prime},y^{\prime})+h(y)\right\}
\]
with $h(y)$ chosen so that\ $\ \left\{  {\Huge -}%
%TCIMACRO{\dint \limits^{x}}%
%BeginExpansion
{\displaystyle\int\limits^{x}}
%EndExpansion%
%TCIMACRO{\dint \limits^{y}}%
%BeginExpansion
{\displaystyle\int\limits^{y}}
%EndExpansion
B_{z}+h(y)\right\}  \boldsymbol{:}$ is independent of\textsf{ }$y$,

then we have (even for $B_{z}(x^{\prime},y^{\prime})\neq0$ for $(x^{\prime
},y^{\prime})\neq(x,y)$)$\boldsymbol{:}$

\bigskip

\textbf{A) \ }$\frac{\partial\Lambda_{2}(x,y)}{\partial y}=A_{y}(x,y)\qquad
$satisfied trivially\qquad$\checkmark$

(because $\left\{  ...\right\}  $ is independent of $y$).

\bigskip

\textbf{B) \ }$\frac{\partial\Lambda_{2}(x,y)}{\partial x}=A_{x}(x,y_{0})+%
%TCIMACRO{\dint \limits_{y_{0}}^{y}}%
%BeginExpansion
{\displaystyle\int\limits_{y_{0}}^{y}}
%EndExpansion
\frac{\partial A_{y}(x,y^{\prime})}{\partial x}dy^{\prime}-%
%TCIMACRO{\dint \limits_{y_{0}}^{y}}%
%BeginExpansion
{\displaystyle\int\limits_{y_{0}}^{y}}
%EndExpansion
B_{z}(x,y^{\prime})dy^{\prime}+\frac{\partial h(y)}{\partial x},$

(the last term being trivially zero, $\frac{\partial h(y)}{\partial x}=0$),
and then with the substitution

$\frac{\partial A_{y}(x,y^{\prime})}{\partial x}=\frac{\partial A_{x}%
(x,y^{\prime})}{\partial y^{\prime}}+B_{z}(x,y^{\prime})$

we obtain

$\frac{\partial\Lambda_{2}(x,y)}{\partial x}=A_{x}(x,y_{0})+%
%TCIMACRO{\dint \limits_{y_{0}}^{y}}%
%BeginExpansion
{\displaystyle\int\limits_{y_{0}}^{y}}
%EndExpansion
\frac{\partial A_{x}(x,y^{\prime})}{\partial y^{\prime}}dy^{\prime}+%
%TCIMACRO{\dint \limits_{y_{0}}^{y}}%
%BeginExpansion
{\displaystyle\int\limits_{y_{0}}^{y}}
%EndExpansion
B_{z}(x,y^{\prime})dy^{\prime}-%
%TCIMACRO{\dint \limits_{y_{0}}^{y}}%
%BeginExpansion
{\displaystyle\int\limits_{y_{0}}^{y}}
%EndExpansion
B_{z}(x,y^{\prime})dy^{\prime}.$

(i) We see that the last two terms of the rhs \textit{cancel each other}, and

(ii) the 2nd term of the rhs is $%
%TCIMACRO{\dint \limits_{y_{0}}^{y}}%
%BeginExpansion
{\displaystyle\int\limits_{y_{0}}^{y}}
%EndExpansion
\frac{\partial A_{x}(x,y^{\prime})}{\partial y^{\prime}}dy^{\prime}%
=A_{x}(x,y)-A_{x}(x,y_{0}).$

Hence finally

$\frac{\partial\Lambda_{2}(x,y)}{\partial x}=A_{x}(x,y).\qquad\checkmark$

\bigskip Once again, all the above are true for any nonzero $B_{z}(x^{\prime
},y^{\prime})$ (in regions $(x^{\prime},y^{\prime})\neq(x,y)$). And a clear
understanding of this proof (through the actual analytical form of $h(y)$) is
given by the \textquotedblleft striped\textquotedblright\ examples of next Section.

\section{Simple Examples$\boldsymbol{:}$ New results shown in explicit form}

To see how the above solutions appear in nontrivial cases (and how they give
completely new results, i.e. \textit{not differing from the usual ones
}(\textit{i.e. from the Dirac phase})\textit{ by a mere constant}) let us
first take examples of striped $B_{z}$-distributions in space$\boldsymbol{:}$

\textbf{(a)} For the case of an extended \textit{vertical} strip - parallel to
the $y$-axis, such as in Fig.1(a) (imagine $t$ replaced by $y$) (i.e. for the
case that the particle has actually passed through nonzero $B_{z}$, hence
through \textit{different} magnetic fields in the two (mapped) systems), then,
for $x$ located outside (and on the right of) the strip, the quantity$\
%TCIMACRO{\dint \limits_{y_{0}}^{y}}%
%BeginExpansion
{\displaystyle\int\limits_{y_{0}}^{y}}
%EndExpansion
dy^{\prime}%
%TCIMACRO{\dint \limits_{x_{0}}^{x}}%
%BeginExpansion
{\displaystyle\int\limits_{x_{0}}^{x}}
%EndExpansion
dx^{\prime}B_{z}(x^{\prime},y^{\prime})$ in\ $\Lambda_{1}$ \textit{is already
independent of}$\ x$\ (since a displacement of the $(x,y)$-corner of the
rectangle to the right, along the $x$-direction, does not change the enclosed
magnetic flux $-$ see Fig. 1(a) for the analogous $(x,t)$-case that will be
discussed in following Sections)$\boldsymbol{.}$ Indeed, in this case the
above quantity (the enclosed flux within the \textquotedblleft observation
rectangle\textquotedblright) does not depend on the $x$-position of the
observation point, but on the positioning of the boundaries of the $B_{z}%
$-distribution in the $x$-direction (better, on the constant width of the
strip) $-$ as the $x$-integral does not give any further contribution when the
dummy variable $x^{\prime}$ goes out of the strip. In fact, in this case the
enclosed flux depends on $y$ as we discuss below (but, again, not on $x$).
Hence, for this case, the function $g(x)$ can be easily determined: it can be
taken as $g(x)=0$ (up to a constant $C$), because then the condition for
$g(x)$ stated in the solution (\ref{static1}) (namely, that the quantity in
brackets must be independent of $x$) is indeed satisfied.%

%TCIMACRO{\FRAME{ftbpF}{3.2681in}{0.9703in}{0in}{}{}{figure1.eps}%
%{\special{ language "Scientific Word";  type "GRAPHIC";
%maintain-aspect-ratio TRUE;  display "USEDEF";  valid_file "F";
%width 3.2681in;  height 0.9703in;  depth 0in;  original-width 3.2232in;
%original-height 0.9375in;  cropleft "0";  croptop "1";  cropright "1";
%cropbottom "0";  filename 'Figure1.eps';file-properties "XNPEU";}}}%
%BeginExpansion
\begin{figure}[ptb]%
\centering
\includegraphics[
height=0.9703in,
width=3.2681in
]%
{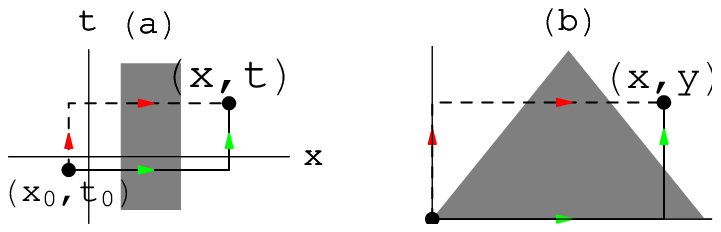}%
\end{figure}
%EndExpansion

We see therefore above that for this setup, the nonlocal term in the solution
\textit{survives} (the quantity in brackets is nonvanishing), but \textit{it
is not constant}$\boldsymbol{:}$ as already noted, this enclosed flux depends
on $y$ (since the enclosed flux \textit{does change} with a displacement of
the $(x,y)$-corner of the rectangle upwards, along the $y$-direction, as the
$y$-integral \textit{is} affected by the positioning of $y$ $-$ the higher the
positioning of the observation point the more flux is enclosed inside the
observation rectangle). Hence, by looking at the alternative solution
$\Lambda_{2}(x,y),$ the quantity$\
%TCIMACRO{\dint \limits_{x_{0}}^{x}}%
%BeginExpansion
{\displaystyle\int\limits_{x_{0}}^{x}}
%EndExpansion
dx^{\prime}%
%TCIMACRO{\dint \limits_{y_{0}}^{y}}%
%BeginExpansion
{\displaystyle\int\limits_{y_{0}}^{y}}
%EndExpansion
dy^{\prime}B_{z}(x^{\prime},y^{\prime})$\ is \textit{dependent on}$\ y$, so
that $h(y)$ must be chosen as $\ h(y)=+%
%TCIMACRO{\dint \limits_{x_{0}}^{x}}%
%BeginExpansion
{\displaystyle\int\limits_{x_{0}}^{x}}
%EndExpansion
dx^{\prime}%
%TCIMACRO{\dint \limits_{y_{0}}^{y}}%
%BeginExpansion
{\displaystyle\int\limits_{y_{0}}^{y}}
%EndExpansion
dy^{\prime}B_{z}(x^{\prime},y^{\prime})$ (up to the same constant $C$)\ in
order to \textit{cancel} this $y$-dependence, so that its own condition stated
in the solution (\ref{static2}) (namely, that the quantity in brackets must be
independent of $y$) is indeed satisfied$\boldsymbol{;}$ as a result, the
quantity in brackets in solution $\Lambda_{2}$ disappears and there is no
nonlocal contribution in $\Lambda_{2}$ (for $C=0$). (Of course, if we had used
a $C\neq0$, the nonlocal contributions would be shared between the two
solutions in a different manner, but without changing the Physics when we take
the \textit{difference} of the two solutions (see below)). [The crucial point
in the above is, once again that, because $B_{z}=0$ at $(x,y)$, any
displacement of this observation point to the right does\textit{ not }change
the flux enclosed inside the \textquotedblleft observation
rectangle\textquotedblright$\boldsymbol{;}$ and this makes the new solutions
(i.e. the functions $g(x)$ and $h(y)$) exist].

With these choices of $h(y)$ and $g(x)$, we already have new results (compared
to the standard ones of the integrals of potentials). I.e. one of the two
solutions, namely $\Lambda_{1}$ \textbf{is} affected nonlocally by the
enclosed flux (and this flux is \textbf{not} constant). Spelled out clearly,
the two results are:%

\[
\Lambda_{1}(x,y)=\Lambda_{1}(x_{0},y_{0})+%
%TCIMACRO{\dint \limits_{x_{0}}^{x}}%
%BeginExpansion
{\displaystyle\int\limits_{x_{0}}^{x}}
%EndExpansion
A_{x}(x^{\prime},y)dx^{\prime}+%
%TCIMACRO{\dint \limits_{y_{0}}^{y}}%
%BeginExpansion
{\displaystyle\int\limits_{y_{0}}^{y}}
%EndExpansion
A_{y}(x_{0},y^{\prime})dy^{\prime}+%
%TCIMACRO{\dint \limits_{y_{0}}^{y}}%
%BeginExpansion
{\displaystyle\int\limits_{y_{0}}^{y}}
%EndExpansion
dy^{\prime}%
%TCIMACRO{\dint \limits_{x_{0}}^{x}}%
%BeginExpansion
{\displaystyle\int\limits_{x_{0}}^{x}}
%EndExpansion
dx^{\prime}B_{z}(x^{\prime},y^{\prime})+C
\]

\[
\Lambda_{2}(x,y)=\Lambda_{2}(x_{0},y_{0})+\int_{x_{0}}^{x}A_{x}(x^{\prime
},y_{0})dx^{\prime}+\int_{y_{0}}^{y}A_{y}(x,y^{\prime})dy^{\prime}+C.
\]
And now it is easy to note that, if we subtract the two solutions $\Lambda
_{1}$ and $\Lambda_{2}$, the result is \textit{zero} (because the line
integrals of the vector potential $\boldsymbol{A}$ in the two solutions are in
opposite senses in the $(x,y)$ plane, hence their difference leads to a
\textit{closed} line integral of $\boldsymbol{A}$, which is in turn equal to
the enclosed magnetic flux, and this flux always happens to be of opposite
sign from that of the enclosed flux that explicitly appears as a
nonlocal\ contribution of the $B_{z}$-fields (i.e. the term that survives in
$\Lambda_{1}$ above)$\boldsymbol{.}$ Hennce, the two solutions are
\textit{equal}. [We of course everywhere assumed, as usual, single-valuedness
of $\Lambda$ at the initial point $(x_{0},y_{0})$, i.e. $\Lambda_{1}%
(x_{0},y_{0})=\Lambda_{2}(x_{0},y_{0});$ matters of multivaluedness of
$\Lambda$ at the observation point $(x,y)$ will be addressed later (Section X)].

The reader should probably note that, formally speaking, the above
\textit{equality} of the two solutions is due to the fact that the
$x$-independent quantity in brackets of the 1st solution (\ref{static1}) is
equal to the function $h(y)$ of the 2nd solution (\ref{static2}), and the
$y$-independent quantity in brackets of the 2nd solution (\ref{static2}) is
equal to the function $g(x)$ of the first solution (\ref{static1}). This will
turn out to be a general behavioral pattern of the two solutions in
simple-connected space, that will be valid for any shape of $B_{z}%
$-distribution, as will be shown in Section X.

This vanishing of $\Lambda_{1}(x,y)-\Lambda_{2}(x,y)$ is a cancellation effect
that is emphasized further (and generally proved) later below (and can be
viewed as a generalization of the Werner \& Brill experimental
observations\cite{WernerBrill} to general delocalized states, as will be fully
discussed, in completely \textit{physical terms}, in Section X). It basically
originates from the single-valuedness of $\Lambda$ at $(x,y)$ for
simple-connected space. This effect is generalized even further in later
Sections (i.e. also to cases of combined spacetime variables $x,y,t$) for the
van Kampen thought-experiment\cite{vanKampen} (where we will have a
combination of spatial multiple-connectivity at an initial instant $t_{0}$,
and simple-connectivity in $(x,t)$ and $(y,t)$ planes).

\textbf{(b) }In the \textquotedblleft dual case\textquotedblright\ of an
extended \textit{horizontal} strip - parallel to the $x$-axis, the proper
choices (for $y$ above the strip) are basically reverse (i.e. we can now take
$h(y)=0$ \ and $g(x)=-%
%TCIMACRO{\dint \limits_{y_{0}}^{y}}%
%BeginExpansion
{\displaystyle\int\limits_{y_{0}}^{y}}
%EndExpansion
dy^{\prime}%
%TCIMACRO{\dint \limits_{x_{0}}^{x}}%
%BeginExpansion
{\displaystyle\int\limits_{x_{0}}^{x}}
%EndExpansion
dx^{\prime}B_{z}(x^{\prime},y^{\prime})$ $\ $(since the flux enclosed in the
rectangle now depends on $x$, but not on $y$), with both choices always up to
a common constant) and once again we can easily see, upon subtraction of the
two solutions, a similar cancellation effect. In this case as well, the
results are again new (a \ nonlocal term survives now in $\Lambda_{2}$). Again
spelled out clearly, these are:%

\[
\Lambda_{1}(x,y)=\Lambda_{1}(x_{0},y_{0})+%
%TCIMACRO{\dint \limits_{x_{0}}^{x}}%
%BeginExpansion
{\displaystyle\int\limits_{x_{0}}^{x}}
%EndExpansion
A_{x}(x^{\prime},y)dx^{\prime}+%
%TCIMACRO{\dint \limits_{y_{0}}^{y}}%
%BeginExpansion
{\displaystyle\int\limits_{y_{0}}^{y}}
%EndExpansion
A_{y}(x_{0},y^{\prime})dy^{\prime}+C
\]

\[
\Lambda_{2}(x,y)=\Lambda_{2}(x_{0},y_{0})+\int_{x_{0}}^{x}A_{x}(x^{\prime
},y_{0})dx^{\prime}+\int_{y_{0}}^{y}A_{y}(x,y^{\prime})dy^{\prime}{\Huge -}%
%TCIMACRO{\dint \limits_{x_{0}}^{x}}%
%BeginExpansion
{\displaystyle\int\limits_{x_{0}}^{x}}
%EndExpansion
dx^{\prime}%
%TCIMACRO{\dint \limits_{y_{0}}^{y}}%
%BeginExpansion
{\displaystyle\int\limits_{y_{0}}^{y}}
%EndExpansion
dy^{\prime}B_{z}(x^{\prime},y^{\prime})+C
\]
(their difference also being zero -- a generalized Werner \& Brill
cancellation (see Section X for further discussion)). Again here the crucial
point is that, because the $B_{z}$-configuration does \textit{not contain the
point} $(x,y)$, a displacement of this observation point upwards does
\textit{not} change the flux inside the \textquotedblleft observation
rectangle\textquotedblright$\boldsymbol{;}$ this makes the new solutions (i.e
the functions $g(x)$ and $h(y)$) exist.

\bigskip

\textbf{(c) }If we want cases that are more involved (i.e. with the nonlocal
contributions appearing nontrivially in \textbf{both} solutions $\Lambda_{1}$
and $\Lambda_{2}$ and with $g(x)$ and $h(y)$ not being \textquotedblleft
immediately visible\textquotedblright), we must consider different shapes of
$B_{z}$-distributions. One such case is a triangular one that is shown in
Fig.1(b) (for simplicity an equilateral triangle, and with the initial point
$(x_{0},y_{0})=(0,0)$) and with the point of observation $(x,y)$ being fairly
close to the triangle's right side as in the Figure. Note that for such a
configuration, the part of the magnetic flux that is inside the
\textquotedblleft observation rectangle\textquotedblright\ (defined by the
right upper corner $(x,y)$) depends on \textbf{both} $x$ \textbf{and} $y$. It
turns out, however, that this $(x$ and $y)-$dependent enclosed flux can be
written as a sum of separate $x$- and $y$-contributions, so that appropriate
$g(x)$ and $h(y)$ can still be found (each one of them must be chosen so that
it only cancels the corresponding variable's dependence of the enclosed flux).
For a homogeneous $B_{z}$ it is a rather straightforward exercise to determine
this enclosed part, i.e. the common area between the observation rectangle and
the equilateral triangle, and from this we can find the appropriate $g(x)$
that will cancel the $x$-dependence, and the appropriate $h(y)$ that will
cancel the $y$-dependence. These appropriate choices turn out to be

\begin{equation}
g(x)=B_{z}\left[  \mathbf{-(}\sqrt{3}ax-\frac{\sqrt{3}}{2}x^{2})+\frac
{\sqrt{3}}{4}a^{2}\right]  +C \label{triangular1}%
\end{equation}
and%

\begin{equation}
h(y)=B_{z}\left[  \mathbf{(}ay-\frac{y^{2}}{\sqrt{3}})-\frac{\sqrt{3}}{4}%
a^{2}\right]  +C \label{triangular2}%
\end{equation}
with $a$ being the side of the equilateral triangle. (We again note that a
physical arbitrariness described by the common constant $C$, does not play any
role when we take the difference of the two solutions (\ref{static1}) and
(\ref{static2})). We should emphasize that expressions (\ref{triangular1}) and
(\ref{triangular2}), if combined with (\ref{static1}) or (\ref{static2}), give
the nontrivial nonlocal contributions of the difference $B_{z}$ of the remote
magnetic fields on $\Lambda$ of each solution (hence on the phase of the
wavefunction of each wavepacket travelling along each path) at the observation
point $(x,y)$, that always lies outside the $B_{z}$-triangle. (We mention
again that in the case of completely spread-out states, the equality of the
two solutions at the observation point essentially demonstrates the uniqueness
(single-valuedness) of the phase in simple-connected space). Further physical
discussion of the above cancellations, and a semiclassical interpretation, is
given later in Section X and in the final Sections of the paper.

Finally, in more \textquotedblleft difficult\textquotedblright\ geometries,
i.e. when the shape of the $B_{z}$-distribution is such that the enclosed flux
does \textit{not}\textbf{ }decouple in a sum of separate $x$- and
$y$-contributions, \ such as cases of circularly shaped $B_{z}$-distributions,
it is advantageous to solve the system (\ref{usualgrad}) directly in
non-Cartesian (i.e. polar) coordinates. The results of such a procedure in
polar coordinates are given in Appendix E (see eqs (\ref{polar1}%
)-(\ref{polarcond2})). A general comment that can be made for general shapes
is that, depending on the geometry of shape of the $B_{z}$-distribution, an
appropriate change of variables (to a new coordinate system) may first be
needed, so that generalized solutions of the system (\ref{usualgrad}) can be
found (namely, so that the enclosed flux inside the \textit{transformed}
observation rectangle (i.e. a slice of an annular section in the case of polar
coordinates) can be written as a sum of separate (transformed) variables), and
then the same methodology (as in the above Cartesian cases) can be followed.

Finally, the reader who may wonder how the usual Aharonov-Bohm result comes
out from the present formulation (that contains the additional nonlocal
terms), must first read Section X (where the most general solutions
(\ref{Lambda(x,y)1}) and (\ref{Lambda(x,y)4}) for this 2-D static case are
derived, containing additional \textquotedblleft
multiplicities\textquotedblright) and then Appendix F that gives a detailed
derivation of the standard Aharonov-Bohm results.

\section{Example of Generalized Solutions in Dynamical Cases \textbf{(with
electric fields, even for cases with }$\boldsymbol{E\neq0}$ in remote
spacetime regions\textbf{)}}

Let us now look at a case with full time-dependence. Although it may be
possible to guess the corresponding generalized results, i.e. for a
spatially-one-dimensional $(x,t)$-problem (by appropriate Euclidian rotation
of the above solutions in 4-D space), let us nevertheless start from the
beginning and give a full physical discussion $-$ as this \textit{is} the case
that actually led us to the above generalized solutions, and a case associated
with a number of misleading arguments (and often incorrect results)
propagating in the literature.

Let us then first focus on the simplest case of one-dimensional quantum
systems, i.e. a single quantum particle of charge $\ q$, but in the presence
of the most general (spatially nonuniform and time-dependent) vector and
scalar potentials, and ask the following question$\boldsymbol{:}$ what is the
gauge function $\Lambda(x,t)$ that takes us from (maps) a system with
potentials $A_{1}(x,t)$ and $\phi_{1}(x,t)$ to a system with potentials
$A_{2}(x,t)$ and $\phi_{2}(x,t)$ (meaning the usual mapping (\ref{UsualPsi})
between the wavefunctions of the two systems through the phase factor
$\frac{q}{\hbar c}\Lambda(x,t)$)? [Once again we should keep in mind that for
this mapping to be possible we \textit{must} assume that at the point $(x,t)$
of observation (or \textquotedblleft measurement\textquotedblright\ of
$\Lambda$ or the wavefunction $\Psi$) \ we have equal electric fields
($E_{i}=-\nabla\phi_{i}-\frac{1}{c}\frac{\partial A_{i}}{\partial t}$), namely%

\begin{equation}
-\frac{\partial\phi_{2}(x,t)}{\partial x}-\frac{1}{c}\frac{\partial
A_{2}(x,t)}{\partial t}=-\frac{\partial\phi_{1}(x,t)}{\partial x}-\frac{1}%
{c}\frac{\partial A_{1}(x,t)}{\partial t} \label{Efield}%
\end{equation}
(so that the $A$'s and $\phi$'s in (\ref{Efield}) can indeed satisfy the basic
system of equations (\ref{gaugetransf}), or equivalently, of the system of
equations (\ref{xt-BasicSystem}) below $-$ as can be seen by taking the
$\frac{1}{c}\frac{\partial}{\partial t}$ of the 1st and the $\frac{\partial
}{\partial x}$ of the 2nd of the system (\ref{xt-BasicSystem}) and adding them
together)$\boldsymbol{.}$ But again, we will \textit{not} exclude the
possibility of the two systems passing through \textit{different }electric
fields in different regions of spacetime, i.e. for $(x^{\prime},t^{\prime
})\neq(x,t)$. In fact, this possibility \textbf{will come out naturally} from
a careful solution of the basic system (\ref{xt-BasicSystem})$\boldsymbol{;}$
it is for example straightforward for the reader to immediately verify that
the results (\ref{LambdaStatic1}) or (\ref{LambdaStatic2}) that will be
derived below (and will contain contributions of electric field-differences
from remote regions of spacetime) indeed satisfy the basic input system of
equations (\ref{xt-BasicSystem}), something that will be explicitly verified
in the next Section].

Returning to the question on the appropriate $\Lambda$ that takes us from the
set $(A_{1},\phi_{1})$ to the set $(A_{2},\phi_{2})$, we note that, in cases
of static vector potentials ($A(x)$'s) \textit{and} spatially uniform scalar
potentials ($\phi(t)$'s) the answer usually given is the well-known%

\begin{equation}
\Lambda(x,t)=\Lambda(x_{0},t_{0})+\int_{x_{0}}^{x}A(x^{\prime})dx^{\prime
}-c\int_{t_{0}}^{t}\phi(t^{\prime})dt^{\prime} \label{LambdaUsual}%
\end{equation}
with $\ A(x)=A_{2}(x)-A_{1}(x)$ \ and $\ \phi(t)=\phi_{2}(t)-\phi_{1}(t)$ (and
it can be viewed as a combination of (\ref{usualABLambda}) and
(\ref{usualelectricABLambda}) (or of (\ref{standard}) and (\ref{standard2})),
being immediately applicable to the description of cases of \textit{combined}
magnetic and electric Aharonov-Bohm effects reviewed in the Introductory Sections).

\bigskip In the most general case (and with the variables$\ x$\ and$\ t$ being
\textbf{completely uncorrelated}), it is often stated in the literature [as in
eq. (57) of Ref.\cite{BrownHolland}, taken for $\mathbf{v}=0$, a very good
example to point to, since that article does not use a path-integral language,
but a canonical formulation with uncorrelated variables] that the appropriate
$\Lambda$ has a form that is a plausible extention of (\ref{LambdaUsual}), namely%

\begin{equation}
\Lambda(x,t)=\Lambda(x_{0},t_{0})+%
%TCIMACRO{\dint \limits_{x_{0}}^{x}}%
%BeginExpansion
{\displaystyle\int\limits_{x_{0}}^{x}}
%EndExpansion
\left[  A_{2}(x^{\prime},t)-A_{1}(x^{\prime},t)\right]  dx^{\prime}-c%
%TCIMACRO{\dint \limits_{t_{0}}^{t}}%
%BeginExpansion
{\displaystyle\int\limits_{t_{0}}^{t}}
%EndExpansion
\left[  \phi_{2}(x,t^{\prime})-\phi_{1}(x,t^{\prime})\right]  dt^{\prime},
\label{BrownHolland}%
\end{equation}
and as already pointed out in Section II, this form is certainly
\ \textit{incorrect} \ for uncorrelated variables $x$ and $t$ \ (the reader
can easily verify that the system of equations (\ref{xt-BasicSystem}) below is
${\large not}$ satisfied by (\ref{BrownHolland}) $-$ see again Section II if
needed, especially the paragraph after eq.(\ref{BasicPDE}))$\boldsymbol{.}$ We
will find in the present work that the correct form consists of two major
modifications$\boldsymbol{:}$ (i) The first leads to the natural appearance of
a \textit{path} that continuously connects initial and final points in
spacetime, a property that (\ref{BrownHolland}) \textit{does not have}
[indeed, if the integration curves of (\ref{BrownHolland}) are drawn in the
$(x,t)$-plane, they do \textit{not} form a continuous path from $(x_{0}%
,t_{0})$ to $(x,t)$]. The reader can immediately see eqs.(\ref{LambdaStatic1})
and (\ref{LambdaStatic2}) that will be given below for the corrected
\textquotedblleft path-forms\textquotedblright\ in the line integrals of
potentials (and these are represented by the red-arrow and green-arrow paths
of Fig.1(a)). (ii) And the second modification is highly
nontrivial$\boldsymbol{:}$ it consists of nonlocal contributions of classical
electric field-differences from remote regions of spacetime. We will discuss
below the consequences of these terms and we will later show that such
nonlocal contributions also appear (in an extended form) in more general
situations, i.e. they are also present in higher spatial dimensionality (and
they then also involve remote magnetic fields in combination with the electric
ones)$\boldsymbol{;}$ these lead to modifications of ordinary Aharonov-Bohm
behaviors or have other important consequences, one of them being a natural
remedy of Causality \textquotedblleft paradoxes\textquotedblright\ in
time-dependent Aharonov-Bohm experiments.

The form (\ref{BrownHolland}) commonly used is of course motivated by the
well-known Wu \& Yang\cite{WuYang} nonintegrable phase factor, that has a
phase equal to \ $\int A_{\mu}dx^{\mu}=\int Adx-c\int\phi dt$, \ a form that
appears naturally within\ the framework of path-integral treatments, or
generally in physical situations where narrow wavepackets are implicitly
assumed for the quantum particle\textbf{:} the integrals appearing in
(\ref{BrownHolland}) are then taken along particle trajectories (hence spatial
and temporal variables \textit{not} being uncorrelated, but being connected in
a particular manner $x(t)$ to produce the path$\boldsymbol{;}$ all integrals
are therefore basically only time-integrals). But even then,
eq.(\ref{BrownHolland}) is valid only when these trajectories are always (in
time) and everywhere (in space) inside identical classical fields for the two
(mapped) systems. Here, however, we will be focusing on what a canonical (and
not a path-integral or other semiclassical) treatment leads to$\boldsymbol{;}$
this will cover the general case of arbitrary wavefunctions that can even be
completely spread-out in space, and will also allow the particle to travel
through different electric fields for the two systems in remote spacetime
regions (e.g. $E_{2}\left(  x,t^{\prime}\right)  \neq E_{1}\left(  x,t\right)
$ \ if $\ \ t^{\prime}<t$ \ etc.).

\bigskip

It is therefore clear that in order to find the appropriate $\ \Lambda(x,t)$
\ that answers the question posed above in full generality will require a
careful solution of the system of PDEs (\ref{gaugetransf}), applied to only
one spatial variable, namely%

\begin{equation}
\frac{\partial\Lambda(x,t)}{\partial x}=A(x,t)\qquad and\qquad-\frac{1}%
{c}\frac{\partial\Lambda(x,t)}{\partial t}=\phi\left(  x,t\right)
\label{xt-BasicSystem}%
\end{equation}
(with $\ A(x,t)=A_{2}(x,t)-A_{1}(x,t)$ \ and $\ \phi\left(  x,t\right)
=\phi_{2}\left(  x,t\right)  -\phi_{1}\left(  x,t\right)  $). This system of
PDEs is solved in detail in Appendix B, and leads to two alternative
solutions, one being%

\begin{equation}
\Lambda(x,t)=\Lambda(x_{0},t_{0})+%
%TCIMACRO{\dint \limits_{x_{0}}^{x}}%
%BeginExpansion
{\displaystyle\int\limits_{x_{0}}^{x}}
%EndExpansion
A(x^{\prime},t)dx^{\prime}-c%
%TCIMACRO{\dint \limits_{t_{0}}^{t}}%
%BeginExpansion
{\displaystyle\int\limits_{t_{0}}^{t}}
%EndExpansion
\phi(x_{0},t^{\prime})dt^{\prime}+\left\{  c%
%TCIMACRO{\dint \limits_{t_{0}}^{t}}%
%BeginExpansion
{\displaystyle\int\limits_{t_{0}}^{t}}
%EndExpansion
dt^{\prime}%
%TCIMACRO{\dint \limits_{x_{0}}^{x}}%
%BeginExpansion
{\displaystyle\int\limits_{x_{0}}^{x}}
%EndExpansion
dx^{\prime}E(x^{\prime},t^{\prime})+g(x)\right\}  +\tau(t_{0})
\label{LambdaStatic1}%
\end{equation}
with $\ g(x)$ chosen\ so that the quantity $\ \left\{  c%
%TCIMACRO{\dint \limits_{t_{0}}^{t}}%
%BeginExpansion
{\displaystyle\int\limits_{t_{0}}^{t}}
%EndExpansion
dt^{\prime}%
%TCIMACRO{\dint \limits_{x_{0}}^{x}}%
%BeginExpansion
{\displaystyle\int\limits_{x_{0}}^{x}}
%EndExpansion
dx^{\prime}E(x^{\prime},t^{\prime})+g(x)\right\}  $ \ is independent of $x$,
and (from an inverted route of integrations) the other solution being%

\begin{equation}
\Lambda(x,t)=\Lambda(x_{0},t_{0})+%
%TCIMACRO{\dint \limits_{x_{0}}^{x}}%
%BeginExpansion
{\displaystyle\int\limits_{x_{0}}^{x}}
%EndExpansion
A(x^{\prime},t_{0})dx^{\prime}-c\int_{t_{0}}^{t}\phi\left(  x,t^{\prime
}\right)  dt^{\prime}+\left\{  -c%
%TCIMACRO{\dint \limits_{x_{0}}^{x}}%
%BeginExpansion
{\displaystyle\int\limits_{x_{0}}^{x}}
%EndExpansion
dx^{\prime}%
%TCIMACRO{\dint \limits_{t_{0}}^{t}}%
%BeginExpansion
{\displaystyle\int\limits_{t_{0}}^{t}}
%EndExpansion
dt^{\prime}E(x^{\prime},t^{\prime})+\hat{g}(t)\right\}  +\chi(x_{0})
\label{LambdaStatic2}%
\end{equation}
with $\ \hat{g}(t)$ \ chosen so that the quantity \ $\left\{  -c%
%TCIMACRO{\dint \limits_{x_{0}}^{x}}%
%BeginExpansion
{\displaystyle\int\limits_{x_{0}}^{x}}
%EndExpansion
dx^{\prime}%
%TCIMACRO{\dint \limits_{t_{0}}^{t}}%
%BeginExpansion
{\displaystyle\int\limits_{t_{0}}^{t}}
%EndExpansion
dt^{\prime}E(x^{\prime},t^{\prime})+\hat{g}(t)\right\}  \ $\ is independent
of$\ t$.

\bigskip In the above $E=(E_{2}-E_{1})$ is the difference of perpendicular
electric fields in the two systems, which can be nonvanishing at regions
remote to the observation point $(x,t)$ (see below). (Note again that at the
point of observation $E(x,t)=0$, signifying the basic fact that the fields in
the two systems are identical at the point of observation $(x,t)$). Solutions
(\ref{LambdaStatic1}) and (\ref{LambdaStatic2}) can be viewed as the (formal)
analogs of (\ref{static1}) and (\ref{static2}) correspondingly, although they
hide in them much richer Physics because of their dynamic character (see
Section IX). (The additional constant last terms will be shown in Section IX
to be related to possible multiplicities of $\Lambda$, and they are zero in
simple-connected spacetimes). Also note again that the integrations of
potentials in (\ref{LambdaStatic1}) and (\ref{LambdaStatic2}) indeed form
paths that continuously connect $(x_{0},t_{0})$ to $(x,t)$ in the $xt$-plane
(the red-arrow and green-arrow paths of Fig.1(a)), a property that the
incorrectly used solution (\ref{BrownHolland}) does \textit{not} have.

\bigskip The reader is once again provided with the direct verification that
(\ref{LambdaStatic1}) or (\ref{LambdaStatic2}) are indeed solutions of the
basic system of PDEs (\ref{xt-BasicSystem}) in the Section that follows.

\section{Verification of solutions and simple dynamical physical examples}

Let us call our first solution \ (eq.(\ref{LambdaStatic1})) for
simple-connected spacetime $\Lambda_{3}$, namely%

\[
\Lambda_{3}(x,t)=\Lambda_{3}(x_{0},t_{0})+%
%TCIMACRO{\dint \limits_{x_{0}}^{x}}%
%BeginExpansion
{\displaystyle\int\limits_{x_{0}}^{x}}
%EndExpansion
A(x^{\prime},t)dx^{\prime}-c%
%TCIMACRO{\dint \limits_{t_{0}}^{t}}%
%BeginExpansion
{\displaystyle\int\limits_{t_{0}}^{t}}
%EndExpansion
\phi(x_{0},t^{\prime})dt^{\prime}+\left\{  c%
%TCIMACRO{\dint \limits_{t_{0}}^{t}}%
%BeginExpansion
{\displaystyle\int\limits_{t_{0}}^{t}}
%EndExpansion
dt^{\prime}%
%TCIMACRO{\dint \limits_{x_{0}}^{x}}%
%BeginExpansion
{\displaystyle\int\limits_{x_{0}}^{x}}
%EndExpansion
dx^{\prime}E(x^{\prime},t^{\prime})+g(x)\right\}
\]

with $g(x)$ chosen so that $\left\{  c%
%TCIMACRO{\dint \limits_{t_{0}}^{t}}%
%BeginExpansion
{\displaystyle\int\limits_{t_{0}}^{t}}
%EndExpansion
dt^{\prime}%
%TCIMACRO{\dint \limits_{x_{0}}^{x}}%
%BeginExpansion
{\displaystyle\int\limits_{x_{0}}^{x}}
%EndExpansion
dx^{\prime}E(x^{\prime},t^{\prime})+g(x)\right\}  $ is independent of $x$.

Verification that it solves the system of PDEs (\ref{xt-BasicSystem}) (even
for $E(x^{\prime},t^{\prime})\neq0$ for $(x^{\prime},t^{\prime})\neq
(x,t)$)$\boldsymbol{:}$

\bigskip

\textbf{A)} $\ \frac{\partial\Lambda_{3}(x,t)}{\partial x}=A(x,t)\qquad
$satisfied trivially\qquad$\checkmark$ \ \ \ \ \ \ \ \ \ \ \ 

(because $\left\{  ....\right\}  $ is independent of $x$).

\bigskip

\bigskip\textbf{B)} \ $-\frac{1}{c}\frac{\partial\Lambda_{3}(x,t)}{\partial
t}=-\frac{1}{c}%
%TCIMACRO{\dint \limits_{x_{0}}^{x}}%
%BeginExpansion
{\displaystyle\int\limits_{x_{0}}^{x}}
%EndExpansion
\frac{\partial A(x^{\prime},t)}{\partial t}dx^{\prime}+\phi(x_{0},t)-%
%TCIMACRO{\dint \limits_{x_{0}}^{x}}%
%BeginExpansion
{\displaystyle\int\limits_{x_{0}}^{x}}
%EndExpansion
E(x^{\prime},t)dx^{\prime}-\frac{1}{c}\frac{\partial g(x)}{\partial t}$
,\ \ \ \ \qquad

(the last term being trivially zero, $\frac{\partial g(x)}{\partial t}=0$ ),
and then with the substitution

$-\frac{1}{c}\frac{\partial A(x^{\prime},t)}{\partial t}=\frac{\partial
\phi(x^{\prime},t)}{\partial x^{\prime}}+E(x^{\prime},t)$

we obtain

$-\frac{1}{c}\frac{\partial\Lambda_{3}(x,t)}{\partial t}=%
%TCIMACRO{\dint \limits_{x_{0}}^{x}}%
%BeginExpansion
{\displaystyle\int\limits_{x_{0}}^{x}}
%EndExpansion
\frac{\partial\phi(x^{\prime},t)}{\partial x^{\prime}}dx^{\prime}+%
%TCIMACRO{\dint \limits_{x_{0}}^{x}}%
%BeginExpansion
{\displaystyle\int\limits_{x_{0}}^{x}}
%EndExpansion
E(x^{\prime},t)dx^{\prime}+\phi(x_{0},t)-%
%TCIMACRO{\dint \limits_{x_{0}}^{x}}%
%BeginExpansion
{\displaystyle\int\limits_{x_{0}}^{x}}
%EndExpansion
E(x^{\prime},t)dx^{\prime}$. \ \ \ 

(i) We see that the 2nd and 4th terms of the rhs \textit{cancel each other}, and

(ii) the 1st term of the rhs is \ $%
%TCIMACRO{\dint \limits_{x_{0}}^{x}}%
%BeginExpansion
{\displaystyle\int\limits_{x_{0}}^{x}}
%EndExpansion
\frac{\partial\phi(x^{\prime},t)}{\partial x^{\prime}}dx^{\prime}%
=\phi(x,t)-\phi(x_{0},t).\qquad$\ 

Hence finally

$-\frac{1}{c}\frac{\partial\Lambda_{3}(x,t)}{\partial t}=\phi(x,t).\qquad
\checkmark$

\bigskip

We have directly shown therefore that the basic system of PDEs
(\ref{xt-BasicSystem}) is indeed satisfied by our \textbf{generalized}
solution $\Lambda_{3}(x,t),$ \textbf{even for any nonzero} $E(x^{\prime
},t^{\prime})$ \ (in regions $(x^{\prime},t^{\prime})\neq(x,t)$). (The reader
is again reminded that always $E(x,t)=0$). Once again, the function $g(x)$
owes its existence to the fact that the spacetime point of observation $(x,t)$
is outside the $E$-distribution (hence the term \textit{nonlocal,} used for
the effect of the field-difference $E$ on the phases), and the reader can
clearly see this in the \textquotedblleft striped\textquotedblright%
\ $E$-distributions of the examples that follow later in this Section.

In a completely analogous way, one can easily see that our alternative
solution (eq.(\ref{LambdaStatic2})) also satisfies the basic system of PDEs
above. Indeed, if we call our second (alternative) solution
(eq.(\ref{LambdaStatic2})) for simple-connected spacetime $\Lambda_{4}$, namely%

\[
\Lambda_{4}(x,t)=\Lambda_{4}(x_{0},t_{0})+%
%TCIMACRO{\dint \limits_{x_{0}}^{x}}%
%BeginExpansion
{\displaystyle\int\limits_{x_{0}}^{x}}
%EndExpansion
A(x^{\prime},t_{0})dx^{\prime}-c\int_{t_{0}}^{t}\phi\left(  x,t^{\prime
}\right)  dt^{\prime}+\left\{  -c%
%TCIMACRO{\dint \limits_{x_{0}}^{x}}%
%BeginExpansion
{\displaystyle\int\limits_{x_{0}}^{x}}
%EndExpansion
dx^{\prime}%
%TCIMACRO{\dint \limits_{t_{0}}^{t}}%
%BeginExpansion
{\displaystyle\int\limits_{t_{0}}^{t}}
%EndExpansion
dt^{\prime}E(x^{\prime},t^{\prime})+\hat{g}(t)\right\}
\]
with $\hat{g}(t)$ chosen so that $\left\{  -c%
%TCIMACRO{\dint \limits_{x_{0}}^{x}}%
%BeginExpansion
{\displaystyle\int\limits_{x_{0}}^{x}}
%EndExpansion
dx^{\prime}%
%TCIMACRO{\dint \limits_{t_{0}}^{t}}%
%BeginExpansion
{\displaystyle\int\limits_{t_{0}}^{t}}
%EndExpansion
dt^{\prime}E(x^{\prime},t^{\prime})+\hat{g}(t)\right\}  $ is independent of
$t$,

then we have (even for $E(x^{\prime},t^{\prime})\neq0$ for $(x^{\prime
},t^{\prime})\neq(x,t)$)$\boldsymbol{:}$

\bigskip

\textbf{A)} $\ -\frac{1}{c}\frac{\partial\Lambda_{4}(x,t)}{\partial t}%
=\phi(x,t)\qquad$satisfied trivially\qquad$\checkmark$ \ \ \ \ \ \ \ \ \ \ \ 

(because $\left\{  ....\right\}  $ is independent of $t$).

\bigskip

\bigskip\textbf{B)} \ $\frac{\partial\Lambda_{4}(x,t)}{\partial x}%
=A(x,t_{0})-c%
%TCIMACRO{\dint \limits_{t_{0}}^{t}}%
%BeginExpansion
{\displaystyle\int\limits_{t_{0}}^{t}}
%EndExpansion
\frac{\partial\phi(x,t^{\prime})}{\partial x}dt^{\prime}-c%
%TCIMACRO{\dint \limits_{t_{0}}^{t}}%
%BeginExpansion
{\displaystyle\int\limits_{t_{0}}^{t}}
%EndExpansion
E(x,t^{\prime})dt^{\prime}+\frac{\partial\hat{g}(t)}{\partial x}$
,\ \ \ \ \qquad

(the last term being trivially zero, $\frac{\partial\hat{g}(t)}{\partial x}=0$
), and then with the substitution

$\frac{\partial\phi(x,t^{\prime})}{\partial x}=-E(x,t^{\prime})-\frac{1}%
{c}\frac{\partial A(x,t^{\prime})}{\partial t^{\prime}}$

we obtain

$\frac{\partial\Lambda_{4}(x,t)}{\partial x}=A(x,t_{0})+c%
%TCIMACRO{\dint \limits_{t_{0}}^{t}}%
%BeginExpansion
{\displaystyle\int\limits_{t_{0}}^{t}}
%EndExpansion
E(x,t^{\prime})dt^{\prime}+%
%TCIMACRO{\dint \limits_{t_{0}}^{t}}%
%BeginExpansion
{\displaystyle\int\limits_{t_{0}}^{t}}
%EndExpansion
\frac{\partial A(x,t^{\prime})}{\partial t^{\prime}}dt^{\prime}-c%
%TCIMACRO{\dint \limits_{t_{0}}^{t}}%
%BeginExpansion
{\displaystyle\int\limits_{t_{0}}^{t}}
%EndExpansion
E(x,t^{\prime})dt^{\prime}$. \ \ \ 

(i) We see that the 2nd and 4th terms of the rhs \textit{cancel each other}, and

(ii) the 3rd term of the rhs is \ $%
%TCIMACRO{\dint \limits_{t_{0}}^{t}}%
%BeginExpansion
{\displaystyle\int\limits_{t_{0}}^{t}}
%EndExpansion
\frac{\partial A(x,t^{\prime})}{\partial t^{\prime}}dt^{\prime}%
=A(x,t)-A(x,t_{0}).\qquad$\ 

Hence finally

$\frac{\partial\Lambda_{4}(x,t)}{\partial x}=A(x,t).\qquad\checkmark$

\bigskip

Once again, all the above are true for any nonzero $E(x^{\prime},t^{\prime})$
(in regions $(x^{\prime},t^{\prime})\neq(x,t)$).

\bigskip

To see again how the above solutions appear in nontrivial cases (and how they
give new results, i.e. not differing from the usual ones by a mere constant)
let us take analogous examples of strips as earlier, but now in spacetime:

\textbf{(a)} For the case of the extended \textit{vertical} strip (parallel to
the $t$-axis) of Fig.1(a) (the case of a one-dimensional capacitor that is
(arbitrarily and variably) charged for all time), then, for $x$ located
outside (and on the right of) the capacitor, the quantity$\ c%
%TCIMACRO{\dint \limits_{t_{0}}^{t}}%
%BeginExpansion
{\displaystyle\int\limits_{t_{0}}^{t}}
%EndExpansion
dt^{\prime}%
%TCIMACRO{\dint \limits_{x_{0}}^{x}}%
%BeginExpansion
{\displaystyle\int\limits_{x_{0}}^{x}}
%EndExpansion
dx^{\prime}E(x^{\prime},t^{\prime})$ in\ $\Lambda_{3}$ is \textit{already
independent of}$\ x$\ (since a displacement of the $(x,t)$-corner of the
rectangle to the right, along the $x$-direction, does not change the enclosed
\textquotedblleft electric flux\textquotedblright, see
Fig.1(a))$\boldsymbol{;}$ hence in this case the function $g(x)$ can be taken
as $g(x)=0$ (up to a constant $C$), because then the condition for $g(x)$
stated in the solution (\ref{LambdaStatic1}) (namely, that the quantity in
brackets must be independent of $x$) is indeed satisfied. (Note again that the
above $x$-independence of the enclosed \textquotedblleft electric
flux\textquotedblright\ is important for the existence of $g(x)$).

So for this setup, the nonlocal term in the solution \textit{survives} (the
quantity in brackets is nonvanishing), but \textit{it is not constant}%
$\boldsymbol{:}$ this enclosed flux depends on $t$ (since the enclosed flux
\textit{does change} with a displacement of the $(x,t)$-corner of the
rectangle upwards, along the $t$-direction). Hence, by looking at the
alternative solution $\Lambda_{4}(x,t),$ the quantity$\ c%
%TCIMACRO{\dint \limits_{x_{0}}^{x}}%
%BeginExpansion
{\displaystyle\int\limits_{x_{0}}^{x}}
%EndExpansion
dx^{\prime}%
%TCIMACRO{\dint \limits_{t_{0}}^{t}}%
%BeginExpansion
{\displaystyle\int\limits_{t_{0}}^{t}}
%EndExpansion
dt^{\prime}E(x^{\prime},t^{\prime})$\ is dependent on$\ t$, so that $\hat
{g}(t)$ must be chosen as $\ \hat{g}(t)=+c%
%TCIMACRO{\dint \limits_{x_{0}}^{x}}%
%BeginExpansion
{\displaystyle\int\limits_{x_{0}}^{x}}
%EndExpansion
dx^{\prime}%
%TCIMACRO{\dint \limits_{t_{0}}^{t}}%
%BeginExpansion
{\displaystyle\int\limits_{t_{0}}^{t}}
%EndExpansion
dt^{\prime}E(x^{\prime},t^{\prime})$ (up to the same constant $C$)\ in order
to \textit{cancel} this $t$-dependence, so that its own condition stated in
the solution (\ref{LambdaStatic2}) (namely, that the quantity in brackets must
be independent of $t$) is indeed satisfied$\boldsymbol{;}$ as a result, the
quantity in brackets in solution $\Lambda_{4}$ disappears and there is no
nonlocal contribution in $\Lambda_{4}$ (for $C=0$). (Once again, if we had
used a $C\neq0$, the nonlocal contributions would be differently shared
between the two solutions, but without changing the Physics when we take the
\textit{difference} of the two solutions). [The reader should once again note
the crucial fact that the point of observation $(x,t)$ is outside the
$E$-distribution, which makes the existence of functions $g(x)$ and $\hat
{g}(t)$ possible].

With these choices of $\hat{g}(t)$ and $g(x)$, we already have new results
(compared to the standard ones of the integrals of potentials). I.e. one of
the two solutions, namely $\Lambda_{3}$ \textbf{is} affected nonlocally by the
enclosed flux (and this flux is \textbf{not} constant). Spelled out clearly,
the two results are:%

\[
\Lambda_{3}(x,t)=\Lambda_{3}(x_{0},t_{0})+%
%TCIMACRO{\dint \limits_{x_{0}}^{x}}%
%BeginExpansion
{\displaystyle\int\limits_{x_{0}}^{x}}
%EndExpansion
A(x^{\prime},t)dx^{\prime}-c%
%TCIMACRO{\dint \limits_{t_{0}}^{t}}%
%BeginExpansion
{\displaystyle\int\limits_{t_{0}}^{t}}
%EndExpansion
\phi(x_{0},t^{\prime})dt^{\prime}+c%
%TCIMACRO{\dint \limits_{t_{0}}^{t}}%
%BeginExpansion
{\displaystyle\int\limits_{t_{0}}^{t}}
%EndExpansion
dt^{\prime}%
%TCIMACRO{\dint \limits_{x_{0}}^{x}}%
%BeginExpansion
{\displaystyle\int\limits_{x_{0}}^{x}}
%EndExpansion
dx^{\prime}E(x^{\prime},t^{\prime})+C
\]

\[
\Lambda_{4}(x,t)=\Lambda_{4}(x_{0},t_{0})+%
%TCIMACRO{\dint \limits_{x_{0}}^{x}}%
%BeginExpansion
{\displaystyle\int\limits_{x_{0}}^{x}}
%EndExpansion
A(x^{\prime},t_{0})dx^{\prime}-c\int_{t_{0}}^{t}\phi\left(  x,t^{\prime
}\right)  dt^{\prime}+C
\]
(and their difference, as mentioned above, is zero - denoting what might be
called a generalized Werner \& Brill cancellation in spacetime).

\bigskip

\textbf{(b) }In the \textquotedblleft dual case\textquotedblright\ of an
extended \textit{horizontal} strip - parallel to the $x$-axis (that
corresponds to a nonzero electric field in all space that has however a finite
duration $T)$, the proper choices (for observation time instant $t>T$) are
basically reverse (i.e. we can now take $\hat{g}(t)=0$ \ and $g(x)=-c%
%TCIMACRO{\dint \limits_{t_{0}}^{t}}%
%BeginExpansion
{\displaystyle\int\limits_{t_{0}}^{t}}
%EndExpansion
dt^{\prime}%
%TCIMACRO{\dint \limits_{x_{0}}^{x}}%
%BeginExpansion
{\displaystyle\int\limits_{x_{0}}^{x}}
%EndExpansion
dx^{\prime}E(x^{\prime},t^{\prime})$ (since the \textquotedblleft electric
flux\textquotedblright\ enclosed in the \textquotedblleft observation
rectangle\textquotedblright\ now depends on $x$, but not on $t$), with both
choices always up to a common constant) and once again we can easily see, upon
subtraction of the two solutions, a similar cancellation effect. In this case
again, the results are also new (a \ nonlocal term survives now in
$\Lambda_{4}$). Again spelled out clearly, these are:%

\[
\Lambda_{3}(x,t)=\Lambda_{3}(x_{0},t_{0})+%
%TCIMACRO{\dint \limits_{x_{0}}^{x}}%
%BeginExpansion
{\displaystyle\int\limits_{x_{0}}^{x}}
%EndExpansion
A(x^{\prime},t)dx^{\prime}-c%
%TCIMACRO{\dint \limits_{t_{0}}^{t}}%
%BeginExpansion
{\displaystyle\int\limits_{t_{0}}^{t}}
%EndExpansion
\phi(x_{0},t^{\prime})dt^{\prime}+C
\]

\[
\Lambda_{4}(x,t)=\Lambda_{4}(x_{0},t_{0})+%
%TCIMACRO{\dint \limits_{x_{0}}^{x}}%
%BeginExpansion
{\displaystyle\int\limits_{x_{0}}^{x}}
%EndExpansion
A(x^{\prime},t_{0})dx^{\prime}-c\int_{t_{0}}^{t}\phi\left(  x,t^{\prime
}\right)  dt^{\prime}-c%
%TCIMACRO{\dint \limits_{x_{0}}^{x}}%
%BeginExpansion
{\displaystyle\int\limits_{x_{0}}^{x}}
%EndExpansion
dx^{\prime}%
%TCIMACRO{\dint \limits_{t_{0}}^{t}}%
%BeginExpansion
{\displaystyle\int\limits_{t_{0}}^{t}}
%EndExpansion
dt^{\prime}E(x^{\prime},t^{\prime})+C
\]
(their difference also being zero -- a generalized Werner \& Brill
cancellation in spacetime).

\bigskip

\textbf{(c) }And again, if we want cases that are more involved (with the
nonlocal contributions appearing nontrivially in \textbf{both} solutions
$\Lambda_{3}$ and $\Lambda_{4}$ and with $g(x)$ and $\hat{g}(t)$ not being
\textquotedblleft immediately visible\textquotedblright) we must again
consider different shapes of $E$-distribution. One such case (the triangular)
was already shown in Fig.1(b) (for the magnetic case, which however is
completely analogous). For such a triangular case the choices of $g(x)$ and
$\hat{g}(t)$ will be different from the above and this will result in
different roles of the nonlocal terms (and these nontrivial results, or more
accurately, their analogs for the magnetic case, were given earlier in closed
analytical form, eqs (\ref{triangular1}) and (\ref{triangular2})). [And even
cases of curved shapes can be addressed more generally (when the shape is such
that the \textquotedblleft flux\textquotedblright\ does \textit{not}\textbf{
}decouple in a sum of separate spatial and temporal contributions), i.e. by
solving the basic system of PDEs directly in polar coordinates (the results
being analogous to the ones given in Appendix E for the magnetic case, see eqs
(\ref{polar1})-(\ref{polarcond2}))].

The reader should note again that, in all the above examples in
simple-connected spacetime, the $x$-independent quantity in brackets of the
1st solution (\ref{LambdaStatic1}) is equal to the function $\hat{g}(t)$ of
the 2nd solution (\ref{LambdaStatic2}), and the $t$-independent quantity in
brackets of the 2nd solution (\ref{LambdaStatic2}) is equal to the function
$g(x)$ of the 1st solution (\ref{LambdaStatic1}). This mathematical pattern is
what leads to the above mentioned cancellations, and it is generally proven
(i.e. for any form of $E$-distribution in the $(x,t)$-plane) in the Section
that follows.

\section{Comments on the general behavior of the $\boldsymbol{(x,t)}%
$-solutions}

Let us first summarize (and prove in generality) some of the behavioral
patterns that we saw in the above examples and then continue on other
properties (i.e. an account of multiplicities of $\Lambda$ in
multiple-connected spacetimes that we left out, which are described by the
constants $\tau(t_{0})$ and $\chi(x_{0})$). First, in (\ref{LambdaStatic1}) or
(\ref{LambdaStatic2}) note the proper appearance and placement of $x_{0}$ and
$t_{0}$ that gives a \textquotedblleft path-sense\textquotedblright\ to the
line integrals of potentials in each solution (with the path consisting of two
straight and perpendicular line segments, continuously connecting the initial
point$\ (x_{0},t_{0})$ to the final point $(x,t)$ for each solution). And
there are naturally two possible paths of this type that connect the initial
point $(x_{0},t_{0})$ with the final point $(x,t)$ (the solution
(\ref{LambdaStatic1}) having a clockwise and the solution (\ref{LambdaStatic2}%
) having a counterclockwise sense, as in Fig.1(a))$\boldsymbol{;}$ with this
construction a natural \textit{observation} \textit{rectangle} is then formed
(see Fig. 1(a)), within which the enclosed \textquotedblleft electric
fluxes\textquotedblright\ (in spacetime) appear to be crucial (showing up as
nonlocal terms of contributions of the electric field difference (recall that
\ $E(x^{\prime},t^{\prime})=E_{2}(x^{\prime},t^{\prime})-E_{1}(x^{\prime
},t^{\prime})$) from regions of time and space that are remote to the
observation point $(x,t)$). The appearance of these nonlocal terms (of the
electric field difference) in $\Lambda(x,t)$ from regions of spacetime
$(x^{\prime},t^{\prime})$ far from the observation point $(x,t)$ seems to have
a direct effect on the wavefunction phases at $(x,t)$ (through the phase
mapping that connects the two quantum systems). The actual manner in which
this happens is of course determined by the form of functions $\ g(x)$ \ or
$\ \hat{g}(t)$ (the existence of which lies in the fact that the spacetime
observation point $(x,t)$ is always outside the $E$%
-distributions)$\boldsymbol{:}$ these functions must be chosen in such a way
that they satisfy their respective conditions, as these are stated after
(\ref{LambdaStatic1}) or (\ref{LambdaStatic2}) respectively. We saw, for
example, that if we have a distribution of $E$ in the $(x,t)$-plane in the
form of an extended \textit{strip} parallel to the $t$-axis, the function
$g(x)$ can be taken as $g(x)=0$ (up to a constant $C$), and that $\hat{g}(t)$
must be chosen as $\ \hat{g}(t)=+c%
%TCIMACRO{\dint \limits_{x_{0}}^{x}}%
%BeginExpansion
{\displaystyle\int\limits_{x_{0}}^{x}}
%EndExpansion
dx^{\prime}%
%TCIMACRO{\dint \limits_{t_{0}}^{t}}%
%BeginExpansion
{\displaystyle\int\limits_{t_{0}}^{t}}
%EndExpansion
dt^{\prime}E(x^{\prime},t^{\prime})$ (up to the same constant $C$)\ in order
to \textit{cancel} the $t$-dependence of the enclosed \textquotedblleft
flux\textquotedblright. Furthermore, with these choices of $\hat{g}(t)$ and
$g(x)$, it is easy to see that, if we subtract the two solutions
(\ref{LambdaStatic1}) and (\ref{LambdaStatic2}), the result is \textit{zero}
(because the line integrals of potentials $A$ and $\phi$ in the two solutions
are in opposite senses in the $(x,t)$ plane, hence their difference leads to a
\textit{closed} line integral, which is in turn equal to the enclosed electric
flux, and this flux always happens to be of opposite sign from that of the
enclosed flux that explicitly appears as a nonlocal contribution of the
$E$-fields (i.e. the term that survives in $\Lambda_{3}$ in case (a) of last
Section)$\boldsymbol{.}$ Such cancellation effects in dynamical cases are
important and will be discussed (and generalized) further in Section XII.

Let us however give here a general proof of the above cancellations (i.e. for
any variable dependence of $E$-distribution). First, by looking at the general
structure of solutions (\ref{LambdaStatic1}) and (\ref{LambdaStatic2}), we
note that in both forms, the last constant terms ($\tau(t_{0})$ and
$\chi(x_{0})$) are only present in cases where $\Lambda$ is expected to be
multivalued (this comes from the definitions of $\tau(t_{0})$ and $\chi
(x_{0})$, and is shown in Appendix C, see eqs (\ref{Taut0}) and (\ref{Chi0}))
and therefore these constant quantities are nonvanishing in cases of motion
only in multiple-connected spacetimes (leading to phenomena of the electric
Aharonov-Bohm type (see the analogous discussion in Appendix F, on the
easier-to-follow magnetic case)). In such multiple-connected cases these last
terms turn out to be simply equal (in absolute value) to the enclosed fluxes
in regions of spacetime that are physically inaccessible to the particle (in
the electric Aharonov-Bohm setup, for example, it turns out that $\tau
(t_{0})=-\chi(x_{0})=$ enclosed \textquotedblleft electric
flux\textquotedblright\ in spacetime). Although such cases can also be covered
by our method below, let us for the moment ignore them (set them to zero) and
focus on cases of motion in simple-connected spacetimes. Then the two
solutions (\ref{LambdaStatic1}) and (\ref{LambdaStatic2}) are actually
\textit{equal} for \textit{any} $E$-distribution. This is rigorously shown in
Appendix C. [It is also shown there that the $x$-independent (hence
$t$-dependent) quantity in brackets of the 1st solution (\ref{LambdaStatic1})
is equal to the function $\hat{g}(t)$ of the 2nd solution (\ref{LambdaStatic2}%
) $-$ and the $t$-independent (hence $x$-dependent) quantity in brackets of
the 2nd solution (\ref{LambdaStatic2}) is equal to the function $g(x)$ of the
1st solution (\ref{LambdaStatic1}). Because of this, it is straightforward to
see (by subtracting the two solutions) the mathematical reason for the
occurence of the cancellations noted earlier, for \textit{any} shape of $E$-distribution].

In spite therefore of the simplicity of the above considered 1-D system, we
are already in a position to draw certain very general conclusions on the
possible physical consequences of the new nonlocal terms of the electric
fields appearing in the solutions (\ref{LambdaStatic1}) and
(\ref{LambdaStatic2}). One can immediately see from the above considerations
(or from the formal proof of Appendix C) that these temporally-nonlocal
contributions$\boldsymbol{\ }$have the tendency of cancelling the
contributions from the $A$- and $\phi$-integrals. This already gives an
indication of cancellations that might also occur in cases of higher spatial
dimensionality (where line-integrals of $A$'s, for example, can be related to
enclosed \textit{magnetic} fluxes). This \textit{is} actually the case in the
van Kampen thought-experiment that will be discussed later in Section XII $-$
although the cancellations there will be slightly more delicate, actually
involving a balance among 3 variables, and with the actual \textit{senses} of
spatial closed line-integrals in the $(x,y)$-plane being nontrivially
important. (Moreover, \textit{instead of lying outside of simple strips, the
spatial point }$(x,y)$\textit{ will in that case lie outside a light-cone},
leading to results that are \textit{causal}, as we shall see).

Finally, with respect to $\tau(t_{0})$\ and $\chi(x_{0})$, we show in the same
Appendix C their already noted properties$\boldsymbol{:}$ ordinarily (in
simple-connectivity) they are zero, or in the most general case (of
multiple-connectivity) they are related to physically inaccessible enclosed
fluxes. [We should also note here that the case of the electric Aharonov-Bohm
setup, with the particles traveling inside distinct equipotential cages with
scalar potentials that last for a finite duration, is the prototype of
\textit{multiple-connectivity in spacetime}, a fact first noted by Iddings and
reported by Noerdlinger\cite{Iddings}. We will see later (Section XII) that
this feature is \textit{not} present in the van Kampen thought-experiment,
hence an electric Aharonov-Bohm argument should not really be invoked in that
case (as van Kampen did) because of this lack of multiple-connectedness in spacetime].

Before, however, leaving this simple $(x,t)$-case, we should finally emphasize
that this (or any other) contribution of electric fields is \textit{not}
present at the level of the basic Lagrangian, and the view holds in the
literature (see e.g. \cite{BrownHome}) that, because of this absence, electric
fields cannot contribute \textit{directly} to the phase of the wavefunctions.
This conclusion originates from the path-integral approach (that is almost
always followed), but, nevertheless, our present work shows that fields
\textit{do} contribute nonlocally. A more general discussion on this issue is
given in the final Section, after discussion of the van Kampen
thought-experiment, and also in connection to related path-integral
approaches\cite{Troudet}.

\section{\bigskip Again on the $\boldsymbol{(x,y)}$-Magnetic Case}

After having discussed fully the simple $(x,t)$-case, let us for completeness
give the analogous (Euclidian-rotated in 4-D spacetime) derivation for
$(x,y)$-variables and briefly discuss the properties of the simpler static
solutions, but now in full generality (also including possible
multi-valuedness of $\Lambda$ in the usual magnetic Aharonov-Bohm cases). We
will simply need to apply the same methodology (of solution of a system of
PDEs) to such static spatially two-dimensional cases (so that now different
(remote) magnetic fields for the two systems, perpendicular to the 2-D space,
will arise). For such cases we need to solve the system of PDEs already shown
in (\ref{usualgradcomps}), namely%

\[
\frac{\partial\Lambda(x,y)}{\partial x}=A_{x}(x,y)\qquad and\qquad
\frac{\partial\Lambda(x,y)}{\partial y}=A_{y}(x,y).
\]
By following then the procedure described in detail in Appendix D, we finally
obtain the following general solution%

\begin{equation}
\Lambda(x,y)=\Lambda(x_{0},y_{0})+%
%TCIMACRO{\dint \limits_{x_{0}}^{x}}%
%BeginExpansion
{\displaystyle\int\limits_{x_{0}}^{x}}
%EndExpansion
A_{x}(x^{\prime},y)dx^{\prime}+%
%TCIMACRO{\dint \limits_{y_{0}}^{y}}%
%BeginExpansion
{\displaystyle\int\limits_{y_{0}}^{y}}
%EndExpansion
A_{y}(x_{0},y^{\prime})dy^{\prime}+\left\{
%TCIMACRO{\dint \limits_{y_{0}}^{y}}%
%BeginExpansion
{\displaystyle\int\limits_{y_{0}}^{y}}
%EndExpansion
dy^{\prime}%
%TCIMACRO{\dint \limits_{x_{0}}^{x}}%
%BeginExpansion
{\displaystyle\int\limits_{x_{0}}^{x}}
%EndExpansion
dx^{\prime}B_{z}(x^{\prime},y^{\prime})+g(x)\right\}  +f(y_{0})
\label{Lambda(x,y)1}%
\end{equation}

\[
with\text{ \ }g(x)\text{ \ }chosen\ \ so\text{ \ }that\text{ \ }\left\{
%TCIMACRO{\dint \limits_{y_{0}}^{y}}%
%BeginExpansion
{\displaystyle\int\limits_{y_{0}}^{y}}
%EndExpansion
dy^{\prime}%
%TCIMACRO{\dint \limits_{x_{0}}^{x}}%
%BeginExpansion
{\displaystyle\int\limits_{x_{0}}^{x}}
%EndExpansion
dx^{\prime}B_{z}(x^{\prime},y^{\prime})+g(x)\right\}  \boldsymbol{:}\text{ is
}\mathsf{independent\ of\ }\ x,
\]
which is basically the example shown earlier in (\ref{static1}) but with
included multiplicities through the extra constant $f(y_{0})$ (that for
simple-connected space can be set to zero)$\boldsymbol{.}$ The result
(\ref{Lambda(x,y)1}) applies to cases where the particle passes through
\textit{different} magnetic fields (recall that $B_{z}={\huge (}%
\boldsymbol{B}_{2}-\boldsymbol{B}_{1}{\huge )}_{z}$) in spatial regions that
are remote to (i.e. do not contain) the observation point $(x,y)$.
Alternatively, by following the reverse route of integrations (see Appendix
D), we finally obtain the following alternative general solution%

\begin{equation}
\Lambda(x,y)=\Lambda(x_{0},y_{0})+\int_{x_{0}}^{x}A_{x}(x^{\prime}%
,y_{0})dx^{\prime}+\int_{y_{0}}^{y}A_{y}(x,y^{\prime})dy^{\prime}+\left\{
{\Huge -}%
%TCIMACRO{\dint \limits_{x_{0}}^{x}}%
%BeginExpansion
{\displaystyle\int\limits_{x_{0}}^{x}}
%EndExpansion
dx^{\prime}%
%TCIMACRO{\dint \limits_{y_{0}}^{y}}%
%BeginExpansion
{\displaystyle\int\limits_{y_{0}}^{y}}
%EndExpansion
dy^{\prime}B_{z}(x^{\prime},y^{\prime})+h(y)\right\}  +\hat{h}(x_{0})
\label{Lambda(x,y)4}%
\end{equation}

\[
with\text{ \ }h(y)\text{ \ }chosen\text{ \ }so\text{ \ }that\text{ \ }\left\{
{\Huge -}%
%TCIMACRO{\dint \limits_{x_{0}}^{x}}%
%BeginExpansion
{\displaystyle\int\limits_{x_{0}}^{x}}
%EndExpansion
dx^{\prime}%
%TCIMACRO{\dint \limits_{y_{0}}^{y}}%
%BeginExpansion
{\displaystyle\int\limits_{y_{0}}^{y}}
%EndExpansion
dy^{\prime}B_{z}(x^{\prime},y^{\prime})+h(y)\right\}  \boldsymbol{:}\text{ is
}\mathsf{independent\ of\ }\ y,
\]
which is basically the example shown earlier in (\ref{static2}) but with
included multiplicities through the extra constant $\hat{h}(x_{0})$. One can
actually show that the two solutions are equivalent (i.e. (\ref{static1}) and
(\ref{static2}) for a simple-connected region are equal\cite{kyriakos}), a
fact that can be proven in a way similar to the $(x,t)$-cases of Section IX
(the actual proof being given in Appendix C). (For the case of
multiple-connectivity of the two-dimensional space, a discussion of the actual
values of the multiplicities $f(y_{0})$ and $\hat{h}(x_{0})$ is given later in
this Section, with the proofs presented in Appendix F).

As we saw in the examples of Section VI, in case of a striped-distribution of
the magnetic field difference $B_{z}$, the functions $g(x)$ and $h(y)$ in
(\ref{Lambda(x,y)1}) and (\ref{Lambda(x,y)4}) (or equivalently in
(\ref{static1}) and (\ref{static2})) have to be chosen in ways that are
compatible with their corresponding constraints (stated after
(\ref{Lambda(x,y)1}) and (\ref{Lambda(x,y)4})) and are completely analogous to
the above discussed $(x,t)$-cases$\boldsymbol{.}$ (In all cases, the fact that
the observation point $(x,y)$ is always outside the nonzero-$B_{z}$ regions is
crucial for the existence of these functions). By then taking the
\textit{difference} of (\ref{static1}) and (\ref{static2}) we obtain that the
\textquotedblleft Aharonov-Bohm phase\textquotedblright\ (the one originating
from the \textit{closed} line integral of $A$'s) is exactly cancelled by the
additional nonlocal term of the magnetic fields (that the particle passed
through). As already mentioned earlier, this is reminiscent of the
cancellation of phases (broadly speaking, a cancellation between the
\textquotedblleft Aharonov-Bohm phase\textquotedblright\ and the semiclassical
phase picked up by the trajectories) observed in the early experiments of
Werner \& Brill\cite{WernerBrill} for particles passing through \textit{full}
(nonvanishing) magnetic fields, and our method seems to provide a very natural
justification$\boldsymbol{:}$ as our results are completely general (and for
delocalized states in a simple-connected region they basically describe the
single-valuedness of $\Lambda$), they are also valid and applicable to cases
of narrow wavepackets (or states that describe semiclassical motion) that pass
through nonvanishing magnetic fields, which \textit{was} the case of the
Werner \& Brill experiments. (A similar cancellation of an electric
Aharonov-Bohm phase also occurs for particles passing through a static
electric field as we saw in Section VIII). We conclude that, for static cases,
and when particles pass through fields, the new nonlocal terms reported in the
present work lead quite generally to a cancellation of Aharonov-Bohm phases
that had earlier been sketchily noticed and only at the semiclassical level.

Since we already mentioned that the deep origin of the above cancellations is
the single-valuedness of $\Lambda$ in simple-connected space, we should add
for completeness that the rigorous proof of the uniqueness at each spatial
point (single-valuedness) of $\Lambda$ for completely delocalized states in
simple-connected space can be given in a directly analogous way to the proof
given in Appendix C for the $(x,t)$-case of Section IX. What is however more
important to point out here is that the above cancellations for semiclassical
trajectories (that pass through a nonzero magnetic field) can alternatively be
understood as a \textit{compatibility} between the Aharonov-Bohm
fringe-displacement and the trajectory-deflection due to the Lorentz force
(i.e. the semiclassical phase picked up due to the optical path difference of
the two deflected trajectories \textit{exactly cancels} (is \textit{opposite
in sign} from) the Aharonov-Bohm phase picked up by the trajectories due to
the enclosed flux). [We may mention that this is also related to the
well-known overall rigid displacement of the single-slit envelope of the
two-slit diffraction pattern, displacement that only occurs if the wavepackets
actually pass through a nonzero field (and not in genuine Aharonov-Bohm
cases)]. These issues are further discussed in the final Section, where some
popular reports in the literature (Feynman\cite{Feynman},
Felsager\cite{Felsager}, Batelaan \& Tonomura\cite{Batelaan}) are given a
minor correction (of a sign). Similarly, and by also including time $t$ (and
by again correcting a sign-error propagating in the standard literature) we
will give an explanation of why certain classical arguments (invoking the past
$t$-dependent history of the experimental set up) seem to be successful (in
giving the correct result for a static Aharonov-Bohm phase).

\bigskip Another point of interest concerning the above found nonlocal
contributions of fields is the plausible question of \textit{what shape }the
field distributions must have (or more accurately, their part enclosed inside
the \textit{observation rectangle}) so that the enclosed flux can be decoupled
to a sum of functions of separate variables, in order for the solutions
obtained above to exist and be immediately applicable (i.e. for the functions
$g(x)$ and $h(y)$ to be directly possible to determine$\boldsymbol{:}$ each of
them must then only \textit{partially} cancel the corresponding $x$
\textit{or} $y$ dependence, respectively). We already provided an example of
such a distribution of a homogeneous$\ B_{z}$ (the triangular one) in Section
VI (see the nontrivial results (\ref{triangular1}) and (\ref{triangular2})).
And as mentioned in Section VI, in cases of circularly shaped distributions
(where the enclosed flux may not be decoupled in $x$ and $y$ terms), it is
advantageous to solve the system directly in polar coordinates. By following a
similar procedure (of solving the system of PDEs resulting from
(\ref{usualgrad})) in polar coordinates $(\rho,\varphi)$, namely%

\[
\frac{\partial\Lambda(\rho,\varphi)}{\partial\rho}=A_{\rho}(\rho
,\varphi)\qquad and\qquad\frac{1}{\rho}\frac{\partial\Lambda(\rho,\varphi
)}{\partial\varphi}=A_{\varphi}(\rho,\varphi)
\]
with steps completely analogous to the above described procedure, one can
obtain analogs of solutions (\ref{Lambda(x,y)1}) and (\ref{Lambda(x,y)4})
given in Appendix E (see eqs (\ref{polar1}) and (\ref{polar2})). [In such a
case, the observation rectangle has now given its place to a slice of an
annular section]. These matters however deserve further investigation, of a
more mathematical type, in applications of the above theory to specific shape-geometries.

Finally, for completeness we discuss the issue of multiplicities (the last
constant terms of (\ref{Lambda(x,y)1}) and (\ref{Lambda(x,y)4})) in case of
spatial multiple-connectivity (such as the standard magnetic Aharonov-Bohm
case, in which we can take $g(x)=0$ \ \textit{and} \ $h(y)=0$, since the
enclosed magnetic flux is independent of both $x$ and $y$). We take up this
issue in detail in Appendix F, where it is proven that\ $\hat{h}(x_{0}%
)=-$\ \ $f(y_{0})=$ enclosed magnetic flux\ (a constant, independent of $x$
and $y$). Since $f(y_{0})$ cancels out the {\Huge \ }$%
%TCIMACRO{\dint \limits_{y_{0}}^{y}}%
%BeginExpansion
{\displaystyle\int\limits_{y_{0}}^{y}}
%EndExpansion
dy^{\prime}%
%TCIMACRO{\dint \limits_{x_{0}}^{x}}%
%BeginExpansion
{\displaystyle\int\limits_{x_{0}}^{x}}
%EndExpansion
dx^{\prime}B_{z}(x^{\prime},y^{\prime})$ term, and $\hat{h}(x_{0})$ cancels
out the {\Huge \ -}$%
%TCIMACRO{\dint \limits_{x_{0}}^{x}}%
%BeginExpansion
{\displaystyle\int\limits_{x_{0}}^{x}}
%EndExpansion
dx^{\prime}%
%TCIMACRO{\dint \limits_{y_{0}}^{y}}%
%BeginExpansion
{\displaystyle\int\limits_{y_{0}}^{y}}
%EndExpansion
dy^{\prime}B_{z}(x^{\prime},y^{\prime})$ term, the two solutions are then
reduced to the usual solutions of mere $A$-integrals along the two paths (i.e.
the standard Dirac phase, with no nonlocal contributions).

\bigskip

\section{\bigskip Full $\boldsymbol{(x,y,t)}$-case}

Finally, let us look at the most general spatially-two-dimensional and
time-dependent case. This combines effects of (perpendicular) magnetic fields
(which, if present only in physically-inaccessible regions, can have
Aharonov-Bohm consequences) with the temporal nonlocalities of electric fields
(parallel to the plane) found in previous Sections. By working again in
Cartesian spatial coordinates, we now have to deal with the full system of PDEs%

\begin{equation}
\frac{\partial\Lambda(x,y,t)}{\partial x}=A_{x}(x,y,t),\qquad\frac
{\partial\Lambda(x,y,t)}{\partial y}=A_{y}(x,y,t),\qquad-\frac{1}{c}%
\frac{\partial\Lambda(x,y,t)}{\partial t}=\phi\left(  x,y,t\right)  .
\label{FullSystem}%
\end{equation}
This exercise is considerably longer than the previous ones but important to
solve, in order to see in what manner the solutions of this system are able to
\textit{combine} the spatial and temporal nonlocal effects found above. There
are now 3!=6 alternative integration routes to follow for solving this system
(and, in addition to this, the results in intermediate steps tend to
proliferate). The corresponding (rather long) procedure for solving the system
(\ref{FullSystem}) is described in detail in Appendix G, and 2 out of the 12
solutions that can be derived turn out to be the most crucial for the
discussion that will follow in the next Section. First, by following steps
similar to the above, the following temporal generalization of
(\ref{Lambda(x,y)4}) is obtained%

\[
\Lambda(x,y,t)=\Lambda(x_{0},y_{0},t)+\int_{x_{0}}^{x}A_{x}(x^{\prime}%
,y_{0},t)dx^{\prime}+\int_{y_{0}}^{y}A_{y}(x,y^{\prime},t)dy^{\prime}+
\]

\begin{equation}
+\left\{  {\Huge -}%
%TCIMACRO{\dint \limits_{x_{0}}^{x}}%
%BeginExpansion
{\displaystyle\int\limits_{x_{0}}^{x}}
%EndExpansion
dx^{\prime}%
%TCIMACRO{\dint \limits_{y_{0}}^{y}}%
%BeginExpansion
{\displaystyle\int\limits_{y_{0}}^{y}}
%EndExpansion
dy^{\prime}B_{z}(x^{\prime},y^{\prime},t)+G(y,t)\right\}  +f(x_{0},t)
\label{Lambda(x,y,t)213}%
\end{equation}

\[
with\text{ \ }G(y,t)\text{ \ }such\text{ }that\text{ \ \ }\left\{  {\Huge -}%
%TCIMACRO{\dint \limits_{x_{0}}^{x}}%
%BeginExpansion
{\displaystyle\int\limits_{x_{0}}^{x}}
%EndExpansion
dx^{\prime}%
%TCIMACRO{\dint \limits_{y_{0}}^{y}}%
%BeginExpansion
{\displaystyle\int\limits_{y_{0}}^{y}}
%EndExpansion
dy^{\prime}B_{z}(x^{\prime},y^{\prime},t)+G(y,t)\right\}  \boldsymbol{:}\text{
\ is }\mathsf{independent\ of\ }\ y,
\]
and from this point on, the third equation of the system (\ref{FullSystem}) is
getting involved to determine the nontrivial effect of scalar potentials on
$G(y,t)\boldsymbol{.}$ Indeed, by combining it with (\ref{Lambda(x,y,t)213})
there results a wealth of patterns, one of them leading finally to our first solution%

\[
\Lambda(x,y,t)=\Lambda(x_{0},y_{0},t_{0})+\int_{x_{0}}^{x}A_{x}(x^{\prime
},y_{0},t)dx^{\prime}+\int_{y_{0}}^{y}A_{y}(x,y^{\prime},t)dy^{\prime}-%
%TCIMACRO{\dint \limits_{x_{0}}^{x}}%
%BeginExpansion
{\displaystyle\int\limits_{x_{0}}^{x}}
%EndExpansion
dx^{\prime}%
%TCIMACRO{\dint \limits_{y_{0}}^{y}}%
%BeginExpansion
{\displaystyle\int\limits_{y_{0}}^{y}}
%EndExpansion
dy^{\prime}B_{z}(x^{\prime},y^{\prime},t)+G(y,t_{0})-
\]

\begin{equation}
-c%
%TCIMACRO{\dint \limits_{t_{0}}^{t}}%
%BeginExpansion
{\displaystyle\int\limits_{t_{0}}^{t}}
%EndExpansion
\phi(x_{0},y_{0},t^{\prime})dt^{\prime}+c%
%TCIMACRO{\dint \limits_{t_{0}}^{t}}%
%BeginExpansion
{\displaystyle\int\limits_{t_{0}}^{t}}
%EndExpansion
dt^{\prime}%
%TCIMACRO{\dint \limits_{x_{0}}^{x}}%
%BeginExpansion
{\displaystyle\int\limits_{x_{0}}^{x}}
%EndExpansion
dx^{\prime}E_{x}(x^{\prime},y,t^{\prime})+c%
%TCIMACRO{\dint \limits_{t_{0}}^{t}}%
%BeginExpansion
{\displaystyle\int\limits_{t_{0}}^{t}}
%EndExpansion
dt^{\prime}%
%TCIMACRO{\dint \limits_{y_{0}}^{y}}%
%BeginExpansion
{\displaystyle\int\limits_{y_{0}}^{y}}
%EndExpansion
dy^{\prime}E_{y}(x_{0},y^{\prime},t^{\prime})+F(x,y)+f(x_{0},t_{0})
\label{LambdaFull1}%
\end{equation}
with the functions $G(y,t_{0})$ \ and $\ F(x,y)$ \ to be chosen in such a way
as to satisfy the following 3 independent conditions$\boldsymbol{:}$%

\begin{equation}
\left\{  G(y,t_{0})-%
%TCIMACRO{\dint \limits_{x_{0}}^{x}}%
%BeginExpansion
{\displaystyle\int\limits_{x_{0}}^{x}}
%EndExpansion
dx^{\prime}%
%TCIMACRO{\dint \limits_{y_{0}}^{y}}%
%BeginExpansion
{\displaystyle\int\limits_{y_{0}}^{y}}
%EndExpansion
dy^{\prime}B_{z}(x^{\prime},y^{\prime},t_{0})\right\}  \boldsymbol{:}%
\ is\text{ \ }\mathsf{independent\ of\ }\ y, \label{Gcondition}%
\end{equation}
which is of course a special case of the condition on $G(y,t)$ above (see
after (\ref{Lambda(x,y,t)213})) applied at $t=t_{0}$, and the other 2 turn out
to be of the form%

\begin{equation}
\left\{  F(x,y)+c%
%TCIMACRO{\dint \limits_{t_{0}}^{t}}%
%BeginExpansion
{\displaystyle\int\limits_{t_{0}}^{t}}
%EndExpansion
dt^{\prime}%
%TCIMACRO{\dint \limits_{x_{0}}^{x}}%
%BeginExpansion
{\displaystyle\int\limits_{x_{0}}^{x}}
%EndExpansion
dx^{\prime}E_{x}(x^{\prime},y,t^{\prime})\right\}  \boldsymbol{:}\ is\text{
\ }\mathsf{independent\ of\ }\ x, \label{F(x,y)condition1}%
\end{equation}

\begin{equation}
\left\{  F(x,y)+c%
%TCIMACRO{\dint \limits_{t_{0}}^{t}}%
%BeginExpansion
{\displaystyle\int\limits_{t_{0}}^{t}}
%EndExpansion
dt^{\prime}%
%TCIMACRO{\dint \limits_{y_{0}}^{y}}%
%BeginExpansion
{\displaystyle\int\limits_{y_{0}}^{y}}
%EndExpansion
dy^{\prime}E_{y}(x,y^{\prime},t^{\prime})\right\}  \boldsymbol{:}\ is\text{
\ }\mathsf{independent\ of\ }\ y. \label{F(x,y)condition2}%
\end{equation}
It is probably important to inform the reader that for the above results the
Faraday's law is crucial (see Appendix G). As for the constant
quantity\ $f(x_{0},t_{0})$ appearing in (\ref{LambdaFull1}), it again
describes possible effects of multiple-connectivity at the instant $t_{0}$
(which are absent for simple-connected spacetimes, but will be crucial in the
discussion of the van Kampen thought-experiment to be discussed in the next Section).

\bigskip Eq. (\ref{LambdaFull1}) is our first solution. It is now crucial to
note that an alternative form of solution (with the functions $G^{\prime}s$
and $F$ satisfying the \textit{same} conditions as above) can be derived (see
Appendix G) which turns out to be%

\[
\Lambda(x,y,t)=\Lambda(x_{0},y_{0},t_{0})+\int_{x_{0}}^{x}A_{x}(x^{\prime
},y_{0},t)dx^{\prime}+\int_{y_{0}}^{y}A_{y}(x,y^{\prime},t)dy^{\prime}-%
%TCIMACRO{\dint \limits_{x_{0}}^{x}}%
%BeginExpansion
{\displaystyle\int\limits_{x_{0}}^{x}}
%EndExpansion
dx^{\prime}%
%TCIMACRO{\dint \limits_{y_{0}}^{y}}%
%BeginExpansion
{\displaystyle\int\limits_{y_{0}}^{y}}
%EndExpansion
dy^{\prime}B_{z}(x^{\prime},y^{\prime},t_{0})+G(y,t_{0})-
\]

\begin{equation}
-c%
%TCIMACRO{\dint \limits_{t_{0}}^{t}}%
%BeginExpansion
{\displaystyle\int\limits_{t_{0}}^{t}}
%EndExpansion
\phi(x_{0},y_{0},t^{\prime})dt^{\prime}+c%
%TCIMACRO{\dint \limits_{t_{0}}^{t}}%
%BeginExpansion
{\displaystyle\int\limits_{t_{0}}^{t}}
%EndExpansion
dt^{\prime}%
%TCIMACRO{\dint \limits_{x_{0}}^{x}}%
%BeginExpansion
{\displaystyle\int\limits_{x_{0}}^{x}}
%EndExpansion
dx^{\prime}E_{x}(x^{\prime},y_{0},t^{\prime})+c%
%TCIMACRO{\dint \limits_{t_{0}}^{t}}%
%BeginExpansion
{\displaystyle\int\limits_{t_{0}}^{t}}
%EndExpansion
dt^{\prime}%
%TCIMACRO{\dint \limits_{y_{0}}^{y}}%
%BeginExpansion
{\displaystyle\int\limits_{y_{0}}^{y}}
%EndExpansion
dy^{\prime}E_{y}(x,y^{\prime},t^{\prime})+F(x,y)+f(x_{0},t_{0}).
\label{LambdaFull2}%
\end{equation}
In this alternative solution we note that, in comparison with
(\ref{LambdaFull1}), the line-integrals of $\ \boldsymbol{E}$ \ have changed
to the \textit{other} alternative \textquotedblleft path\textquotedblright%
\ (note the difference in the placement of the coordinates of the initial
point $(x_{0},y_{0})$ in the arguments of $E_{x}$ and $E_{y}$ compared to
(\ref{LambdaFull1})) and they happen to have the same sense as the
$\boldsymbol{A}$-integrals, while simultaneously the magnetic flux difference
shows up with its value at the initial time $t_{0}$ rather than at $t$. This
alternative form will be shown to be useful in cases where we want to directly
compare physical situations in the present (at time $t$) and in the past (at
time $t_{0}$), and the above noted change of sense of $\boldsymbol{E}%
$-integrals (compared to (\ref{LambdaFull1})) will be crucial in the
discussion that follows in the next Section. (It is also important here to
note that, in the form (\ref{LambdaFull2}), the electric fields have already
incorporated the effect of radiated $B_{z}$-fields in space (through the
Maxwell's equations, see Appendix G), and this is why at the end only the
$B_{z}$ at $t_{0}$ appears explicitly).

Once again the reader can directly verify that (\ref{LambdaFull1}) or
(\ref{LambdaFull2}) indeed satisfy the basic input system (\ref{FullSystem}).
(This verification is considerably more tedious than the earlier ones but
rather straightforward).

\bigskip But a last mathematical step remains$\boldsymbol{:}$ in order to
discuss the van Kampen case, namely an enclosed (and physically inaccessible)
magnetic flux (which however is \textit{time-dependent})\textit{, }it is
important to have the analogous forms through a reverse route of integrations
(see Appendix G), where at the end we will have the reverse \textquotedblleft
path\textquotedblright\ of $\boldsymbol{A}$-integrals (so that by taking the
\textit{difference} of the resulting solution and the above solution
(\ref{LambdaFull1}) (or (\ref{LambdaFull2})) will lead to the \textit{closed}
line integral of $\boldsymbol{A}$\textbf{ }which will be immediately related
to the van Kampen's magnetic flux (at the instant $t$)). By following then the
reverse route, and by applying a similar strategy at every intermediate step,
we finally obtain the following solution (the spatially \textquotedblleft
dual\textquotedblright\ of (\ref{LambdaFull1})), namely%

\[
\Lambda(x,y,t)=\Lambda(x_{0},y_{0},t_{0})+\int_{x_{0}}^{x}A_{x}(x^{\prime
},y,t)dx^{\prime}+\int_{y_{0}}^{y}A_{y}(x_{0},y^{\prime},t)dy^{\prime}+%
%TCIMACRO{\dint \limits_{x_{0}}^{x}}%
%BeginExpansion
{\displaystyle\int\limits_{x_{0}}^{x}}
%EndExpansion
dx^{\prime}%
%TCIMACRO{\dint \limits_{y_{0}}^{y}}%
%BeginExpansion
{\displaystyle\int\limits_{y_{0}}^{y}}
%EndExpansion
dy^{\prime}B_{z}(x^{\prime},y^{\prime},t)+\hat{G}(x,t_{0})-
\]

\begin{equation}
-c%
%TCIMACRO{\dint \limits_{t_{0}}^{t}}%
%BeginExpansion
{\displaystyle\int\limits_{t_{0}}^{t}}
%EndExpansion
\phi(x_{0},y_{0},t^{\prime})dt^{\prime}+c%
%TCIMACRO{\dint \limits_{t_{0}}^{t}}%
%BeginExpansion
{\displaystyle\int\limits_{t_{0}}^{t}}
%EndExpansion
dt^{\prime}%
%TCIMACRO{\dint \limits_{x_{0}}^{x}}%
%BeginExpansion
{\displaystyle\int\limits_{x_{0}}^{x}}
%EndExpansion
dx^{\prime}E_{x}(x^{\prime},y_{0},t^{\prime})+c%
%TCIMACRO{\dint \limits_{t_{0}}^{t}}%
%BeginExpansion
{\displaystyle\int\limits_{t_{0}}^{t}}
%EndExpansion
dt^{\prime}%
%TCIMACRO{\dint \limits_{y_{0}}^{y}}%
%BeginExpansion
{\displaystyle\int\limits_{y_{0}}^{y}}
%EndExpansion
dy^{\prime}E_{y}(x,y^{\prime},t^{\prime})+F(x,y)+\hat{h}(y_{0},t_{0})
\label{LambdaFull4}%
\end{equation}
with the functions $\hat{G}(x,t_{0})$ \ and $\ F(x,y)$ \ to be chosen in such
a way as to satisfy the following 3 independent conditions$\boldsymbol{:}$%

\begin{equation}
\left\{  \hat{G}(x,t_{0})+%
%TCIMACRO{\dint \limits_{y_{0}}^{y}}%
%BeginExpansion
{\displaystyle\int\limits_{y_{0}}^{y}}
%EndExpansion
dy^{\prime}%
%TCIMACRO{\dint \limits_{x_{0}}^{x}}%
%BeginExpansion
{\displaystyle\int\limits_{x_{0}}^{x}}
%EndExpansion
dx^{\prime}B_{z}(x^{\prime},y^{\prime},t_{0})\right\}  \boldsymbol{:}%
\ is\ \ \mathsf{independent\ of\ }\ x, \label{Gcondition3}%
\end{equation}

\bigskip%
\begin{equation}
\left\{  F(x,y)+c%
%TCIMACRO{\dint \limits_{t_{0}}^{t}}%
%BeginExpansion
{\displaystyle\int\limits_{t_{0}}^{t}}
%EndExpansion
dt^{\prime}%
%TCIMACRO{\dint \limits_{x_{0}}^{x}}%
%BeginExpansion
{\displaystyle\int\limits_{x_{0}}^{x}}
%EndExpansion
dx^{\prime}E_{x}(x^{\prime},y,t^{\prime})\right\}  \boldsymbol{:}%
\ is\ \ \mathsf{independent\ of\ }\ x, \label{F(x,y)condition3}%
\end{equation}

\bigskip%
\begin{equation}
\left\{  F(x,y)+c%
%TCIMACRO{\dint \limits_{t_{0}}^{t}}%
%BeginExpansion
{\displaystyle\int\limits_{t_{0}}^{t}}
%EndExpansion
dt^{\prime}%
%TCIMACRO{\dint \limits_{y_{0}}^{y}}%
%BeginExpansion
{\displaystyle\int\limits_{y_{0}}^{y}}
%EndExpansion
dy^{\prime}E_{y}(x,y^{\prime},t^{\prime})\right\}  \boldsymbol{:}%
\ is\ \ \mathsf{independent\ of\ }\ y, \label{F(x,y)condition4}%
\end{equation}
where again for the above results the Faraday's law is crucial (see Appendix
G). The corresponding analog of the alternative form (\ref{LambdaFull2})
(where $B_{z}$ appears at $t_{0}$) is more important (for the discussion of
the next Section) and turns out to be%

\[
\Lambda(x,y,t)=\Lambda(x_{0},y_{0},t_{0})+\int_{x_{0}}^{x}A_{x}(x^{\prime
},y,t)dx^{\prime}+\int_{y_{0}}^{y}A_{y}(x_{0},y^{\prime},t)dy^{\prime}+%
%TCIMACRO{\dint \limits_{x_{0}}^{x}}%
%BeginExpansion
{\displaystyle\int\limits_{x_{0}}^{x}}
%EndExpansion
dx^{\prime}%
%TCIMACRO{\dint \limits_{y_{0}}^{y}}%
%BeginExpansion
{\displaystyle\int\limits_{y_{0}}^{y}}
%EndExpansion
dy^{\prime}B_{z}(x^{\prime},y^{\prime},t_{0})+\hat{G}(x,t_{0})-
\]

\begin{equation}
-c%
%TCIMACRO{\dint \limits_{t_{0}}^{t}}%
%BeginExpansion
{\displaystyle\int\limits_{t_{0}}^{t}}
%EndExpansion
\phi(x_{0},y_{0},t^{\prime})dt^{\prime}+c%
%TCIMACRO{\dint \limits_{t_{0}}^{t}}%
%BeginExpansion
{\displaystyle\int\limits_{t_{0}}^{t}}
%EndExpansion
dt^{\prime}%
%TCIMACRO{\dint \limits_{x_{0}}^{x}}%
%BeginExpansion
{\displaystyle\int\limits_{x_{0}}^{x}}
%EndExpansion
dx^{\prime}E_{x}(x^{\prime},y,t^{\prime})+c%
%TCIMACRO{\dint \limits_{t_{0}}^{t}}%
%BeginExpansion
{\displaystyle\int\limits_{t_{0}}^{t}}
%EndExpansion
dt^{\prime}%
%TCIMACRO{\dint \limits_{y_{0}}^{y}}%
%BeginExpansion
{\displaystyle\int\limits_{y_{0}}^{y}}
%EndExpansion
dy^{\prime}E_{y}(x_{0},y^{\prime},t^{\prime})+F(x,y)+\hat{h}(y_{0},t_{0})
\label{LambdaFIN}%
\end{equation}
with $\hat{G}(x,t_{0})$ and $F(x,y)$ following the same 3 conditions above.
The constant term $\hat{h}(y_{0},t_{0})$ again describes possible
multiplicities at the instant $t_{0}\boldsymbol{;}$ it is absent for
simple-connected spacetimes, but will be crucial in the discussion of the van
Kampen thought-experiment.

In (\ref{LambdaFull4}) (and in (\ref{LambdaFIN})), note the \textquotedblleft
alternative paths\textquotedblright\ (compared to solution (\ref{LambdaFull1})
(and (\ref{LambdaFull2}))) of line integrals of $\boldsymbol{A}$'s (or of
$\boldsymbol{E}$'s). But the most crucial element for what follows is the need
to \textit{exclusively} use the forms (\ref{LambdaFull2}) and (\ref{LambdaFIN}%
) (where $B_{z}$ only appears at $t_{0}$), and the fact that, within each
solution, the sense of $\boldsymbol{A}$-integrals is the \textit{same} as the
sense of the $\boldsymbol{E}$-integrals. (This is \textit{not} true in the
other solutions where $B_{z}(..,t)$ appears, as the reader can directly see).
These facts will be crucial to the discussion that follows, which briefly
addresses the so called \textquotedblleft van Kampen paradox\textquotedblright.

\section{\bigskip The van Kampen thought-experiment -- how the above solutions
enforce Causality}

In an early work\cite{vanKampen} van Kampen considered a genuine Aharonov-Bohm
case, with a magnetic flux (physically inaccessible to the particle) which,
however, is time-dependent$\boldsymbol{:}$ van Kampen envisaged turning on the
flux very late, or equivalently, observing the interference of the two
wavepackets (on a distant screen) very early, earlier than the time it takes
light to travel the distance to the screen, hence using the (instantaneous
nature of the) Aharonov-Bohm phase to transmit information (on the existence
of a confined magnetic flux somewhere in space) \textit{superluminally}.
Indeed, the Aharonov-Bohm phase at any later instant $t$ is determined by
differences of $\frac{q}{\hbar c}\Lambda(\mathbf{r},t)$, \ with $\ \Lambda
(\mathbf{r},t)=\int_{\mathbf{r}_{0}}^{\mathbf{r}}\boldsymbol{A}(\mathbf{r}%
^{\prime},t)\boldsymbol{.}d\mathbf{r}^{\prime}+$ $const.$ (which basically
results as a special case (but in higher dimensionality) of the incorrect
expression (\ref{BrownHolland}) (or (\ref{wrong})) in the temporal gauge
$\phi=0$, the constant being $\Lambda(\mathbf{r}_{0},t_{0})$). However, let us
for this case utilize instead our results (\ref{LambdaFull2}) and
(\ref{LambdaFIN}) above, where we have the additional appearance of the
nonlocal $E$-terms (and of the $B_{z}$-term at $t_{0}$).

In order to be slightly more general than van Kampen, let us for example
assume that the inaccessible magnetic flux had the value $\Phi(t_{0})$ at
$t_{0}$, and then it started changing with time. By using a narrow wavepacket
picture like van Kampen, we can then subtract (\ref{LambdaFull2}) and
(\ref{LambdaFIN}) in order to find the phase difference at a time $t$ that is
smaller than the time required for light to reach the observation point
$(x,y)$ (i.e. $t<$ $\frac{L}{c}$, with $L$ the corresponding distance). For a
spatially-confined magnetic flux $\Phi(t)$, the functions $G,$ $\hat{G}$ and
$F$ in the above solutions can then all be taken zero$\boldsymbol{:}$
\textbf{(i)} their conditions are all satisfied for a flux $\Phi(t)$ that is
not spatially-extended (hence, from (\ref{Gcondition}) and (\ref{Gcondition3})
we obtain $G=\hat{G}=0$ since the integrals in brackets are all independent of
$x$ and $y$), and \textbf{(ii)} for $t<$ $\frac{L}{c}$, the integrals of
$E_{x}$ and $E_{y}$ in conditions (\ref{F(x,y)condition1}) and
(\ref{F(x,y)condition2}) (or in (\ref{F(x,y)condition3}) and
(\ref{F(x,y)condition4})) are already independent of \textit{both} $x$ and $y$
(since $E_{x}(x,y,t^{\prime})=E_{y}(x,y,t^{\prime})=0$ for all $t^{\prime
}<t<\frac{L}{c}$, with $(x,y)$ the observation point (since at instant $t$,
the $\boldsymbol{E}$-field has not yet reached the spatial point $(x,y)$ of
the screen), and therefore all integrations of $E_{x}$ and $E_{y}$ with
respect to $x^{\prime}$ and $y^{\prime}$ will be contributing only up to a
light-cone (see Fig.2) and they will therefore give results that are
\textit{independent of the integration upper limits }$x$\textit{ and }$y$ $-$
basically a generalization of the striped cases that we saw earlier but now to
the case of 3 spatio-temporal variables (with now the spatial point $(x,y)$
being outside the light-cone defined by $t$ (see Fig.2; in this Figure the
initial spatial point $(x_{0},y_{0})$, taken for simplicity at $(0,0)$, has
been supposed to be in the area of the inaccessible flux $\Phi(t)$, so that,
for $R=\sqrt{(x-x_{0})^{2}+(y-y_{0})^{2}}$, we have indeed that $R\sim L$,
hence that $ct$ (which is $<L$, see above) is also $<R$, as written on the
Figure)))$\boldsymbol{;}$ we therefore rigorously obtain $F=0$). Moreover, the
multiplicities $(f$ and $\hat{h})$ \ lead to cancellation of the $B_{z}$-terms
(at $t_{0}$) in exactly the same manner as outlined in the static case earlier
(at the end of Section X). By choosing then the temporal gauge $\phi=0$ like
van Kampen$,$ we have for the difference (\ref{LambdaFull2}) $-$
(\ref{LambdaFIN}) at the point and instant of observation the following result%
%TCIMACRO{\FRAME{ftbpF}{3.0614in}{2.2952in}{0pt}{}{}{figure2.eps}%
%{\special{ language "Scientific Word";  type "GRAPHIC";
%maintain-aspect-ratio TRUE;  display "USEDEF";  valid_file "F";
%width 3.0614in;  height 2.2952in;  depth 0pt;  original-width 6.3027in;
%original-height 4.7115in;  cropleft "0";  croptop "1";  cropright "1";
%cropbottom "0";  filename 'Figure2.eps';file-properties "XNPEU";}}}%
%BeginExpansion
\begin{figure}[ptb]%
\centering
\includegraphics[
height=2.2952in,
width=3.0614in
]%
{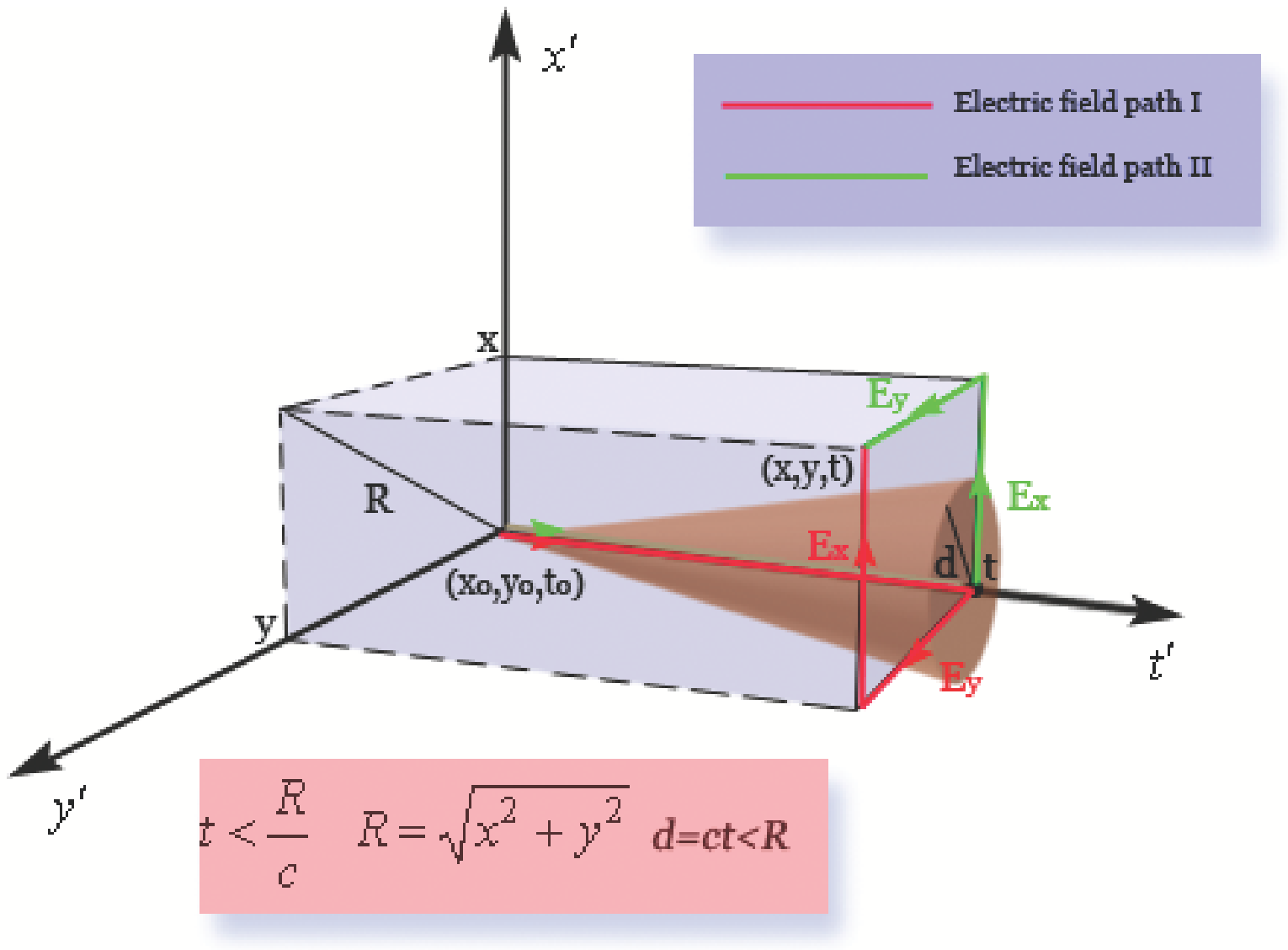}%
\end{figure}
%EndExpansion
%

\[
\Delta\Lambda(x,y,t)=\int_{x_{0}}^{x}A_{x}(x^{\prime},y_{0},t)dx^{\prime}%
+\int_{y_{0}}^{y}A_{y}(x,y^{\prime},t)dy^{\prime}-\int_{x_{0}}^{x}%
A_{x}(x^{\prime},y,t)dx^{\prime}-\int_{y_{0}}^{y}A_{y}(x_{0},y^{\prime
},t)dy^{\prime}+
\]

\begin{equation}
+c%
%TCIMACRO{\dint \limits_{t_{0}}^{t}}%
%BeginExpansion
{\displaystyle\int\limits_{t_{0}}^{t}}
%EndExpansion
dt^{\prime}\left\{
%TCIMACRO{\dint \limits_{x_{0}}^{x}}%
%BeginExpansion
{\displaystyle\int\limits_{x_{0}}^{x}}
%EndExpansion
dx^{\prime}E_{x}(x^{\prime},y_{0},t^{\prime})+%
%TCIMACRO{\dint \limits_{y_{0}}^{y}}%
%BeginExpansion
{\displaystyle\int\limits_{y_{0}}^{y}}
%EndExpansion
dy^{\prime}E_{y}(x,y^{\prime},t^{\prime})-%
%TCIMACRO{\dint \limits_{x_{0}}^{x}}%
%BeginExpansion
{\displaystyle\int\limits_{x_{0}}^{x}}
%EndExpansion
dx^{\prime}E_{x}(x^{\prime},y,t^{\prime})-%
%TCIMACRO{\dint \limits_{y_{0}}^{y}}%
%BeginExpansion
{\displaystyle\int\limits_{y_{0}}^{y}}
%EndExpansion
dy^{\prime}E_{y}(x_{0},y^{\prime},t^{\prime})\right\}  . \label{DeltaLambda}%
\end{equation}
In (\ref{DeltaLambda}) the sum of the four $A$-integrals gives the
\textit{closed} line-integral of vector $\boldsymbol{A}$ around the
\textit{observation rectangle }at time $t$ (in the positive sense) and it is
equal to the instantaneous magnetic flux $\Phi(t)$ (that leads to the
\textquotedblleft usual\textquotedblright\ magnetic Aharonov-Bohm
phase)$\boldsymbol{;}$ the sum of the four $E$-integrals inside the brackets
in the last terms (originating from our nonlocal contributions) gives the
\textit{closed} line-integral of vector $\boldsymbol{E}$ around the same
rectangle at any arbitrary $t^{\prime}$, and in the same (positive) sense
(something we wouldn't have if we had taken the first type of solutions,
(\ref{LambdaFull1}) and (\ref{LambdaFull4}) $-$ this signifying the importance
of taking the right form, the one that contains $B_{z}$ at $t_{0}$ (with the
$t$-propagation of $B_{z}$ \textit{having already been incorporated} in the
$E_{x}$ and $E_{y}$ terms of (\ref{LambdaFull2}) and (\ref{LambdaFIN}))). By
denoting therefore the closed loop integral (around the rectangle) as $%
%TCIMACRO{\doint }%
%BeginExpansion
{\displaystyle\oint}
%EndExpansion
$ always in the positive sense (and with the understanding that the
rectangle's upper right corner is the spatial point of observation $(x,y)$),
(\ref{DeltaLambda}) reads%

\begin{equation}
\Delta\Lambda(x,y,t)=%
%TCIMACRO{\doint }%
%BeginExpansion
{\displaystyle\oint}
%EndExpansion
\mathbf{A}(\mathbf{r}^{\prime},t)\boldsymbol{.}d\mathbf{r}^{\prime}+c%
%TCIMACRO{\dint \limits_{t_{0}}^{t}}%
%BeginExpansion
{\displaystyle\int\limits_{t_{0}}^{t}}
%EndExpansion
dt^{\prime}%
%TCIMACRO{\doint }%
%BeginExpansion
{\displaystyle\oint}
%EndExpansion
\mathbf{E}(\mathbf{r}^{\prime},t^{\prime})\boldsymbol{.}d\mathbf{r}^{\prime}
\label{DeltaLambdaBrief}%
\end{equation}
which, with $%
%TCIMACRO{\doint }%
%BeginExpansion
{\displaystyle\oint}
%EndExpansion
\mathbf{A}(\mathbf{r}^{\prime},t)\boldsymbol{.}d\mathbf{r}^{\prime}=\Phi(t)$
\ the instantaneous enclosed magnetic flux and with the help of Faraday's law
$%
%TCIMACRO{\doint }%
%BeginExpansion
{\displaystyle\oint}
%EndExpansion
\mathbf{E}(\mathbf{r}^{\prime},t^{\prime})\boldsymbol{.}d\mathbf{r}^{\prime
}=-\frac{1}{c}\frac{d\Phi(t^{\prime})}{dt^{\prime}},$ \ gives%

\begin{equation}
\Delta\Lambda(x,y,t)=\Phi(t)-{\huge (}\Phi(t)-\Phi(t_{0}){\huge )}=\Phi
(t_{0}). \label{DeltaLambdaFinal}%
\end{equation}
Although $\Delta\Lambda$ is generally $t$-dependent, we obtain the intuitive
(causal) result that, for $t<\frac{L}{c}$ (i.e. if the physical information
has not yet reached the screen), the phase-difference turns out to be
$t$-independent, and leads to the magnetic Aharonov-Bohm\ phase that we
\textit{would} observe at $t_{0}$. The new nonlocal terms have conspired in
such a way as to\textit{ exactly cancel }the Causality-violating Aharonov-Bohm
phase (that would be proportional to the instantaneous $\Phi(t)$).

This gives an honest resolution of the \textquotedblleft van Kampen
paradox\textquotedblright\ within a canonical formulation, without using any
vague electric Aharonov-Bohm effect argument as was done by van Kampen (since
in the gauge chosen $(\phi=0)$ there are no scalar potentials\ -- and, most
importantly, \textit{there is no multiple-connectivity in }$(x,t)$%
\textit{-plane} as in the electric Aharonov-Bohm case\cite{Iddings}). In this
van Kampen thought experiment the particle actually passes through the
electric and magnetic fields that are radiated outside the confined magnetic
flux, and the electric type of phase that recovers Causality is actually an
example of our nonlocal terms. The recovery of Causality is the result of the
action of these new nonlocal terms, in a type of \textquotedblleft generalized
Werner \& Brill cancellation in spacetime\textquotedblright\ (the earlier
strips having now given their place to a light-cone). An additional physical
element (in comparison to van Kampen's electric phase interpretation) is that,
for the above cancellation, it is not only the $E$-fields but also the
$t$-propagation in space of the $B_{z}$-fields (the full \textquotedblleft
radiation field\textquotedblright) that plays a role.

Finally, a number of other forms of solutions can be obtained that result from
different ordering of integrations of the system (\ref{FullSystem}) (a full
list of 12 different (but quite long) results is available, directly
verifiable that they satisfy the system (\ref{FullSystem})). The reader can
follow the strategies of solution suggested here and derive the forms that are
appropriate to particular physical cases of interest that may be different
from the above magnetic case, some potential candidates being the
\textquotedblleft electric analog\textquotedblright\ of the van Kampen
thought-experiment, or its \textit{bound state analog} in nanorings. For the
former we pay particular attention below$\boldsymbol{;}$ for the latter, and
especially for 1-D nanorings (or other nanoscopic devices) driven by a
$t$-dependent magnetic flux, the new nonlocal terms are expected to be of
relevance if they are included in standard treatments\cite{LuanTang}, and the
effects are expected to appear in the PetaHertz range. (Similarly we might
expect a nontrivial role in cases of quantal astrophysical objects due to the
large distances involved (hence retardation effects being more pronounced)).

For the \textquotedblleft electric analog\textquotedblright\ of the van Kampen
case, we note that, although this has never really been discussed in the
literature (in such terminology), nevertheless, it has been essentially
briefly mentioned in Appendix B of Peshkin\cite{Peshkin} (where the point is
made about what happens when \textit{first} the particle exits the cages, and
\textit{only then} we switch on the outside electric field, together with the
comment of the author that the results must be \textquotedblleft consistent
with ordinary ideas about Causality\textquotedblright$\boldsymbol{;}$ Peshkin
correctly states: \textquotedblleft One cannot wait for the electron to pass
and only later switch on the field to cause a physical
effect\textquotedblright). As our new nonlocal terms seem to be especially
suited for addressing such Causality issues, let us slightly expand on this
point$\boldsymbol{:}$ in this most authoritative (and carefully written)
review of the Aharonov-Bohm effect in the literature, Peshkin uses (for the
electric effect) a solution-form (his eq.(B.5) together with (B.6)) based on
(\ref{BrownHolland}), i.e. the \textquotedblleft standard
result\textquotedblright\ (but applied to a spatially-dependent scalar
potential) $-$ but he clearly states that it is an approximation (and actually
later in the review, he states that this form cannot be a solution for all
$t$). Indeed, from the present work we learn that Peshkin's eqs (B.5) and
(B.6) do \textit{not} give the solution when the scalar potential depends on
spatial variables (because the spatial variables inside the potential will
give $-$ through its nonzero gradient $-$ an extra vector potential (that will
result from $\nabla\Lambda$), hence an extra minimal substitution in the
Hamiltonian $H$, violating therefore the mapping between two pre-determined
systems that we want to achieve). As we saw in the present work, the correct
solution for all $t$ and in all space consists of additional nonlocal terms of
the appropriate form. If we view the form (B.5) and (B.6) of ref.
\cite{Peshkin} as an \textit{ansatz}, then it is understandable why a
\textit{condition} (Peshkin's eq.(B.8), and later (B.9)) needs to be
\textit{enforced} on the electric field outside the cages (in order for the
extra (annoying) terms (that show up from expansion of the squared minimal
substitution) to vanish and for (B.5) to be a solution). And then Peshkin
notes that the extra condition cannot always be satisfied $-$ \textit{it must
fail} for some times (hence (B.5) is not really the solution for all times),
drawing from this a correct conclusion, namely that \textquotedblleft the
electron must traverse some region where the electric field \textit{has
been}\textquotedblright\ (earlier). However, the causal feature pointed out
above, although mentioned in words, is not dealt with quantitatively. From our
present work, it turns out that the total \textquotedblleft radiation
field\textquotedblright\ outside the cages is crucial in recovering Causality,
in a similar way as in the case presented above in this Section for the usual
(magnetic) version of the van Kampen experiment. In this \textquotedblleft
electric analog\textquotedblright\ that we are discussing now, the
causally-offending part of the electric Aharonov-Bohm phase difference will be
cancelled by a magnetic type of phase, that originates from the magnetic field
that is associated with the $t$-dependence of the electric field
$\boldsymbol{E}$ outside the cages.

It should be re-emphasized that the correct quantitative physical behavior of
the above system for all times comes out from the treatment shown in detail in
the present work, with no enforced constraints, but with conditions that come
out naturally from the solution of the PDEs. The results that are derived from
this careful procedure give the full solutions (correct for all space and for
all $t$)$\boldsymbol{:}$ Peshkin's ansatz (B.6) turns out (from an honest and
careful solution of the full PDEs) to be augmented by nonlocal terms of the
electric fields, and these directly influence the phases of wavefunctions (by
always respecting Causality, with no need of enforced conditions) -- and can
even include the contributions of vector potentials and magnetic fields
(through nonlocal magnetic terms in space) associated with the $t$-variation
of the electric field outside the cages, that Peshkin has omitted (as he
actually admits in the beginning of his Appendix). As already mentioned, the
total \textquotedblleft radiation field\textquotedblright\ outside the cages
is crucial in recovering Causality, in a way similar to what was presented
above for the usual (magnetic) version of the van Kampen experiment. We
conclude that our (exact) results accomplish precisely what Peshkin has in
mind in his discussion (on Causality), but in a direct and fully quantitative
manner, and with \textit{no ansatz} based on an incorrect form.

\section{Physical Discussion}

\bigskip In attempting to evaluate in a broader sense the crucial nonlocal
influences found in all the above physical examples, we should probably first
reemphasize that at the level of the basic Lagrangian $L(\mathbf{r}%
,\mathbf{v},t)=\frac{1}{2}m\mathbf{v}^{2}+\frac{q}{c}\mathbf{v}\boldsymbol{.}%
\mathbf{A}(\mathbf{r},t)-q\phi(\mathbf{r},t)$ \ there are no fields present,
and the view holds in the literature\cite{BrownHome} that electric or magnetic
fields cannot contribute \textit{directly} to the phase of quantum
wavefunctions. This view originates from the path-integral treatments widely
used (where the Lagrangian determines directly the phases of Propagators),
but, nevertheless, our canonical formulation treatment shows that fields
\textit{do} contribute nonlocally, and they are actually crucial in recovering
Relativistic Causality. Moreover, path-integral discussions\cite{Troudet} of
the van Kampen case use wave (retarded)-solutions for the vector potentials
$\mathbf{A}$ (hence they are treated in Lorenz gauge, which is not
sufficiently general$\boldsymbol{:}$ even if $\mathbf{A}$ has not yet reached
the screen, we can always add a constant $\mathbf{A}$ (a pure gauge) over all
space, and there are no more retarded wave-solutions for the potentials, the
proposed path-integral resolution of the paradox\cite{Troudet} being,
therefore, at least incomplete). Our results are gauge-invariant and take
advantage of only the retardation of \textit{fields} $\mathbf{E}$ and
$\mathbf{B}$\textbf{ }(true in \textit{any} gauge), and \textit{not }of
potentials. In addition, Troudet\cite{Troudet} clearly (and correctly) states
that his treatment is good for not highly-delocalized states in space, and
that in case of delocalization the proper treatment \textquotedblleft would be
much more complicated, and would require a much more complete
analysis\textquotedblright. It is clear that we have provided such a complete
analysis in the present work. It should be added that in a recent Compendium
of Quantum Physics\cite{Compendium}, the \textquotedblleft van Kampen
paradox\textquotedblright\ still seems to be thought of as remarkable. It is
fair to state that the present work has provided a natural and general
resolution, and most importantly, through nonlocal and Relativistically causal
propagation of wavefunction phases in the Schr\"{o}dinger Picture\textbf{
}(this point being expanded further below, at the end of the article).

At several places in this paper we have pointed out a number of
\textquotedblleft misconceptions\textquotedblright\ in the literature (mostly
on the uncritical use of the (standard) Dirac phases even for $t$-dependent
vector potentials and spatially-dependent scalar potentials, which is plainly
incorrect for uncorrelated variables), and we have explicitly provided their
\textquotedblleft healing\textquotedblright\ through appropriate nonlocal
field-terms. It should be re-emphasized here that this is not a merely
marginal misconception, but it appears all over the place in the literature
(due to the Feynman path-integral bias)$\boldsymbol{;}$ it is even stated by
Feynman himself in volume II of his \textit{Lectures on Physics}%
\cite{Feynman}, namely\textit{\ }that the simple phase factor $\int%
^{x}\boldsymbol{A}\cdot d\mathbf{r}^{\prime}-c\int^{t}\phi dt^{\prime}$ is
valid generally, i.e. even for $t$-dependent fields. Similarly, this erroneous
generalization is also explicitly stated in the review on Aharonov-Bohm
effects of Erlichson\cite{Erlichson} that has given a very balanced view of
earlier controversy, and also elsewhere $-$ the books of
Silverman\cite{Silverman} being the clearest case that we are aware of with a
careful wording about (\ref{BrownHolland}) being only restrictedly valid (for
$t$-independent $\boldsymbol{A}$'s and $\mathbf{r}$-independent $\phi$'s),
although even there the nonlocal terms have been missed. We believe that the
above misconceptions (and the overlooking of the nonlocal terms) are the basic
reason why \textquotedblleft it appears that no exact theoretical treatment
has been given\textquotedblright\ (for the electric Aharonov-Bohm effect), as
correctly stated by Peshkin in his Appendix B of Ref.\cite{Peshkin}.

And let us now come to a second type of misconception, that has appeared only
in semiclassical conditions $-$ but is essential to mention here, as it is
another example that exhibits the merits of our approach (and the deeper
physical understanding that our results can lead to). What we learn from the
generalized Werner \& Brill cancellations pointed out rather emphatically in
the present work is that, at the point of observation, the nonlocal terms of
classical remote fields have the tendency to contribute a phase \textit{of
opposite sign} to the \textquotedblleft Aharonov-Bohm phase\textquotedblright%
\ (of potentials). We want to point out to the reader that, for semiclassical
trajectories, this is actually descriptive of the compatibility (or
consistency) of the Aharonov-Bohm fringe-displacement and the associated
trajectory-deflection due to the classical (Lorentz) forces. Let us for
example look at Fig.15-8 of Feynman\cite{Feynman2}, or at Fig.2.16 of
Felsager\cite{Felsager}, where, classical trajectories are deflected after
they pass through a strip of a homogeneous magnetic field that is placed on
the right of a standard double-slit experimental apparatus (see also our own
Fig.3). Both authors determine the semiclassical phase picked up by the
trajectories (that have been deflected by the Lorentz force) and they find
that they are consistent with the Aharonov-Bohm phase (picked up due to the
flux enclosed by the same trajectories). However, it is not very difficult to
see that the two phases \textit{have opposite sign }(they are \textit{not
equal} as implied by the authors). The reader is also invited to carry out a
similar exercise, with particles passing through an analogous homogeneous
\textit{electric} field on the right of the double-slit apparatus, with the
field being \textit{parallel} to the screen and being switched on for a finite
duration $T\boldsymbol{:}$ it then turns out again that the semiclassical
phase picked up is \textit{opposite} to the electric Aharonov-Bohm type of
phase (in case this is not immediately clear, a quantitative discussion is
given further below). Similarly, in the recent review of Batelaan \&
Tonomura\cite{Batelaan}, their Fig.2 contains visual information that is very
relevant to our discussion: it is a quite descriptive picture of the
wavefronts associated to the classical trajectories, where the authors state
that \textquotedblleft the phase shift calculated in terms of the Lorentz
force is the same as that predicted by the Aharonov-Bohm effect in terms of
the vector potential $A$ circling the magnetic bar\textquotedblright. The
reader, however, should notice once more that the sign of the classical
phase-difference is really opposite to the sign of the Aharonov-Bohm phase.
The phases are \textit{not equal} as stated, but opposite (see below for a
detailed proof). All the above examples may be viewed as a manifestation \ of
the cancellations that have been derived in the present work (for general
quantum states), but here they are just special cases for semiclassical
trajectories. (We could also restate here that these cancellations have to do
with the known rigid displacement of the \textquotedblleft single-slit
envelope\textquotedblright\ of the two-slit diffraction pattern in a
double-slit experiment, when the particle actually passes through an
additional strip of a magnetic field that has been placed on the right of the
apparatus).%
%TCIMACRO{\FRAME{ftbpF}{3.0614in}{2.2952in}{0pt}{}{}{figure3.eps}%
%{\special{ language "Scientific Word";  type "GRAPHIC";
%maintain-aspect-ratio TRUE;  display "USEDEF";  valid_file "F";
%width 3.0614in;  height 2.2952in;  depth 0pt;  original-width 6.3027in;
%original-height 4.7115in;  cropleft "0";  croptop "1";  cropright "1";
%cropbottom "0";  filename 'Figure3.eps';file-properties "XNPEU";}}}%
%BeginExpansion
\begin{figure}[ptb]%
\centering
\includegraphics[
height=2.2952in,
width=3.0614in
]%
{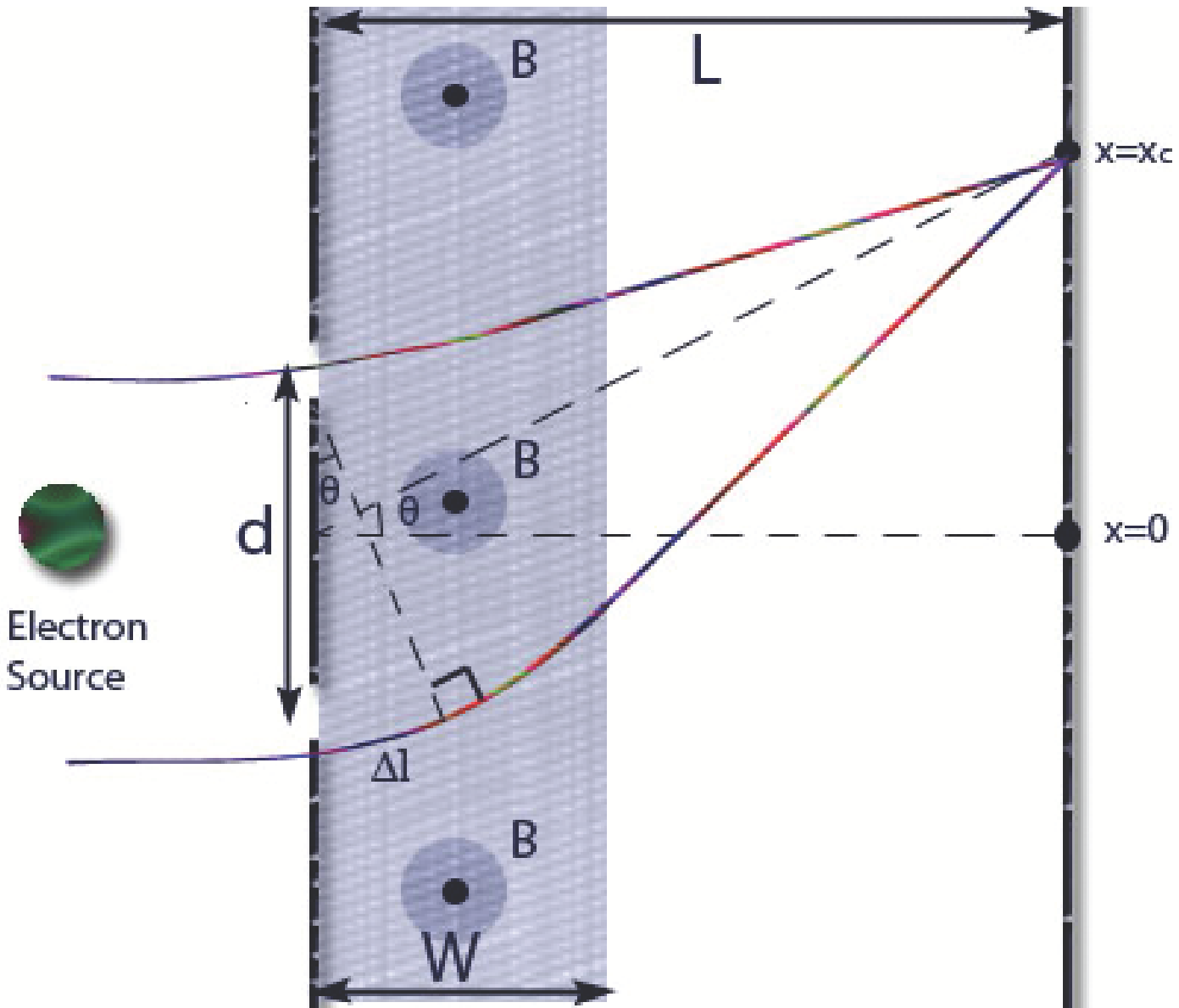}%
\end{figure}
%EndExpansion

In case that the reader does not easily see the signs of the relevant phase
differences, we provide below elementary proofs of the opposite signs argued
above for semiclassical trajectories (and for the special case of small
deflections, as usually done in elementary discussions of the standard
double-slit setup). First, our Fig.3 demonstrates, by way of an example, the
spatial point where the new position of the \textit{central fringe} is now
located (after the classical trajectories have been deflected by the
additional magnetic strip $B$). It is shown below why the two phases
(semiclassical and Aharonov-Bohm) must indeed be opposite (not equal) so that,
in this new fringe position, the \textit{total} phase difference (i.e. the sum
of the above two) is again zero (as it actually \textit{should} be for the
central fringe). Indeed, if $d$ is the distance between the two slits, and $W$
the width of the magnetic strip (assumed to be $W<<L$ so that the deflections
are very small), we have that the \textquotedblleft Aharonov-Bohm
phase\textquotedblright\ enclosed between the two classical trajectories (of a
particle of charge $q$) is
\begin{equation}
\Delta\varphi^{AB}=2\pi\frac{q}{e}\frac{\Phi}{\Phi_{0}}, \label{ABphase}%
\end{equation}
with $\Phi_{0}=\frac{hc}{e}$ the flux quantum, and $\Phi\thickapprox BWd$ the
enclosed flux between the two trajectories (always for small
trajectory-deflections), with the deflection originating from the presence of
the magnetic strip $B$ and the associated Lorentz forces. On the other hand,
the semiclassical phase generally picked up by a trajectory of length $l$ is
$\varphi^{semi}=\frac{2\pi}{\lambda}l$, with $\lambda=\frac{h}{\Pi}$ being the
de Broglie wavelength (and $\Pi$ being the classical kinematic momentum $mv$,
with $v$ the speed of the particle, taken almost constant (as usually done)
due to the small deflections). The semiclassical phase difference between the
2 classical trajectories is therefore $\Delta\varphi^{semi}=\frac{2\pi
}{\lambda}\Delta l$ (with $\Delta l$ the difference between the 2
semiclassical paths, which in Fig.3 is $\Delta l\thickapprox d\sin
\theta\thickapprox d\frac{x_{c}}{L}$, with $x_{c}$ being the (displaced)
position of the central fringe on the screen, and $L$ the distance between the
slit-plane and the screen (note that in Fig.3 we have electrons (hence
$q=-e<0$), the deflections are therefore upward, and we have considered the
semiclassical phase difference between the lower trajectory and the upper
trajectory (the lower one has a longer path, hence it picks up a higher phase,
hence $\Delta\varphi^{semi}>0\qquad\checkmark$))). We have therefore
\begin{equation}
\Delta\varphi^{semi}=\frac{2\pi}{\lambda}d\frac{x_{c}}{L}. \label{Semiphase}%
\end{equation}
Now, the Lorentz force (exerted only during the passage through the thin
magnetic strip, hence only during a time interval $\Delta t=\frac{W}{v}$) has
a component parallel to the screen (let us call it $x$-component) that is
given by
\begin{equation}
F_{x}=\frac{q}{c}(\boldsymbol{v}\times\boldsymbol{B})_{x}=-\frac{q}%
{c}vB=-\frac{BWq}{c\frac{W}{v}}=-\frac{BWq}{c\Delta t} \label{Fx}%
\end{equation}
which shows that there is a change of kinematic momentum (parallel to the
screen) $\Delta\Pi_{x}=-\frac{BWq}{c},$ or, equivalently, a change of parallel speed%

\begin{equation}
\Delta v_{x}=-\frac{BWq}{mc} \label{Dvx}%
\end{equation}
which is the speed of the central fringe's motion (i.e. its displacement over
time along the screen). Although this has been caused by the presence of the
thin deflecting magnetic strip, this displacement is occuring uniformly during
a time interval $t=\frac{L}{v},$ and this time interval must satisfy%

\begin{equation}
\Delta v_{x}=\frac{x_{c}}{t} \label{Dvx2}%
\end{equation}
(as, for small displacements, the beams travel most of the time in uniform
motion, i.e. $\Delta t<<t$). We therefore have that the central fringe
displacement must be $x_{c}=\Delta v_{x}t=-\frac{BWq}{cm}\frac{L}{v},$ and
noting that $mv=\Pi=\frac{h}{\lambda}$, we finally have%

\begin{equation}
x_{c}=-\frac{BWqL\lambda}{hc} \label{xc}%
\end{equation}
(and we note that this displacement is indeed upward (positive) for a negative
charge ($q<0$)). By susbstituting (\ref{xc}) into (\ref{Semiphase}), the
lengths $L$ and $\lambda$ cancel out, and we finally have $\Delta
\varphi^{semi}=-2\pi\frac{q}{e}\frac{BWd}{\frac{hc}{e}},$ which with
$\frac{hc}{e}=\Phi_{0}$ the flux quantum, and $BWd\thickapprox\Phi$ the
enclosed flux (always for small trajectory-deflections) gives (through
comparison with (\ref{ABphase})) our final proof that%

\begin{equation}
\Delta\varphi^{semi}=-2\pi\frac{q}{e}\frac{\Phi}{\Phi_{0}}=-\Delta\varphi
^{AB}. \label{proof}%
\end{equation}
We note therefore that the semiclassical phase difference (between two
trajectories) picked up due to the Lorentz force (exerted on them) is indeed
opposite to the Aharonov-Bohm phase due to the magnetic flux enclosed between
the same trajectories.%

%TCIMACRO{\FRAME{ftbpF}{3.0614in}{2.2952in}{0in}{}{}{figure4.eps}%
%{\special{ language "Scientific Word";  type "GRAPHIC";
%maintain-aspect-ratio TRUE;  display "USEDEF";  valid_file "F";
%width 3.0614in;  height 2.2952in;  depth 0in;  original-width 6.3027in;
%original-height 4.7115in;  cropleft "0";  croptop "1";  cropright "1";
%cropbottom "0";  filename 'Figure4.eps';file-properties "XNPEU";}}}%
%BeginExpansion
\begin{figure}[ptb]%
\centering
\includegraphics[
height=2.2952in,
width=3.0614in
]%
{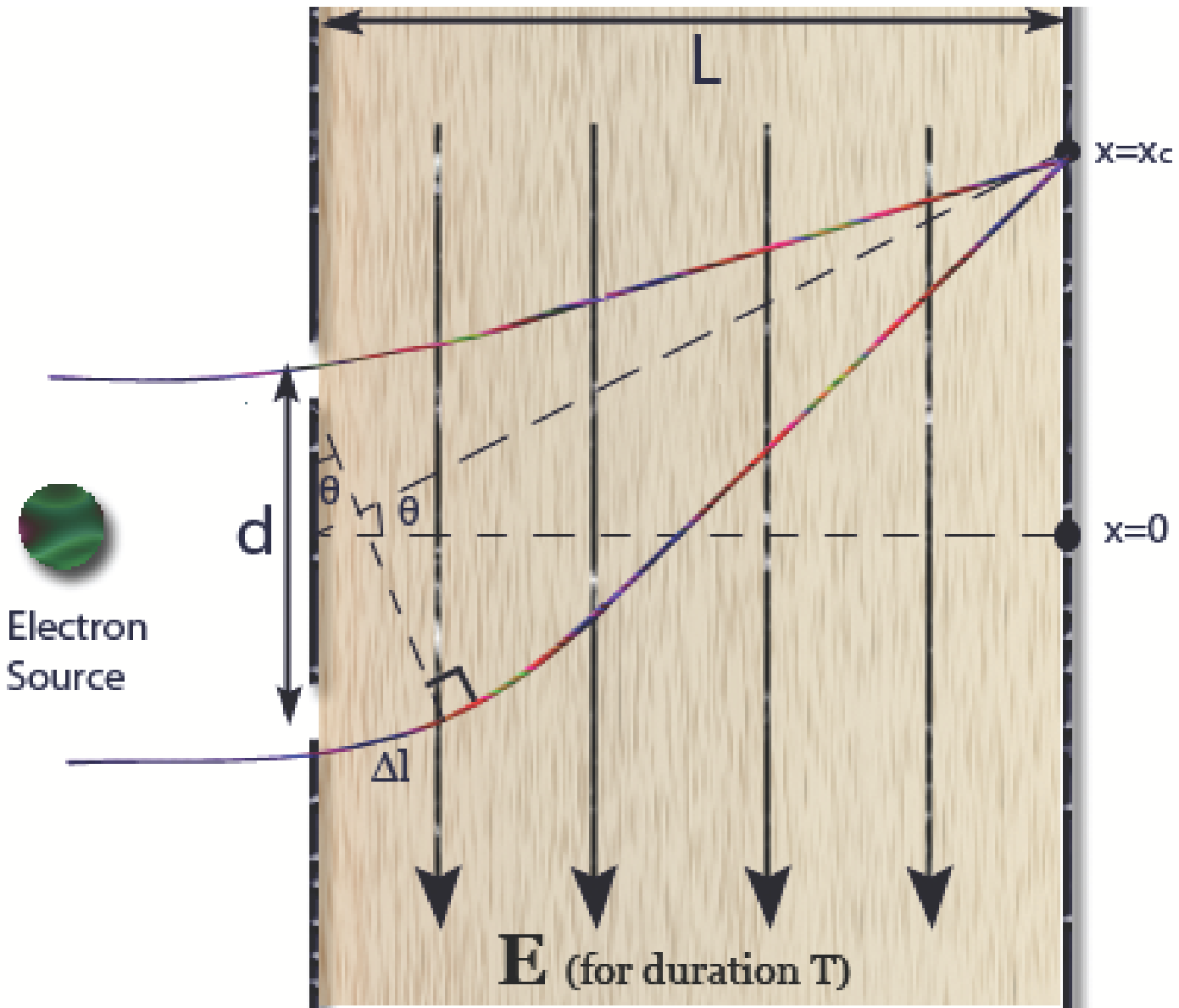}%
\end{figure}
%EndExpansion

A corresponding electric case is shown in our Fig.4 (discussed here for the
first time, to the best of our knowledge), where now an additional electric
field $E$ (pointing downwards) is present everywhere in space, but for only a
finite time duration $T$ (which we take to be much shorter than the time of
travel $t=\frac{L}{v}$, $T<<t$). In this case therefore the electric Lorentz
force $qE$ is exerted on the trajectories only during the small time interval
$\Delta t=T$ (note the difference with the above magnetic case$\boldsymbol{;}$
we now have an electric strip in \textit{time} rather than a magnetic strip in
space as we had earlier). Let us then follow an analogous calculation as above
but now adopted to this electric case. The electric type of Aharonov-Bohm
phase is%

\begin{equation}
\Delta\varphi^{AB}=-2\pi\frac{q}{e}\frac{cT\Delta V}{\Phi_{0}}, \label{ABele}%
\end{equation}
with $\Delta V$ being the electric potential difference between the two
trajectories, hence $\Delta V\thickapprox Ed$ (again for small
trajectory-deflections). On the other hand, the semiclassical phase difference
between the two trajectories is again given by (\ref{Semiphase}), but the
position $x_{c}$ of the central fringe must now be determined by the electric
field force $qE\boldsymbol{:}$ The change of kinematic momentum (always
parallel to the screen) is now $\Delta\Pi_{x}=qET$, hence the analog of
(\ref{Dvx}) is now%

\begin{equation}
\Delta v_{x}=\frac{qET}{m} \label{Dvxnew}%
\end{equation}
which if combined with (\ref{Dvx2}) (that is obviously valid for this case as
well, again for small deflections, due to the $\Delta t=T<<t$), and always
with $t=\frac{L}{v}$, gives that the central fringe displacement in this case
must be $x_{c}=\Delta v_{x}t=\frac{qET}{m}\frac{L}{v}$, and using again
$mv=\Pi=\frac{h}{\lambda}$, we finally have the following analog of (\ref{xc})%

\begin{equation}
x_{c}=\frac{qETL\lambda}{h}. \label{xcnew}%
\end{equation}
(Note again that for a negative charge and a negative electric field (i.e.
pointing downwards) the central fringe displacement is indeed upwards). By
substituting (\ref{xcnew}) into (\ref{Semiphase}), the lengths $L$ and
$\lambda$ again cancel out, and we finally have $\Delta\varphi^{semi}=2\pi
d\frac{qETL\lambda}{h}=2\pi\frac{q}{e}\frac{EdcT}{\frac{hc}{e}}$, which with
$\frac{hc}{e}=\Phi_{0}$ the flux quantum, and through comparison with
(\ref{ABele}) leads once again to our final proof that%

\begin{equation}
\Delta\varphi^{semi}=-\Delta\varphi^{AB}. \label{proofnew}%
\end{equation}
We note therefore that even in the electric case, the semiclassical phase
difference (between two trajectories) picked up due to the Lorentz force
(exerted on them) is once again opposite to the electric Aharonov-Bohm phase
due to the electric flux (in spacetime) enclosed between the same trajectories.

We should point out once again, however, that although the above elementary
considerations apply to semiclassical motion of narrow wavepackets, in this
paper we have given a more general understanding of the above opposite signs
that applies to general (even completely delocalized) states, and that
originates from our generalized Werner \& Brill cancellations.

In a slightly different vein, we should also point out that the above
cancellations give a justification of why certain semiclassical arguments that
focus on the history of the experimental set up (usually based on Faraday's
law for a $t$-dependent magnetic flux) seem to give at the end a result that
is consistent with the result of a static Aharonov-Bohm arrangement. However,
there is a again an opposite sign that seems to have been largely unnoticed in
such arguments as well (i.e. see the simplest possible argument in
Silverman\cite{Silverman2}, where in his eq.(1.34) there should be an extra
minus sign). Our above observation essentially describes the fact that,
\textit{if we had actually used} a $t$-dependent magnetic flux (with its final
value being the actual value of our static flux), then the induced electric
field (viewed now as a nonlocal term of the present work) would have cancelled
the static Aharonov-Bohm phase. Of course now, this $t$-dependent experimental
set up has not been used (the flux is static) and we obtain the usual magnetic
Aharonov-Bohm phase, but the above argument (of a \textquotedblleft potential
experiment\textquotedblright\ that \textit{could have been carried out}) takes
the \textquotedblleft mystery\textquotedblright\ away of why such
history-based arguments generally work $-$ although \textit{they have to be
corrected with a sign}. The above also gives a rather natural account of the
\textquotedblleft dynamical nonlocality\textquotedblright%
\ character\cite{Popescu} attributed to the various Aharonov-Bohm phenomena
(magnetic, electric or combined), although $-$ in the present work $-$ this
dynamical quantum nonlocality seems to simultaneously respect Causality, a
rather pleasing characteristic of this theory that, as far as we are aware, is
reported here for the first time.

Finally, coming back to an even broader significance of the new solutions, one
may wonder about possible consequences of the nonlocal terms if these are
included in more general physical models that have a gauge structure (in
Condensed Matter or High Energy Physics). It is also worth mentioning that if
one follows the same \textquotedblleft unconventional\textquotedblright%
\ method (of solution of PDEs) with the Maxwell's equations for the electric
and magnetic fields (rather than with the PDEs for the potentials that give
$\Lambda$), the corresponding nonlocal terms can be derived, and one can then
see that these essentially demonstrate the causal propagation of the radiation
electric and magnetic fields outside physically inaccessible confined sources
(i.e. solenoids or electric cages). Although this is of course widely known at
the level of classical fields, a major conclusion that can be drawn from the
present work (at the level of gauge transformations) is that \textit{a
corresponding Causality also exists at the level of quantum mechanical phases}
as well, and this is enforced by the nonlocal terms in $t$-dependent cases. It
strongly indicates that the nonlocal terms found in this work at the level of
quantum mechanical phases reflect a causal propagation of wavefunction phases
\textbf{in the Schr\"{o}dinger Picture} (at least one part of them, the one
containing the fields, which competes with the Aharonov-Bohm types of phases
containing the potentials). This is an entirely new concept (given the
\textit{local nature} but also the \textit{nonrelativistic character} of the
Schr\"{o}dinger equation) and deserves to be further explored. It would indeed
be worth investigating possible applications of the above results (of nonlocal
phases of wavefunctions, solutions of the local Schr\"{o}dinger equation) in
$t$-dependent single- \textit{vs} double-slit experiments recently discussed
by the group of Aharonov\cite{Tollaksen} who use a completely different method
(with modular variables in the Heisenberg picture). One should also note
recent work\cite{He}, that rightly emphasizes that Physics cannot currently
predict how we dynamically go from the single-slit diffraction pattern to the
double-slit diffraction pattern (whether it is in a gradual and causal manner
or not) and where a relevant experiment is proposed to decide on (measure)
exactly this. Application of our nonlocal terms in such questions in analogous
experiments (i.e. by introducing (finite) scalar potentials on slits in a
$t$-dependent way) provides a completely new formulation for addressing causal
issues of this type, and is currently under investigation. Furthermore,
\textit{SU(2)} generalizations would be an obviously interesting extension of
the above \textit{U(1)} theory, and such generalizations are rather formally
direct and not difficult to make (an immediate physically interesting question
being whether the new nonlocal terms might have a nontrivial impact on i.e.
spin-$\frac{1}{2}$-states, since these terms would act asymmetrically on
opposite spins). Finally, it is worth noting that, if $E$'s were substituted
by gravitational fields and $B$'s by Coriolis force fields arising in
non-inertial frames of reference, the above nonlocalities (and their apparent
causal nature) could possibly have an interesting story to tell about quantum
mechanical phase behavior in a Relativistic/Gravitational framework.

\bigskip

\bigskip Graduate students Kyriakos Kyriakou and Georgios Konstantinou of the
University of Cyprus and Areg Ghazaryan of Yerevan State University are
acknowledged for having carefully reproduced all (extremely long) results
(most of them not reported here). Georgios Konstantinou is also acknowledged
for having drawn Figures 2, 3 and 4. Dr. Cleopatra Christoforou of the
Department of Mathematics and Statistics of the University of Cyprus is
acknowledged for a discussion concerning the mathematical method followed.

\bigskip

\bigskip%
\[
\]

\subsection{Appendix A}

The time-dependent Schr\"{o}dinger equation for a particle (of mass $m$ and
charge $q$) moving in the set of potentials ($\mathbf{A}_{1}(\mathbf{r}%
,t),\phi_{1}(\mathbf{r},t)$) is%

\begin{equation}
\frac{\lbrack\mathbf{-i}\hbar\nabla-\frac{q}{c}\mathbf{A}_{1}(\mathbf{r}%
,t)]^{2}}{2m}\Psi_{1}(\mathbf{r},t)+q\phi_{1}(\mathbf{r},t)\Psi_{1}%
(\mathbf{r},t)=i\hbar\frac{\partial}{\partial t}\Psi_{1}(\mathbf{r},t)
\label{Schr1}%
\end{equation}
and the one for the same particle moving in the set of potentials
($\mathbf{A}_{2}(\mathbf{r},t),\phi_{2}(\mathbf{r},t)$) is%

\begin{equation}
\frac{\lbrack\mathbf{-i}\hbar\nabla-\frac{q}{c}\mathbf{A}_{2}(\mathbf{r}%
,t)]^{2}}{2m}\Psi_{2}(\mathbf{r},t)+q\phi_{2}(\mathbf{r},t)\Psi_{2}%
(\mathbf{r},t)=i\hbar\frac{\partial}{\partial t}\Psi_{2}(\mathbf{r},t).
\label{Schr2}%
\end{equation}

Below we recall the general proof that solutions of the two above equations
are formally connected through%

\[
\Psi_{2}(\mathbf{r},t)=e^{i\frac{q}{\hbar c}\Lambda(\mathbf{r},t)}\Psi
_{1}(\mathbf{r},t),
\]
which is eq.(\ref{Basic1}) of the text, with the function $\Lambda
(\mathbf{r},t)$ satisfying the system of PDEs (\ref{Basic11}), namely%

\[
\nabla\Lambda(\mathbf{r,t})=\mathbf{A}_{2}(\mathbf{r},t)-\mathbf{A}%
_{1}(\mathbf{r},t)\qquad and\qquad-\frac{1}{c}\frac{\partial\Lambda
(\mathbf{r},t)}{\partial t}=\phi_{2}\left(  \mathbf{r},t\right)  -\phi
_{1}(\mathbf{r},t).
\]

Indeed, it is an obvious vector identity that%

\begin{equation}
\lbrack\mathbf{-i}\hbar\nabla-\frac{q}{c}(\mathbf{A}_{1}(\mathbf{r}%
,t)+\nabla\Lambda(\mathbf{r,t}))]e^{i\frac{q}{\hbar c}\Lambda(\mathbf{r}%
,t)}\Psi_{1}(\mathbf{r},t)=e^{i\frac{q}{\hbar c}\Lambda(\mathbf{r}%
,t)}[\mathbf{-i}\hbar\nabla-\frac{q}{c}\mathbf{A}_{1}(\mathbf{r},t)]\Psi
_{1}(\mathbf{r},t), \label{identity1}%
\end{equation}
which, if applied once more (but now on the new function $\hat{Y}$
$(\mathbf{r},t)=[\mathbf{-i}\hbar\nabla-\frac{q}{c}\mathbf{A}_{1}%
(\mathbf{r},t)]\Psi_{1}(\mathbf{r},t)$ in place of the single $\Psi
_{1}(\mathbf{r},t)$) gives the well-known generalization%

\begin{equation}
\lbrack\mathbf{-i}\hbar\nabla-\frac{q}{c}(\mathbf{A}_{1}(\mathbf{r}%
,t)+\nabla\Lambda(\mathbf{r,t}))]^{2}e^{i\frac{q}{\hbar c}\Lambda
(\mathbf{r},t)}\Psi_{1}(\mathbf{r},t)=e^{i\frac{q}{\hbar c}\Lambda
(\mathbf{r},t)}[\mathbf{-i}\hbar\nabla-\frac{q}{c}\mathbf{A}_{1}%
(\mathbf{r},t)]^{2}\Psi_{1}(\mathbf{r},t) \label{identity2}%
\end{equation}
(and repeated application would of course give a similar identity for any
positive integer power). In addition, we trivially have%

\begin{equation}
i\hbar\frac{\partial}{\partial t}[e^{i\frac{q}{\hbar c}\Lambda(\mathbf{r}%
,t)}\Psi_{1}(\mathbf{r},t)]=e^{i\frac{q}{\hbar c}\Lambda(\mathbf{r},t)}%
i\hbar\frac{\partial}{\partial t}\Psi_{1}(\mathbf{r},t)-e^{i\frac{q}{\hbar
c}\Lambda(\mathbf{r},t)}\frac{q}{c}\frac{\partial\Lambda(\mathbf{r}%
,t)}{\partial t}\Psi_{1}(\mathbf{r},t). \label{identity3}%
\end{equation}
One trivially notes then that, indeed, if $\Psi_{2}(\mathbf{r},t)$ is
substituted by $e^{i\frac{q}{\hbar c}\Lambda(\mathbf{r},t)}\Psi_{1}%
(\mathbf{r},t)$, $\mathbf{A}_{2}(\mathbf{r},t)$ is substituted by
$\mathbf{A}_{1}(\mathbf{r},t)+\nabla\Lambda(\mathbf{r,t})$ and $\phi
_{2}\left(  \mathbf{r},t\right)  $ is substituted by $\phi_{1}\left(
\mathbf{r},t\right)  -\frac{1}{c}\frac{\partial\Lambda(\mathbf{r},t)}{\partial
t}$, then the left-hand-side (lhs) of (\ref{Schr2}) becomes $e^{i\frac
{q}{\hbar c}\Lambda(\mathbf{r},t)}\ast\lbrack$lhs of (\ref{Schr1})$-\frac
{q}{c}\frac{\partial\Lambda(\mathbf{r},t)}{\partial t}\Psi_{1}(\mathbf{r}%
,t)]$, while the right-hand-side of (\ref{Schr2}) is the above
(\ref{identity3}), and this equality is obviously satisfied (after
cancellation of the $\frac{\partial\Lambda}{\partial t}$-additive term, and
then of the common global phase factor from both sides) if $\Psi
_{1}(\mathbf{r},t)$ satisfies (\ref{Schr1}).\qquad$\checkmark$

\subsection{\bigskip Appendix B}

We present here the full derivation of the solutions of the system of PDEs
(\ref{gaugetransf}), which if applied to only one spatial variable is
(\ref{xt-BasicSystem}), namely%

\[
\frac{\partial\Lambda(x,t)}{\partial x}=A(x,t)\qquad and\qquad-\frac{1}%
{c}\frac{\partial\Lambda(x,t)}{\partial t}=\phi\left(  x,t\right)
\]
(with $\ A(x,t)=A_{2}(x,t)-A_{1}(x,t)$ \ and $\ \phi\left(  x,t\right)
=\phi_{2}\left(  x,t\right)  -\phi_{1}\left(  x,t\right)  $). The system is
underdetermined in the sense that we only have knowledge of $\Lambda$ at an
initial point $(x_{0},t_{0})$ and with no further boundary conditions (hence
multiplicities of solutions are generally expected, and these are discussed
separately below and mainly in the text). Let us first look for unique
(single-valued) solutions (i.e. with $\Lambda$ being a \textit{function }on
the $(x,t)$-plane\textit{,} in the sense of Elementary Analysis) and let us
integrate the\textit{\ first} of (\ref{xt-BasicSystem}) -- without dropping
terms that may at first sight appear redundant -- to obtain%

\begin{equation}
\Lambda(x,t)-\Lambda(x_{0},t)=\int_{x_{0}}^{x}A(x^{\prime},t)dx^{\prime}%
+\tau(t). \label{1stintegration}%
\end{equation}

By then substituting this to the second of (\ref{xt-BasicSystem}) (and
assuming that interchanges of derivatives and integrals are allowed, i.e.
covering cases of potentials with discontinuous first derivatives, something
that corresponds to the physical case of discontinuous magnetic fields $-$ a
case very often discussed in the literature), we obtain%

\begin{equation}
\phi\left(  x,t\right)  =-\frac{1}{c}%
%TCIMACRO{\dint \limits_{x_{0}}^{x}}%
%BeginExpansion
{\displaystyle\int\limits_{x_{0}}^{x}}
%EndExpansion
\frac{\partial A(x^{\prime},t)}{\partial t}dx^{\prime}-\frac{1}{c}%
\frac{\partial\tau(t)}{\partial t}-\frac{1}{c}\frac{\partial\Lambda(x_{0}%
,t)}{\partial t}, \label{Phi}%
\end{equation}
which if integrated gives%

\begin{equation}
\tau(t)=\tau(t_{0})+\Lambda(x_{0},t_{0})-\Lambda(x_{0},t)-%
%TCIMACRO{\dint \limits_{t_{0}}^{t}}%
%BeginExpansion
{\displaystyle\int\limits_{t_{0}}^{t}}
%EndExpansion
dt^{\prime}%
%TCIMACRO{\dint \limits_{x_{0}}^{x}}%
%BeginExpansion
{\displaystyle\int\limits_{x_{0}}^{x}}
%EndExpansion
dx^{\prime}\frac{\partial A(x^{\prime},t^{\prime})}{\partial t^{\prime}}%
-c\int_{t_{0}}^{t}\phi\left(  x,t^{\prime}\right)  dt^{\prime}+g(x)
\label{Tau}%
\end{equation}
with $g(x)$ to be chosen in such a way that the entire right-hand-side of
(\ref{Tau}) is only a function of$\ t$, as it should be (hence independent
of$\ x$). Finally, by substituting \ \ $\frac{\partial A(x^{\prime},t^{\prime
})}{\partial t^{\prime}}$ \ with \ \ $-c\left(  E(x^{\prime},t^{\prime}%
)+\frac{\partial\phi(x^{\prime},t^{\prime})}{\partial x^{\prime}}\right)  $,
\ (where $E(x^{\prime},t^{\prime})=E_{2}(x^{\prime},t^{\prime})-E_{1}%
(x^{\prime},t^{\prime})$), carrying out the integration with respect to
$x^{\prime}$, and by demanding that $\tau(t)$ \ be independent of $\ x$, we
finally obtain the following general solution%

\[
\Lambda(x,t)=\Lambda(x_{0},t_{0})+%
%TCIMACRO{\dint \limits_{x_{0}}^{x}}%
%BeginExpansion
{\displaystyle\int\limits_{x_{0}}^{x}}
%EndExpansion
A(x^{\prime},t)dx^{\prime}-c%
%TCIMACRO{\dint \limits_{t_{0}}^{t}}%
%BeginExpansion
{\displaystyle\int\limits_{t_{0}}^{t}}
%EndExpansion
\phi(x_{0},t^{\prime})dt^{\prime}+\left\{  c%
%TCIMACRO{\dint \limits_{t_{0}}^{t}}%
%BeginExpansion
{\displaystyle\int\limits_{t_{0}}^{t}}
%EndExpansion
dt^{\prime}%
%TCIMACRO{\dint \limits_{x_{0}}^{x}}%
%BeginExpansion
{\displaystyle\int\limits_{x_{0}}^{x}}
%EndExpansion
dx^{\prime}E(x^{\prime},t^{\prime})+g(x)\right\}  +\tau(t_{0})
\]
with $\ g(x)$ \ chosen in such a way that the quantity $\ \ \left\{  c%
%TCIMACRO{\dint \limits_{t_{0}}^{t}}%
%BeginExpansion
{\displaystyle\int\limits_{t_{0}}^{t}}
%EndExpansion
dt^{\prime}%
%TCIMACRO{\dint \limits_{x_{0}}^{x}}%
%BeginExpansion
{\displaystyle\int\limits_{x_{0}}^{x}}
%EndExpansion
dx^{\prime}E(x^{\prime},t^{\prime})+g(x)\right\}  $ \ \ is independent of $x$.
This is solution (\ref{LambdaStatic1}) of the text.

\bigskip

Here it should be noted that, if we had first integrated the \textit{second}
of (\ref{xt-BasicSystem}) we would have%

\begin{equation}
\Lambda(x,t)-\Lambda(x,t_{0})=-c\int_{t_{0}}^{t}\phi(x,t^{\prime})dt^{\prime
}+\chi(x) \label{2ndintegration}%
\end{equation}
and then from the first of (\ref{xt-BasicSystem}) we would get%

\begin{equation}
A\left(  x,t\right)  =-c%
%TCIMACRO{\dint \limits_{t_{0}}^{t}}%
%BeginExpansion
{\displaystyle\int\limits_{t_{0}}^{t}}
%EndExpansion
\frac{\partial\phi(x,t^{\prime})}{\partial x}dt^{\prime}+\frac{\partial
\chi(x)}{\partial x}+\frac{\partial\Lambda(x,t_{0})}{\partial x},
\label{Aintermediate}%
\end{equation}
which after integration would give%

\begin{equation}
\chi(x)=\chi(x_{0})+\Lambda(x_{0},t_{0})-\Lambda(x,t_{0})+c%
%TCIMACRO{\dint \limits_{x_{0}}^{x}}%
%BeginExpansion
{\displaystyle\int\limits_{x_{0}}^{x}}
%EndExpansion
dx^{\prime}%
%TCIMACRO{\dint \limits_{t_{0}}^{t}}%
%BeginExpansion
{\displaystyle\int\limits_{t_{0}}^{t}}
%EndExpansion
dt^{\prime}\frac{\partial\phi(x^{\prime},t^{\prime})}{\partial x^{\prime}%
}+\int_{x_{0}}^{x}A\left(  x^{\prime},t\right)  dx^{\prime}+\hat{g}(t)
\label{chi}%
\end{equation}
with $\hat{g}(t)$ to be chosen in such a way that the entire right-hand-side
of (\ref{chi}) is only a function of$\ x$, as it should be (hence independent
of$\ t$). Finally, by substituting \ \ $\frac{\partial\phi(x^{\prime
},t^{\prime})}{\partial x^{\prime}}$ \ with \ \ $-\left(  E(x^{\prime
},t^{\prime})+\frac{1}{c}\frac{\partial A(x^{\prime},t^{\prime})}{\partial
t^{\prime}}\right)  $, carrying out the integration with respect to
$t^{\prime}$, and by demanding that $\chi(x)$ \ be independent of $\ t$, we
would finally obtain the following general solution%

\[
\Lambda(x,t)=\Lambda(x_{0},t_{0})+%
%TCIMACRO{\dint \limits_{x_{0}}^{x}}%
%BeginExpansion
{\displaystyle\int\limits_{x_{0}}^{x}}
%EndExpansion
A(x^{\prime},t_{0})dx^{\prime}-c\int_{t_{0}}^{t}\phi\left(  x,t^{\prime
}\right)  dt^{\prime}+\left\{  -c%
%TCIMACRO{\dint \limits_{x_{0}}^{x}}%
%BeginExpansion
{\displaystyle\int\limits_{x_{0}}^{x}}
%EndExpansion
dx^{\prime}%
%TCIMACRO{\dint \limits_{t_{0}}^{t}}%
%BeginExpansion
{\displaystyle\int\limits_{t_{0}}^{t}}
%EndExpansion
dt^{\prime}E(x^{\prime},t^{\prime})+\hat{g}(t)\right\}  +\chi(x_{0})
\]
with $\ \hat{g}(t)$ \ chosen in such a way that the quantity \ $\ \left\{  -c%
%TCIMACRO{\dint \limits_{x_{0}}^{x}}%
%BeginExpansion
{\displaystyle\int\limits_{x_{0}}^{x}}
%EndExpansion
dx^{\prime}%
%TCIMACRO{\dint \limits_{t_{0}}^{t}}%
%BeginExpansion
{\displaystyle\int\limits_{t_{0}}^{t}}
%EndExpansion
dt^{\prime}E(x^{\prime},t^{\prime})+\hat{g}(t)\right\}  \ \ $\ \ is
independent of$\ t$. This is solution (\ref{LambdaStatic2}) of the text.

\bigskip Solutions (\ref{LambdaStatic1}) and (\ref{LambdaStatic2}) can be
viewed as the (formal) analogs of (\ref{static1}) and (\ref{static2})
correspondingly, although they hide in them much richer Physics because of
their dynamic character (see Section IX). (The additional constant last terms
were shown in Section IX to be related to possible multiplicities of $\Lambda
$, and they are zero in simple-connected spacetimes).

\bigskip The reader was provided with the direct verification (i.e. proof by
\textquotedblleft going backwards\textquotedblright) that (\ref{LambdaStatic1}%
) or (\ref{LambdaStatic2}) are indeed solutions of the basic system of PDEs
(\ref{xt-BasicSystem}) in Section VIII.

\subsection{Appendix C}

We provide here a general proof of the generalized Werner \& Brill
cancellations in simple-connected spacetime, namely, that solutions
(\ref{LambdaStatic1}) and (\ref{LambdaStatic2}) are equivalent. By looking
first at the general structure of solutions (\ref{LambdaStatic1}) and
(\ref{LambdaStatic2}), we note that in both forms, the last constant terms
($\tau(t_{0})$ and $\chi(x_{0})$) are only present in cases where $\Lambda$ is
expected to be multivalued (this comes from the definitions of $\tau(t_{0})$
and $\chi(x_{0})$, see discussion below) and therefore these constant
quantities are nonvanishing in cases of motion only in multiple-connected
spacetimes (leading to phenomena of the Aharonov-Bohm type (see the analogous
discussion in Appendix F, on the easier-to-follow magnetic case)). In such
multiple-connected cases these last terms are simply equal (in absolute value)
to the enclosed fluxes in regions of spacetime that are physically
inaccessible to the particle (in the electric Aharonov-Bohm setup, for
example, it turns out that $\tau(t_{0})=-\chi(x_{0})=$ enclosed
\textquotedblleft electric flux\textquotedblright\ in spacetime, see below for
the proof). Although such cases can also be covered by our method below, let
us for the moment ignore them (set them to zero) and focus on cases of motion
in simple-connected spacetimes. Then the two solutions (\ref{LambdaStatic1})
and (\ref{LambdaStatic2}) are actually equal as is now shown$\boldsymbol{:}$
since $\left\{  c%
%TCIMACRO{\dint \limits_{t_{0}}^{t}}%
%BeginExpansion
{\displaystyle\int\limits_{t_{0}}^{t}}
%EndExpansion
dt^{\prime}%
%TCIMACRO{\dint \limits_{x_{0}}^{x}}%
%BeginExpansion
{\displaystyle\int\limits_{x_{0}}^{x}}
%EndExpansion
dx^{\prime}E(x^{\prime},t^{\prime})+g(x)\right\}  $ is independent of $x$, its
$x$-derivative is zero which leads to $g^{\prime}(x)=-c%
%TCIMACRO{\dint \limits_{t_{0}}^{t}}%
%BeginExpansion
{\displaystyle\int\limits_{t_{0}}^{t}}
%EndExpansion
dt^{\prime}E(x,t^{\prime})$, with a general solution $\ \ g(x)=g(x_{0})-c%
%TCIMACRO{\dint \limits_{x_{0}}^{x}}%
%BeginExpansion
{\displaystyle\int\limits_{x_{0}}^{x}}
%EndExpansion
dx^{\prime}%
%TCIMACRO{\dint \limits_{t_{0}}^{t}}%
%BeginExpansion
{\displaystyle\int\limits_{t_{0}}^{t}}
%EndExpansion
dt^{\prime}E(x^{\prime},t^{\prime})+C(t),$ \ \ and with a $C(t)$ such that the
right-hand-side is only a function of$\ x$, hence independent of
$t\boldsymbol{;}$ but this is exactly the form of (\ref{LambdaStatic2}), if we
identify $C(t)$ with $\hat{g}(t)$ (and $g(x_{0})$ with $\chi(x_{0})$). This
can be easily seen if we note that substitution of $E(x^{\prime},t^{\prime})$
with \ $\ -\frac{\partial\phi(x^{\prime},t^{\prime})}{\partial x^{\prime}%
}-\frac{1}{c}\frac{\partial A(x^{\prime},t^{\prime})}{\partial t^{\prime}}$
\ and two integrations carried out finally interchange the forms of the 1st
solution (\ref{LambdaStatic1}) from $\left(
%TCIMACRO{\dint \limits_{x_{0}}^{x}}%
%BeginExpansion
{\displaystyle\int\limits_{x_{0}}^{x}}
%EndExpansion
A(x^{\prime},t)dx^{\prime}-c%
%TCIMACRO{\dint \limits_{t_{0}}^{t}}%
%BeginExpansion
{\displaystyle\int\limits_{t_{0}}^{t}}
%EndExpansion
\phi(x_{0},t^{\prime})dt^{\prime}\right)  $\ \ to $\left(
%TCIMACRO{\dint \limits_{x_{0}}^{x}}%
%BeginExpansion
{\displaystyle\int\limits_{x_{0}}^{x}}
%EndExpansion
A(x^{\prime},t_{0})dx^{\prime}-c%
%TCIMACRO{\dint \limits_{t_{0}}^{t}}%
%BeginExpansion
{\displaystyle\int\limits_{t_{0}}^{t}}
%EndExpansion
\phi(x,t^{\prime})dt^{\prime}\right)  $ of the 2nd solution
(\ref{LambdaStatic2}).

\bigskip The above could alternatively be proven if in (\ref{Tau}), instead of
substituting \ $\frac{\partial A(x^{\prime},t^{\prime})}{\partial t^{\prime}}$
\ in terms of the electric field difference, we had merely interchanged the
ordering of integrations in the 1st integral term. This would then immediately
take us to the 2nd solution (\ref{LambdaStatic2}), with automatically
identifying the $t$-independent (hence $x$-dependent) quantity \ $\left\{  -c%
%TCIMACRO{\dint \limits_{x_{0}}^{x}}%
%BeginExpansion
{\displaystyle\int\limits_{x_{0}}^{x}}
%EndExpansion
dx^{\prime}%
%TCIMACRO{\dint \limits_{t_{0}}^{t}}%
%BeginExpansion
{\displaystyle\int\limits_{t_{0}}^{t}}
%EndExpansion
dt^{\prime}E(x^{\prime},t^{\prime})+\hat{g}(t)\right\}  $ \ of the 2nd
solution (\ref{LambdaStatic2}) with the function$\ g(x)$ of the 1st solution
(\ref{LambdaStatic1}). (In a similar way, one can prove the identification of
the $x$-independent (hence $t$-dependent) quantity $\ \left\{  c%
%TCIMACRO{\dint \limits_{t_{0}}^{t}}%
%BeginExpansion
{\displaystyle\int\limits_{t_{0}}^{t}}
%EndExpansion
dt^{\prime}%
%TCIMACRO{\dint \limits_{x_{0}}^{x}}%
%BeginExpansion
{\displaystyle\int\limits_{x_{0}}^{x}}
%EndExpansion
dx^{\prime}E(x^{\prime},t^{\prime})+g(x)\right\}  $ \ of the 1st solution
(\ref{LambdaStatic1}) with the function $\hat{g}(t)$ of the 2nd solution
(\ref{LambdaStatic2})). This is the deep mathematical cause of the generalized
Werner \& Brill cancellations of the text in this electric case.

Finally, with respect to $\tau(t_{0})$\ and $\chi(x_{0})$, let us give an
example to see why ordinarily (in simple-connectivity) they are zero, or in
the most general case they are related to physically inaccessible enclosed
fluxes. Starting from (\ref{1stintegration}), where $\tau(t)$ was first
introduced, we have that%

\begin{equation}
\tau(t_{0})=\Lambda(x,t_{0})-\Lambda(x_{0},t_{0})-\int_{x_{0}}^{x}A(x^{\prime
},t_{0})dx^{\prime}, \label{Taut0}%
\end{equation}
which should be independent of $\ x$ \ (and \textit{it is} as can easily be
proven, since its $x$-derivative gives $\frac{\partial\Lambda(x,t_{0}%
)}{\partial x}-A(x,t_{0})$ \ which is zero, as $\Lambda(x,t)$ satisfies by
assumption the first equation of the system (\ref{xt-BasicSystem}) of PDEs
(evaluated at $t=t_{0}$)). We can therefore determine its value by taking the
limit $x\rightarrow x_{0}$ \ in (\ref{Taut0}), which is zero, unless there is
a multivaluedness of $\Lambda$ at the point $(x_{0},t_{0})$. This happens for
example for $A$ having a $\delta$-function form (a case however which we leave
out, otherwise the assumed interchanges might not be allowed) or in cases that
there is a \textquotedblleft memory\textquotedblright\ that the system has
multiplicities in $\Lambda$, i.e. in Aharonov-Bohm configurations (with
enclosed and inaccessible fluxes in space-time), hence the value of
$\tau(t_{0})$ being expected to be equal to the enclosed \textquotedblleft
electric flux\textquotedblright$\boldsymbol{:}$ the limit $x\rightarrow x_{0}$
(for fixed $t_{0}$) in the path sense of solution (\ref{LambdaStatic1}) that
is needed then to determine $\tau(t_{0})$ is equivalent to making an entire
closed trip around the observation rectangle in the \textit{positive} sense,
landing on the initial point $(x_{0},t_{0})$. This, from (\ref{Taut0}), gives
that $\tau(t_{0})$ = enclosed \textquotedblleft electric
flux\textquotedblright. A similar argument applied for%

\begin{equation}
\chi(x_{0})=\Lambda(x_{0},t)-\Lambda(x_{0},t_{0})+c\int_{t_{0}}^{t}\phi
(x_{0},t^{\prime})dt^{\prime} \label{Chi0}%
\end{equation}
leads to the value of $\chi(x_{0})$ being equal to \textit{minus} the enclosed
\textquotedblleft electric flux\textquotedblright\ (a corresponding limit
$t\rightarrow t_{0}$ (for fixed $x_{0}$) in the path sense of solution
(\ref{LambdaStatic2}) is now equivalent to making an entire trip around the
rectangle in the \textit{negative} sense, landing again on the same initial
point $(x_{0},t_{0})$). This, from (\ref{Chi0}), gives that $\chi(x_{0})$ =
$-$enclosed \textquotedblleft electric flux\textquotedblright. If these values
are actually substituted in (\ref{LambdaStatic1}) (with $g(x)=0$) and in
(\ref{LambdaStatic2}) (with $\hat{g}(t)=0$) they give the correct electric
Aharonov-Bohm result (where effectively there are no nonlocal contributions,
and only the line-integrals of $A$ and $\phi$ contribute to the phase). [The
above choice $\ g(x)=\hat{g}(t)=0$ \ is a natural one, made because, in this
Aharonov-Bohm case, the enclosed \textquotedblleft electric
flux\textquotedblright\ is independent of both $x$ and $t]$. A more detailed
discussion of such multiplicities in the standard \textit{static} magnetic
Aharonov-Bohm case is given in Appendix F.

\bigskip%

\[
\]

\subsection{\bigskip Appendix D}

After having discussed fully the simple $(x,t)$-case, let us for completeness
give the analogous (Euclidian-rotated in 4-D spacetime) derivation for
$(x,y)$-variables and briefly discuss the properties of the simpler static
solutions, but now in full generality (also including possible
multi-valuedness of $\Lambda$ in the usual magnetic Aharonov-Bohm cases). We
will simply need to apply the same methodology (of solution of a system of
PDEs) to the system already shown in (\ref{usualgradcomps}), namely%

\[
\frac{\partial\Lambda(x,y)}{\partial x}=A_{x}(x,y)\qquad and\qquad
\frac{\partial\Lambda(x,y)}{\partial y}=A_{y}(x,y).
\]
By first integrating the 1st of this (again without dropping any terms that
may appear redundant) we obtain the analog of (\ref{1stintegration}), namely%

\begin{equation}
\Lambda(x,y)-\Lambda(x_{0},y)=\int_{x_{0}}^{x}A_{x}(x^{\prime},y)dx^{\prime
}+f(y) \label{1stintegration(x,y)}%
\end{equation}
and by then substituting the result to the 2nd we have%

\begin{equation}
A_{y}\left(  x,y\right)  =%
%TCIMACRO{\dint \limits_{x_{0}}^{x}}%
%BeginExpansion
{\displaystyle\int\limits_{x_{0}}^{x}}
%EndExpansion
\frac{\partial A_{x}(x^{\prime},y)}{\partial y}dx^{\prime}+f^{\prime}%
(y)+\frac{\partial\Lambda(x_{0},y)}{\partial y} \label{Ay}%
\end{equation}
which if integrated leads to%

\begin{equation}
f(y)=f(y_{0})-\Lambda(x_{0},y)+\Lambda(x_{0},y_{0})-%
%TCIMACRO{\dint \limits_{y_{0}}^{y}}%
%BeginExpansion
{\displaystyle\int\limits_{y_{0}}^{y}}
%EndExpansion
dy^{\prime}%
%TCIMACRO{\dint \limits_{x_{0}}^{x}}%
%BeginExpansion
{\displaystyle\int\limits_{x_{0}}^{x}}
%EndExpansion
dx^{\prime}\frac{\partial A_{x}(x^{\prime},y^{\prime})}{\partial y^{\prime}%
}+\int_{y_{0}}^{y}A_{y}\left(  x,y^{\prime}\right)  dy^{\prime}+g(x)
\label{f(y)}%
\end{equation}
with $g(x)$ to be chosen in such a way that the entire right-hand-side of
(\ref{f(y)}) is only a function of $\ y$, as it should be (hence independent
of $\ x$). Finally, by substituting \ \ $\frac{\partial A_{x}(x^{\prime
},y^{\prime})}{\partial y^{\prime}}$ \ with \ \ $\frac{\partial A_{y}%
(x^{\prime},y^{\prime})}{\partial x^{\prime}}-B_{z}(x^{\prime},y^{\prime})$,
carrying out the integration with respect to $x^{\prime}$, and by demanding
that $f(y)$ \ be independent of $\ x$, we finally obtain the following general solution%

\[
\Lambda(x,y)=\Lambda(x_{0},y_{0})+%
%TCIMACRO{\dint \limits_{x_{0}}^{x}}%
%BeginExpansion
{\displaystyle\int\limits_{x_{0}}^{x}}
%EndExpansion
A_{x}(x^{\prime},y)dx^{\prime}+%
%TCIMACRO{\dint \limits_{y_{0}}^{y}}%
%BeginExpansion
{\displaystyle\int\limits_{y_{0}}^{y}}
%EndExpansion
A_{y}(x_{0},y^{\prime})dy^{\prime}+\left\{
%TCIMACRO{\dint \limits_{y_{0}}^{y}}%
%BeginExpansion
{\displaystyle\int\limits_{y_{0}}^{y}}
%EndExpansion
dy^{\prime}%
%TCIMACRO{\dint \limits_{x_{0}}^{x}}%
%BeginExpansion
{\displaystyle\int\limits_{x_{0}}^{x}}
%EndExpansion
dx^{\prime}B_{z}(x^{\prime},y^{\prime})+g(x)\right\}  +f(y_{0})
\]

\[
with\text{ \ }g(x)\text{ \ }chosen\ \ so\text{ \ }that\text{ \ }\left\{
%TCIMACRO{\dint \limits_{y_{0}}^{y}}%
%BeginExpansion
{\displaystyle\int\limits_{y_{0}}^{y}}
%EndExpansion
dy^{\prime}%
%TCIMACRO{\dint \limits_{x_{0}}^{x}}%
%BeginExpansion
{\displaystyle\int\limits_{x_{0}}^{x}}
%EndExpansion
dx^{\prime}B_{z}(x^{\prime},y^{\prime})+g(x)\right\}  \boldsymbol{:}\text{ is
}\mathsf{independent\ of\ }\ x,
\]
which is eq.(\ref{Lambda(x,y)1}) of the text, or eq.(\ref{static1}) but with
included multiplicities through the extra $f(y_{0})$ (which for
simple-connected space can be set to zero)$\boldsymbol{.}$ The result
(\ref{Lambda(x,y)1}) applies to cases where the particle passes through
\textit{different} magnetic fields (recall that $B_{z}={\huge (}%
\boldsymbol{B}_{2}-\boldsymbol{B}_{1}{\huge )}_{z}$) in spatial regions that
are remote to the observation point $(x,y)$. Alternatively, by following the
reverse route (first integrating the 2nd equation of the basic system
(\ref{usualgradcomps})) we would obtain%

\begin{equation}
\Lambda(x,y)-\Lambda(x,y_{0})=\int_{y_{0}}^{y}A_{y}(x,y^{\prime})dy^{\prime
}+\hat{h}(x) \label{2ndintegration(x,y)}%
\end{equation}
and by then substituting the result to the 1st we would have%

\begin{equation}
A_{x}\left(  x,y\right)  =%
%TCIMACRO{\dint \limits_{y_{0}}^{y}}%
%BeginExpansion
{\displaystyle\int\limits_{y_{0}}^{y}}
%EndExpansion
\frac{\partial A_{y}(x,y^{\prime})}{\partial x}dy^{\prime}+\hat{h}^{\prime
}(x)+\frac{\partial\Lambda(x,y_{0})}{\partial x} \label{Ax}%
\end{equation}
which if integrated would lead to%

\begin{equation}
\hat{h}(x)=\hat{h}(x_{0})-\Lambda(x,y_{0})+\Lambda(x_{0},y_{0})-%
%TCIMACRO{\dint \limits_{x_{0}}^{x}}%
%BeginExpansion
{\displaystyle\int\limits_{x_{0}}^{x}}
%EndExpansion
dx^{\prime}%
%TCIMACRO{\dint \limits_{y_{0}}^{y}}%
%BeginExpansion
{\displaystyle\int\limits_{y_{0}}^{y}}
%EndExpansion
dy^{\prime}\frac{\partial A_{y}(x^{\prime},y^{\prime})}{\partial x^{\prime}%
}+\int_{x_{0}}^{x}A_{x}\left(  x^{\prime},y\right)  dx^{\prime}+h(y)
\label{h(x)}%
\end{equation}
with $h(y)$ to be chosen in such a way that the entire right-hand-side of
(\ref{h(x)}) is only a function of $\ x$, as it should be (hence independent
of $\ y$). Finally, by substituting \ \ $\frac{\partial A_{y}(x^{\prime
},y^{\prime})}{\partial x^{\prime}}$ \ with \ \ $\frac{\partial A_{x}%
(x^{\prime},y^{\prime})}{\partial y^{\prime}}+B_{z}(x^{\prime},y^{\prime})$,
carrying out the integration with respect to $y^{\prime}$, and by demanding
that $\hat{h}(x)$ \ be independent of $\ y$, we would finally obtain the
following general solution%

\[
\Lambda(x,y)=\Lambda(x_{0},y_{0})+\int_{x_{0}}^{x}A_{x}(x^{\prime}%
,y_{0})dx^{\prime}+\int_{y_{0}}^{y}A_{y}(x,y^{\prime})dy^{\prime}+\left\{
{\Huge -}%
%TCIMACRO{\dint \limits_{x_{0}}^{x}}%
%BeginExpansion
{\displaystyle\int\limits_{x_{0}}^{x}}
%EndExpansion
dx^{\prime}%
%TCIMACRO{\dint \limits_{y_{0}}^{y}}%
%BeginExpansion
{\displaystyle\int\limits_{y_{0}}^{y}}
%EndExpansion
dy^{\prime}B_{z}(x^{\prime},y^{\prime})+h(y)\right\}  +\hat{h}(x_{0})
\]

\[
with\text{ \ }h(y)\text{ \ }chosen\text{ \ }so\text{ \ }that\text{ \ }\left\{
{\Huge -}%
%TCIMACRO{\dint \limits_{x_{0}}^{x}}%
%BeginExpansion
{\displaystyle\int\limits_{x_{0}}^{x}}
%EndExpansion
dx^{\prime}%
%TCIMACRO{\dint \limits_{y_{0}}^{y}}%
%BeginExpansion
{\displaystyle\int\limits_{y_{0}}^{y}}
%EndExpansion
dy^{\prime}B_{z}(x^{\prime},y^{\prime})+h(y)\right\}  \boldsymbol{:}\text{ is
}\mathsf{independent\ of\ }\ y,
\]
which is eq.(\ref{Lambda(x,y)4}) of the text, or eq.(\ref{static2}) but with
included multiplicities through the extra $\hat{h}(x_{0})$. One can actually
show that the two solutions are equivalent (i.e. (\ref{static1}) and
(\ref{static2}) for a simple-connected space are equal\cite{kyriakos}), a fact
that can be proven in a way similar to the $(x,t)$-cases of Appendix C. (For
the case of multiple-connectivity of the two-dimensional space, a brief
discussion of the actual values of the multiplicities $f(y_{0})$ and $\hat
{h}(x_{0})$ has been given at the end of Section X and is presented in more
detail in Appendix F).

\subsection{Appendix E}

\bigskip We here provide the spatial solutions in polar coordinates. By
following a similar procedure (of solving the system of PDEs resulting from
(\ref{usualgrad})) in polar coordinates $(\rho,\varphi)$, namely%

\[
\frac{\partial\Lambda(\rho,\varphi)}{\partial\rho}=A_{\rho}(\rho
,\varphi)\qquad and\qquad\frac{1}{\rho}\frac{\partial\Lambda(\rho,\varphi
)}{\partial\varphi}=A_{\varphi}(\rho,\varphi)
\]
with steps completely analogous to those of Appendix D, one can obtain the
following analogs of solutions (\ref{Lambda(x,y)1}) and (\ref{Lambda(x,y)4}), namely%

\begin{equation}
\Lambda(\rho,\varphi)=\Lambda(\rho_{0},\varphi_{0})+%
%TCIMACRO{\dint \limits_{\rho_{0}}^{\rho}}%
%BeginExpansion
{\displaystyle\int\limits_{\rho_{0}}^{\rho}}
%EndExpansion
A_{\rho}(\rho^{\prime},\varphi)d\rho^{\prime}+%
%TCIMACRO{\dint \limits_{\varphi_{0}}^{\varphi}}%
%BeginExpansion
{\displaystyle\int\limits_{\varphi_{0}}^{\varphi}}
%EndExpansion
\rho_{0}A_{\varphi}(\rho_{0},\varphi^{\prime})d\varphi^{\prime}+\left\{
%TCIMACRO{\dint \limits_{\varphi_{0}}^{\varphi}}%
%BeginExpansion
{\displaystyle\int\limits_{\varphi_{0}}^{\varphi}}
%EndExpansion
d\varphi^{\prime}%
%TCIMACRO{\dint \limits_{\rho_{0}}^{\rho}}%
%BeginExpansion
{\displaystyle\int\limits_{\rho_{0}}^{\rho}}
%EndExpansion
\rho^{\prime}d\rho^{\prime}B_{z}(\rho^{\prime},\varphi^{\prime})+g(\rho
)\right\}  +f(\varphi_{0}) \label{polar1}%
\end{equation}

\begin{equation}
with\ \ g(\rho)\ \ chosen\text{ \ }so\text{ \ }that\ \ \left\{
%TCIMACRO{\dint \limits_{\varphi_{0}}^{\varphi}}%
%BeginExpansion
{\displaystyle\int\limits_{\varphi_{0}}^{\varphi}}
%EndExpansion
d\varphi^{\prime}%
%TCIMACRO{\dint \limits_{\rho_{0}}^{\rho}}%
%BeginExpansion
{\displaystyle\int\limits_{\rho_{0}}^{\rho}}
%EndExpansion
\rho^{\prime}d\rho^{\prime}B_{z}(\rho^{\prime},\varphi^{\prime})+g(\rho
)\right\}  \boldsymbol{:}\text{ is }\mathsf{independent\ of\ }\ \rho,
\label{polarcond1}%
\end{equation}

and%

\begin{equation}
\Lambda(\rho,\varphi)=\Lambda(\rho_{0},\varphi_{0})+%
%TCIMACRO{\dint \limits_{\rho_{0}}^{\rho}}%
%BeginExpansion
{\displaystyle\int\limits_{\rho_{0}}^{\rho}}
%EndExpansion
A_{\rho}(\rho^{\prime},\varphi_{0})d\rho^{\prime}+%
%TCIMACRO{\dint \limits_{\varphi_{0}}^{\varphi}}%
%BeginExpansion
{\displaystyle\int\limits_{\varphi_{0}}^{\varphi}}
%EndExpansion
\rho A_{\varphi}(\rho,\varphi^{\prime})d\varphi^{\prime}+\left\{  -%
%TCIMACRO{\dint \limits_{\rho_{0}}^{\rho}}%
%BeginExpansion
{\displaystyle\int\limits_{\rho_{0}}^{\rho}}
%EndExpansion
\rho^{\prime}d\rho^{\prime}%
%TCIMACRO{\dint \limits_{\varphi_{0}}^{\varphi}}%
%BeginExpansion
{\displaystyle\int\limits_{\varphi_{0}}^{\varphi}}
%EndExpansion
d\varphi^{\prime}B_{z}(\rho^{\prime},\varphi^{\prime})+h(\varphi)\right\}
+\hat{h}(\rho_{0}) \label{polar2}%
\end{equation}

\begin{equation}
with\ \ h(\varphi)\ \ chosen\text{ \ }so\text{ \ }that\ \ \left\{  -%
%TCIMACRO{\dint \limits_{\rho_{0}}^{\rho}}%
%BeginExpansion
{\displaystyle\int\limits_{\rho_{0}}^{\rho}}
%EndExpansion
\rho^{\prime}d\rho^{\prime}%
%TCIMACRO{\dint \limits_{\varphi_{0}}^{\varphi}}%
%BeginExpansion
{\displaystyle\int\limits_{\varphi_{0}}^{\varphi}}
%EndExpansion
d\varphi^{\prime}\in(\rho^{\prime},\varphi^{\prime})+h(\varphi)\right\}
\boldsymbol{:}\text{ is }\mathsf{independent\ of\ }\ \varphi,
\label{polarcond2}%
\end{equation}
and in these, the proper choices of $g(\rho)$ and $h(\varphi)$ will again be
determined by their corresponding conditions, depending on the actual shape of
the $B_{z}$-distribution and the positioning of initial and final points
$(\rho_{0},\varphi_{0})$ and $(\rho,\varphi).$ [Furthermore, the observation
rectangle has now given its place to a slice of an annular section].

\subsection{Appendix F}

We here discuss the issue of multiplicities (the last terms of
(\ref{Lambda(x,y)1}) and (\ref{Lambda(x,y)4})) in case of spatial
multiple-connectivity (such as the standard magnetic Aharonov-Bohm case, in
which we can take $g(x)=0$ \ \textit{and} \ $h(y)=0$, since the enclosed
magnetic flux is independent of both $x$ and $y$).

For the Aharonov-Bohm setting we will have to deal with multiple-connected
space and with a (static) magnetic flux $\Phi$ being contained only in the
physically inaccessible region.\ In such a case we know that the
$\Lambda(\mathbf{r})$ that solves (\ref{usualgrad}) is not single-valued. How
is this fact (and the standard result (\ref{usualABLambda})) compatible with
the new formulation? To answer this in full generality we will consider two
separate cases that arise naturally (pertaining to the issue of what the dummy
variables $(x^{\prime},y^{\prime})$ inside the $B_{z}$-terms of our results
(i.e. of (\ref{static1}) and (\ref{static2})) actually represent). First, if
the variables $x$ and $y$ everywhere in the text always denote only
coordinates of the region that is physically accessible to the particle, then
$B_{z}$ is everywhere vanishing, this effectively reducing (\ref{static1}) and
(\ref{static2}) to%

\[
\Lambda(x,y)=\Lambda(x_{0},y_{0})+\int_{x_{0}}^{x}A_{x}(x^{\prime
},y)dx^{\prime}+\int_{y_{0}}^{y}A_{y}(\mathbf{x}_{\mathbf{0}},y^{\prime
})dy^{\prime}+C
\]

\[
\Lambda(x,y)=\Lambda(x_{0},y_{0})+\int_{x_{0}}^{x}A_{x}(x^{\prime}%
,\mathbf{y}_{\mathbf{0}})dx^{\prime}+\int_{y_{0}}^{y}A_{y}(x,y^{\prime
})dy^{\prime}+C
\]
with $C$ a common constant; these are the standard results (the Dirac phases)
along the two alternative paths discussed in the text (the red and green paths
of Fig.1) that (through their difference) lead to the magnetic Aharonov-Bohm
effect ($\Lambda$ being no longer single-valued and the difference of the two
solutions giving the enclosed (and physically inaccessible) $\Phi).$ Let us
however be even more general and let us decide to use the variables $x$ and
$y$ to \textit{also} denote coordinates of the physically inaccessible
region$\boldsymbol{;}$ this would be the case, if, for example, we had
previously started with that region being accessible (i.e. through a
penetrable scalar potential) and at the end we followed a limiting procedure
(i.e. of this scalar potential going to infinity) so that this region would
become in the limit impenetrable and therefore inaccessible. In such a case
the variables $x$ and $y$ would now contain \textit{remnants} of the
previously allowed values (but currently not allowed for the description of
particle coordinates) such as the values of the dummy variables $x^{\prime}$
and $y^{\prime}$ in the $B_{z}$-terms of (\ref{static1}) and (\ref{static2}%
)$\boldsymbol{;}$ such values would therefore still be present in the
expressions giving $\Lambda$ (even though these dummy variables $x^{\prime}$
and $y^{\prime}$ would now describe an inaccessible region). In other words,
the inaccessible $B_{z}$ is still formally present in the problem and it shows
up explicitly in the generalized gauge functions of the new formulation. How
does this formulation then lead to the standard Aharonov-Bohm result in such a
limiting case (essentially a case of smoothly-induced spatial multiple-connectivity)?

Before we answer this, the reader should probably be reminded that our
formulation only deals with wavefunction-phases$\boldsymbol{;}$ questions
therefore of rigid (vanishing) boundary conditions (on the boundary of the
inaccessible region) that apply to (and must be imposed on) the entire
wavefunction, and mostly on its modulus, can only be addressed indirectly (and
as we will see, through a \textquotedblleft memory\textquotedblright\ that the
phases have of their multivaluedness, whenever the space is
multiple-connected). To see this, we need the generalized results (eqs
(\ref{Lambda(x,y)1}) and (\ref{Lambda(x,y)4})) that contain the additional
\textquotedblleft multiplicities\textquotedblright. These most general results
(for multiple-connected space) were derived in Section IX and have the form%

\[
\Lambda(x,y)=\Lambda(x_{0},y_{0})+%
%TCIMACRO{\dint \limits_{x_{0}}^{x}}%
%BeginExpansion
{\displaystyle\int\limits_{x_{0}}^{x}}
%EndExpansion
A_{x}(x^{\prime},y)dx^{\prime}+%
%TCIMACRO{\dint \limits_{y_{0}}^{y}}%
%BeginExpansion
{\displaystyle\int\limits_{y_{0}}^{y}}
%EndExpansion
A_{y}(x_{0},y^{\prime})dy^{\prime}+\left\{
%TCIMACRO{\dint \limits_{y_{0}}^{y}}%
%BeginExpansion
{\displaystyle\int\limits_{y_{0}}^{y}}
%EndExpansion
dy^{\prime}%
%TCIMACRO{\dint \limits_{x_{0}}^{x}}%
%BeginExpansion
{\displaystyle\int\limits_{x_{0}}^{x}}
%EndExpansion
dx^{\prime}B_{z}(x^{\prime},y^{\prime})+g(x)\right\}  +f(y_{0})
\]
and%
\[
\Lambda(x,y)=\Lambda(x_{0},y_{0})+\int_{x_{0}}^{x}A_{x}(x^{\prime}%
,y_{0})dx^{\prime}+\int_{y_{0}}^{y}A_{y}(x,y^{\prime})dy^{\prime}+\left\{
{\Huge -}%
%TCIMACRO{\dint \limits_{x_{0}}^{x}}%
%BeginExpansion
{\displaystyle\int\limits_{x_{0}}^{x}}
%EndExpansion
dx^{\prime}%
%TCIMACRO{\dint \limits_{y_{0}}^{y}}%
%BeginExpansion
{\displaystyle\int\limits_{y_{0}}^{y}}
%EndExpansion
dy^{\prime}B_{z}(x^{\prime},y^{\prime})+h(y)\right\}  +\hat{h}(x_{0})
\]
with the functions $g(x)$ and $h(y)$ satisfying the same conditions as in
(\ref{static1}) and (\ref{static2}). We note the extra appearance of the new
constant terms $f(y_{0})$ and $\hat{h}(x_{0})$ (the \textquotedblleft
multiplicities\textquotedblright) and these are \textquotedblleft
defined\textquotedblright\ (see (\ref{1stintegration(x,y)}) and
(\ref{2ndintegration(x,y)}) where the functions $f$ and $\hat{h}$ were first
introduced) by%
\[
f(y_{0})=\Lambda(x,y_{0})-\Lambda(x_{0},y_{0})-\int_{x_{0}}^{x}A_{x}%
(x^{\prime},y_{0})dx^{\prime}%
\]
and%
\[
\hat{h}(x_{0})=\Lambda(x_{0},y)-\Lambda(x_{0},y_{0})-\int_{y_{0}}^{y}%
A_{y}(x_{0},y^{\prime})dy^{\prime}.
\]
Let us then identify proper choices for the functions $g(x)$ and $h(y)$ and
for the constants $f(y_{0})$ and $\hat{h}(x_{0})$ in the above case of spatial
multiple-connectivity (such as the standard magnetic Aharonov-Bohm case, with
a non-extended (and static) magnetic flux in the forbidden
region)$\boldsymbol{:}$ First, we can always take $g(x)=0$ and\ $h(y)=0$
(always up to a common constant as discussed earlier), since the enclosed
magnetic flux is (in this Aharonov-Bohm case) independent of both $x$ and $y$
-- the conditions of $g(x)$ and $h(y)$ being then automatically satisfied.
Second, let us look more closely at the above \textquotedblleft
definitions\textquotedblright\ of $f(y_{0})$ and $\hat{h}(x_{0})\boldsymbol{:}%
$

We first note that $f(y_{0})$ must be independent of $x$, and this is indeed
true as is apparent by formally taking the derivative of the above definition
of $f(y_{0})$ with respect to $x\boldsymbol{;}$ we then have $\frac{\partial
f(y_{0})}{\partial x}=$ $\frac{\partial\Lambda(x,y_{0})}{\partial x}%
-A_{x}(x,y_{0})$ which is indeed zero (as $\Lambda(x,y)$ satisfies by
assumption the first equation of the system (\ref{usualgradcomps}) of PDEs
(evaluated at $y=y_{0}$)), showing that $\frac{\partial f(y_{0})}{\partial
x}=0$ and that $f(y_{0})$ does not really depend on the variable $x$ that
appears in its definition. We can therefore determine its value by taking the
limit $x\rightarrow x_{0}$ (for fixed $y_{0}$)$\boldsymbol{:}$ we see from the
above that this limit is simply equal to $\lim_{x\rightarrow x_{0}}%
\Lambda(x,y_{0})-\Lambda(x_{0},y_{0})$ [as mentioned earlier, we leave out
cases where $A_{x}$ has a $\delta$-function form, so that interchanges of all
integrals are allowed in our earlier derivations (in Appendix D)], and this
difference is nonzero only when there is a multivaluedness of $\Lambda$ at the
point $(x_{0},y_{0})$, as \textit{is }actually our case. The limit
$x\rightarrow x_{0}$ (for fixed $y_{0}$) in the path-sense of solution
(\ref{static1}) (or of (\ref{Lambda(x,y)1})) that is then needed here in order
to determine $f(y_{0})$, is equivalent to making an entire closed trip around
the observation rectangle in the \textit{negative} sense, landing on the
initial point $(x_{0},y_{0}),$ this therefore giving the value $f(y_{0})=$
minus enclosed magnetic flux $=-\Phi$ (which is indeed a constant independent
of $x$ and $y$, as it \textit{should} be).\ By following a completely
symmetric argument for the above definition of $\hat{h}(x_{0})$ (and by now
taking the limit $y\rightarrow y_{0}$ (for fixed $x_{0}$), that is now
equivalent to going around the loop in the \textit{positive} sense, landing
again on the initial point $(x_{0},y_{0})$) we obtain\ $\hat{h}(x_{0})=+\Phi$.
If these values of $f(y_{0})$ and $\hat{h}(x_{0})$ are finally substituted in
the above most general solutions (eqs (\ref{Lambda(x,y)1}) and
(\ref{Lambda(x,y)4})) together with $g(x)=h(y)=0$, then we note that
$f(y_{0})$ cancels out the {\Huge \ }$%
%TCIMACRO{\dint \limits_{y_{0}}^{y}}%
%BeginExpansion
{\displaystyle\int\limits_{y_{0}}^{y}}
%EndExpansion
dy^{\prime}%
%TCIMACRO{\dint \limits_{x_{0}}^{x}}%
%BeginExpansion
{\displaystyle\int\limits_{x_{0}}^{x}}
%EndExpansion
dx^{\prime}B_{z}(x^{\prime},y^{\prime})$ term (which is here just equal to the
inaccessible flux $\Phi$), and $\hat{h}(x_{0})$ cancels out the \ {\Huge -}$%
%TCIMACRO{\dint \limits_{x_{0}}^{x}}%
%BeginExpansion
{\displaystyle\int\limits_{x_{0}}^{x}}
%EndExpansion
dx^{\prime}%
%TCIMACRO{\dint \limits_{y_{0}}^{y}}%
%BeginExpansion
{\displaystyle\int\limits_{y_{0}}^{y}}
%EndExpansion
dy^{\prime}B_{z}(x^{\prime},y^{\prime})$ term, and the two solutions are then
once again reduced to the usual solutions of mere $A$-integrals along the two
paths (i.e. the standard Dirac phase, with no nonlocal contributions) -- their
difference giving the closed loop integral of $\boldsymbol{A},$ hence the
inaccessible flux and, finally, the well-known magnetic Aharonov-Bohm result.
One should note here that the standard result in the new formulation requires
some effort and it is only derived indirectly (due to the fact that we only
deal with phases and not the moduli of wavefunctions, on which boundary
conditions are normally imposed), and it basically comes from the
\textquotedblleft memory\textquotedblright\ of the multivaluedness that the
\textquotedblleft gauge function\textquotedblright\ $\Lambda$ carries (due to
the multiple-connectivity of space).

\subsection{Appendix G}

We here present the method of solution of the system (\ref{FullSystem}), and
provide a detailed derivation of solutions (\ref{LambdaFull1}),
(\ref{LambdaFull2}), (\ref{LambdaFull4}) and (\ref{LambdaFIN}) of the text.
Starting with the \textit{second} of (\ref{FullSystem}), and by integrating it
we obtain the expected generalization of (\ref{2ndintegration(x,y)}), namely%

\begin{equation}
\Lambda(x,y,t)-\Lambda(x,y_{0},t)=\int_{y_{0}}^{y}A_{y}(x,y^{\prime
},t)dy^{\prime}+f(x,t) \label{1stintegrationFull}%
\end{equation}
which if substituted to the first of (\ref{FullSystem}) gives (after
integration over $x^{\prime}$) \ a $t$-generalization of (\ref{h(x)}), namely%

\begin{equation}
f(x,t)=f(x_{0},t)-\Lambda(x,y_{0},t)+\Lambda(x_{0},y_{0},t)-%
%TCIMACRO{\dint \limits_{x_{0}}^{x}}%
%BeginExpansion
{\displaystyle\int\limits_{x_{0}}^{x}}
%EndExpansion
dx^{\prime}%
%TCIMACRO{\dint \limits_{y_{0}}^{y}}%
%BeginExpansion
{\displaystyle\int\limits_{y_{0}}^{y}}
%EndExpansion
dy^{\prime}\frac{\partial A_{y}(x^{\prime},y^{\prime},t)}{\partial x^{\prime}%
}+\int_{x_{0}}^{x}A_{x}\left(  x^{\prime},y,t\right)  dx^{\prime}+G(y,t)
\label{medium}%
\end{equation}
with $G(y,t)$ to be chosen in such a way that the entire right-hand-side of
(\ref{medium}) is only a function of $\ x$ and $t$, as it should be (hence
independent of $\ y$). Finally, by substituting \ \ $\frac{\partial
A_{y}(x^{\prime},y^{\prime},t)}{\partial x^{\prime}}$ \ with \ \ $\frac
{\partial A_{x}(x^{\prime},y^{\prime},t)}{\partial y^{\prime}}+B_{z}%
(x^{\prime},y^{\prime},t)$, carrying out the integration with respect to
$y^{\prime}$, and by demanding that $\ f(x,t)$ \ be independent of $\ y$, we
obtain the following temporal generalization of (\ref{Lambda(x,y)4})%

\[
\Lambda(x,y,t)=\Lambda(x_{0},y_{0},t)+\int_{x_{0}}^{x}A_{x}(x^{\prime}%
,y_{0},t)dx^{\prime}+\int_{y_{0}}^{y}A_{y}(x,y^{\prime},t)dy^{\prime}+
\]

\[
+\left\{  {\Huge -}%
%TCIMACRO{\dint \limits_{x_{0}}^{x}}%
%BeginExpansion
{\displaystyle\int\limits_{x_{0}}^{x}}
%EndExpansion
dx^{\prime}%
%TCIMACRO{\dint \limits_{y_{0}}^{y}}%
%BeginExpansion
{\displaystyle\int\limits_{y_{0}}^{y}}
%EndExpansion
dy^{\prime}B_{z}(x^{\prime},y^{\prime},t)+G(y,t)\right\}  +f(x_{0},t)
\]

\[
with\text{ \ }G(y,t)\text{ \ }such\text{ }that\text{ \ \ }\left\{  {\Huge -}%
%TCIMACRO{\dint \limits_{x_{0}}^{x}}%
%BeginExpansion
{\displaystyle\int\limits_{x_{0}}^{x}}
%EndExpansion
dx^{\prime}%
%TCIMACRO{\dint \limits_{y_{0}}^{y}}%
%BeginExpansion
{\displaystyle\int\limits_{y_{0}}^{y}}
%EndExpansion
dy^{\prime}B_{z}(x^{\prime},y^{\prime},t)+G(y,t)\right\}  \boldsymbol{:}\text{
\ is }\mathsf{independent\ of\ }\ y.
\]
This is eq.(\ref{Lambda(x,y,t)213}) of the text. From this point on, the third
equation of the system (\ref{FullSystem}) is getting involved to determine the
nontrivial effect of scalar potentials on $G(y,t)\boldsymbol{;}$ by combining
it with (\ref{Lambda(x,y,t)213}) there results a wealth of
patterns$\boldsymbol{:}$ integration with respect to $t^{\prime}$ leads to%

\[
G(y,t)=G(y,t_{0})-\Lambda(x_{0},y_{0},t)+\Lambda(x_{0},y_{0},t_{0}%
)-f(x_{0},t)+f(x_{0},t_{0})-c%
%TCIMACRO{\dint \limits_{t_{0}}^{t}}%
%BeginExpansion
{\displaystyle\int\limits_{t_{0}}^{t}}
%EndExpansion
\phi(x,y,t^{\prime})dt^{\prime}-
\]

\begin{equation}
-\left[
%TCIMACRO{\dint \limits_{t_{0}}^{t}}%
%BeginExpansion
{\displaystyle\int\limits_{t_{0}}^{t}}
%EndExpansion
dt^{\prime}%
%TCIMACRO{\dint \limits_{x_{0}}^{x}}%
%BeginExpansion
{\displaystyle\int\limits_{x_{0}}^{x}}
%EndExpansion
dx^{\prime}\frac{\partial A_{x}(x^{\prime},y_{0},t^{\prime})}{\partial
t^{\prime}}+%
%TCIMACRO{\dint \limits_{t_{0}}^{t}}%
%BeginExpansion
{\displaystyle\int\limits_{t_{0}}^{t}}
%EndExpansion
dt^{\prime}%
%TCIMACRO{\dint \limits_{y_{0}}^{y}}%
%BeginExpansion
{\displaystyle\int\limits_{y_{0}}^{y}}
%EndExpansion
dy^{\prime}\frac{\partial A_{y}(x,y^{\prime},t^{\prime})}{\partial t^{\prime}%
}\right]  +%
%TCIMACRO{\dint \limits_{t_{0}}^{t}}%
%BeginExpansion
{\displaystyle\int\limits_{t_{0}}^{t}}
%EndExpansion
dt^{\prime}%
%TCIMACRO{\dint \limits_{x_{0}}^{x}}%
%BeginExpansion
{\displaystyle\int\limits_{x_{0}}^{x}}
%EndExpansion
dx^{\prime}%
%TCIMACRO{\dint \limits_{y_{0}}^{y}}%
%BeginExpansion
{\displaystyle\int\limits_{y_{0}}^{y}}
%EndExpansion
dy^{\prime}\frac{\partial B_{z}(x^{\prime},y^{\prime},t^{^{\prime}})}{\partial
t^{\prime}}+F(x,y) \label{med1}%
\end{equation}
with $F(x,y)$ to be chosed in such a way that the entire right-hand-side of
(\ref{med1}) is only a function of $(y,t),$ as it should be, hence independent
of $x.$ In (\ref{med1}) there are two possible ways to determine the term in
brackets, and another two ways to determine the term containing $B_{z}.$ The
easiest to follow (the one that more directly leads to the final
\textit{conditions} that the functions $F(x,y)$ and $G(y,t_{0})$ are required
to satisfy) is$\boldsymbol{:}$ (i) to substitute $\frac{\partial
A_{x}(x^{\prime},y_{0},t^{\prime})}{\partial t^{\prime}}$ \ with
\ \ $-c\left(  E_{x}(x^{\prime},y_{0},t^{\prime})+\frac{\partial\phi
(x^{\prime},y_{0},t^{\prime})}{\partial x^{\prime}}\right)  $ (and similarly
for $\frac{\partial A_{y}(x,y^{\prime},t^{\prime})}{\partial t^{\prime}}$),
and \ (ii) to use the proviso that magnetic and electric fields are connected
through the Faraday's law of Induction, namely \ $\frac{\partial
B_{z}(x^{\prime},y^{\prime},t^{\prime})}{\partial t^{\prime}}=-c\left(
\frac{\partial E_{y}(x^{\prime},y^{\prime},t^{\prime})}{\partial x^{\prime}%
}-\frac{\partial E_{x}(x^{\prime},y^{\prime},t^{\prime})}{\partial y^{\prime}%
}\right)  .$ These substitutions lead to cancellations of several intermediate
quantities in (\ref{Lambda(x,y,t)213}) and (\ref{med1}) and lead to the final
result (which is eq.(\ref{LambdaFull1}) of the text), namely%

\[
\Lambda(x,y,t)=\Lambda(x_{0},y_{0},t_{0})+\int_{x_{0}}^{x}A_{x}(x^{\prime
},y_{0},t)dx^{\prime}+\int_{y_{0}}^{y}A_{y}(x,y^{\prime},t)dy^{\prime}-%
%TCIMACRO{\dint \limits_{x_{0}}^{x}}%
%BeginExpansion
{\displaystyle\int\limits_{x_{0}}^{x}}
%EndExpansion
dx^{\prime}%
%TCIMACRO{\dint \limits_{y_{0}}^{y}}%
%BeginExpansion
{\displaystyle\int\limits_{y_{0}}^{y}}
%EndExpansion
dy^{\prime}B_{z}(x^{\prime},y^{\prime},t)+G(y,t_{0})-
\]

\[
-c%
%TCIMACRO{\dint \limits_{t_{0}}^{t}}%
%BeginExpansion
{\displaystyle\int\limits_{t_{0}}^{t}}
%EndExpansion
\phi(x_{0},y_{0},t^{\prime})dt^{\prime}+c%
%TCIMACRO{\dint \limits_{t_{0}}^{t}}%
%BeginExpansion
{\displaystyle\int\limits_{t_{0}}^{t}}
%EndExpansion
dt^{\prime}%
%TCIMACRO{\dint \limits_{x_{0}}^{x}}%
%BeginExpansion
{\displaystyle\int\limits_{x_{0}}^{x}}
%EndExpansion
dx^{\prime}E_{x}(x^{\prime},y,t^{\prime})+c%
%TCIMACRO{\dint \limits_{t_{0}}^{t}}%
%BeginExpansion
{\displaystyle\int\limits_{t_{0}}^{t}}
%EndExpansion
dt^{\prime}%
%TCIMACRO{\dint \limits_{y_{0}}^{y}}%
%BeginExpansion
{\displaystyle\int\limits_{y_{0}}^{y}}
%EndExpansion
dy^{\prime}E_{y}(x_{0},y^{\prime},t^{\prime})+F(x,y)+f(x_{0},t_{0})
\]
with the functions $G(y,t_{0})$ \ and $\ F(x,y)$ \ to be chosen in such a way
as to satisfy the following 3 independent conditions$\boldsymbol{:}$%

\[
\left\{  G(y,t_{0})-%
%TCIMACRO{\dint \limits_{x_{0}}^{x}}%
%BeginExpansion
{\displaystyle\int\limits_{x_{0}}^{x}}
%EndExpansion
dx^{\prime}%
%TCIMACRO{\dint \limits_{y_{0}}^{y}}%
%BeginExpansion
{\displaystyle\int\limits_{y_{0}}^{y}}
%EndExpansion
dy^{\prime}B_{z}(x^{\prime},y^{\prime},t_{0})\right\}  \boldsymbol{:}%
\ is\ \mathsf{independent\ of\ }\ y,
\]
which is eq.(\ref{Gcondition}) of the text (and a special case of the
condition on $G(y,t)$ above (see after (\ref{Lambda(x,y,t)213}))), and the
other 2 turn out to be of the form%

\[
\left\{  F(x,y)+c%
%TCIMACRO{\dint \limits_{t_{0}}^{t}}%
%BeginExpansion
{\displaystyle\int\limits_{t_{0}}^{t}}
%EndExpansion
dt^{\prime}%
%TCIMACRO{\dint \limits_{x_{0}}^{x}}%
%BeginExpansion
{\displaystyle\int\limits_{x_{0}}^{x}}
%EndExpansion
dx^{\prime}E_{x}(x^{\prime},y,t^{\prime})\right\}  \boldsymbol{:}%
\ is\ \mathsf{independent\ of\ }\ x,
\]

\[
\left\{  F(x,y)+c%
%TCIMACRO{\dint \limits_{t_{0}}^{t}}%
%BeginExpansion
{\displaystyle\int\limits_{t_{0}}^{t}}
%EndExpansion
dt^{\prime}%
%TCIMACRO{\dint \limits_{y_{0}}^{y}}%
%BeginExpansion
{\displaystyle\int\limits_{y_{0}}^{y}}
%EndExpansion
dy^{\prime}E_{y}(x,y^{\prime},t^{\prime})\right\}  \boldsymbol{:}%
\ is\ \mathsf{independent\ of\ }\ y,
\]
which are eqs (\ref{F(x,y)condition1}) and (\ref{F(x,y)condition2}) of the
text. It should be noted (for the reader who wants to follow all the steps)
that the final condition (\ref{F(x,y)condition2}) does \textit{not} come out
\textit{directly} as the other two$\boldsymbol{;}$ because the function
$G(y,t)$ has disappeared from the final form (\ref{LambdaFull1}), one needs to
\textit{separately} impose the condition above for $G(y,t)$ \ {\Huge (}namely
$\left\{  {\Huge -}%
%TCIMACRO{\dint \limits_{x_{0}}^{x}}%
%BeginExpansion
{\displaystyle\int\limits_{x_{0}}^{x}}
%EndExpansion
dx^{\prime}%
%TCIMACRO{\dint \limits_{y_{0}}^{y}}%
%BeginExpansion
{\displaystyle\int\limits_{y_{0}}^{y}}
%EndExpansion
dy^{\prime}B_{z}(x^{\prime},y^{\prime},t)+G(y,t)\right\}  \boldsymbol{:}$
$\mathsf{independent\ of\ }\ y${\Huge )} directly on the form (\ref{med1}%
)$\boldsymbol{;}$ and in so doing, it is advantageous to interchange
integrations (namely, do the $t^{\prime}$-integral first) \ in the $B_{z}%
$-term of (\ref{med1}), so that $%
%TCIMACRO{\dint \limits_{t_{0}}^{t}}%
%BeginExpansion
{\displaystyle\int\limits_{t_{0}}^{t}}
%EndExpansion
dt^{\prime}%
%TCIMACRO{\dint \limits_{x_{0}}^{x}}%
%BeginExpansion
{\displaystyle\int\limits_{x_{0}}^{x}}
%EndExpansion
dx^{\prime}%
%TCIMACRO{\dint \limits_{y_{0}}^{y}}%
%BeginExpansion
{\displaystyle\int\limits_{y_{0}}^{y}}
%EndExpansion
dy^{\prime}\frac{\partial B_{z}(x^{\prime},y^{\prime},t^{^{\prime}})}{\partial
t^{\prime}}=%
%TCIMACRO{\dint \limits_{x_{0}}^{x}}%
%BeginExpansion
{\displaystyle\int\limits_{x_{0}}^{x}}
%EndExpansion
dx^{\prime}%
%TCIMACRO{\dint \limits_{y_{0}}^{y}}%
%BeginExpansion
{\displaystyle\int\limits_{y_{0}}^{y}}
%EndExpansion
dy^{\prime}{\huge (}B_{z}(x^{\prime},y^{\prime},t)-B_{z}(x^{\prime},y^{\prime
},t_{0}){\huge )}$, and then impose the (less stringent) condition
(\ref{Gcondition}) on $G(y,t_{0})\boldsymbol{;}$ by following this strategy,
after a number of cancellations of intermediate quantities one finally obtains
the 3rd condition (\ref{F(x,y)condition2}) on $F(x,y).$ (As for the constant
quantity\ $f(x_{0},t_{0})$ appearing in (\ref{LambdaFull1}), this again
describes possible effects of multiple-connectivity at the instant $t_{0}$
(which are absent for simple-connected spacetimes, but were crucial in the
discussion of the van Kampen thought-experiment of the text)).

\bigskip Eq. (\ref{LambdaFull1}) is our first solution. It is now crucial to
note that an alternative form of solution (with the functions $G^{\prime}s$
and $F$ satisfying the \textit{same} conditions as above) can be derived if,
in the term in brackets of (\ref{med1}) we merely interchange integrations,
leaving therefore $A$'s everywhere rather than introducing electric
fields$\boldsymbol{;}$ following at the same time the above strategy of
changing the ordering of integrations in the $B_{z}$-term as well (without
therefore using Faraday's law) this alternative form of solution turns out to be%

\[
\Lambda(x,y,t)=\Lambda(x_{0},y_{0},t_{0})+\int_{x_{0}}^{x}A_{x}(x^{\prime
},y_{0},t)dx^{\prime}+\int_{y_{0}}^{y}A_{y}(x,y^{\prime},t)dy^{\prime}-%
%TCIMACRO{\dint \limits_{x_{0}}^{x}}%
%BeginExpansion
{\displaystyle\int\limits_{x_{0}}^{x}}
%EndExpansion
dx^{\prime}%
%TCIMACRO{\dint \limits_{y_{0}}^{y}}%
%BeginExpansion
{\displaystyle\int\limits_{y_{0}}^{y}}
%EndExpansion
dy^{\prime}B_{z}(x^{\prime},y^{\prime},t_{0})+G(y,t_{0})-
\]

\[
-c%
%TCIMACRO{\dint \limits_{t_{0}}^{t}}%
%BeginExpansion
{\displaystyle\int\limits_{t_{0}}^{t}}
%EndExpansion
\phi(x_{0},y_{0},t^{\prime})dt^{\prime}+c%
%TCIMACRO{\dint \limits_{t_{0}}^{t}}%
%BeginExpansion
{\displaystyle\int\limits_{t_{0}}^{t}}
%EndExpansion
dt^{\prime}%
%TCIMACRO{\dint \limits_{x_{0}}^{x}}%
%BeginExpansion
{\displaystyle\int\limits_{x_{0}}^{x}}
%EndExpansion
dx^{\prime}E_{x}(x^{\prime},y_{0},t^{\prime})+c%
%TCIMACRO{\dint \limits_{t_{0}}^{t}}%
%BeginExpansion
{\displaystyle\int\limits_{t_{0}}^{t}}
%EndExpansion
dt^{\prime}%
%TCIMACRO{\dint \limits_{y_{0}}^{y}}%
%BeginExpansion
{\displaystyle\int\limits_{y_{0}}^{y}}
%EndExpansion
dy^{\prime}E_{y}(x,y^{\prime},t^{\prime})+F(x,y)+f(x_{0},t_{0}),
\]
and is eq.(\ref{LambdaFull2}) of the text. In this alternative solution we
note that, in comparison with (\ref{LambdaFull1}), the line-integrals of
$\ \boldsymbol{E}$ \ have changed to the \textit{other} alternative
\textquotedblleft path\textquotedblright\ (note the difference in the
placement of the coordinates of the initial point $(x_{0},y_{0})$ in the
arguments of $E_{x}$ and $E_{y}$) and they happen to have the same sense as
the $\boldsymbol{A}$-integrals, while simultaneously the magnetic flux
difference shows up with its value at the initial time $t_{0}$ rather than at
$t$. This alternative form is useful in cases where we want to directly
compare physical situations in the present (at time $t$) and in the past (at
time $t_{0}$), and the above noted change of sense of $\boldsymbol{E}%
$-integrals (compared to (\ref{LambdaFull1})) was crucial in the discussion of
the text.

Once again the reader can directly verify that (\ref{LambdaFull1}) or
(\ref{LambdaFull2}) indeed satisfy the basic input system (\ref{FullSystem}).
(This verification is considerably more tedious than the ones of the main text
but straightforward, and is not shown here).

\bigskip But in order to make the above formalism useful for the van Kampen
case, namely an enclosed (and physically inaccessible) magnetic flux (which
however is \textit{time-dependent})\textit{, }it is important to have the
analogous forms through a reverse route, namely starting with (integrating)
the \textit{first} of (\ref{FullSystem}) and then substituting the result to
the second$\boldsymbol{;}$ in this way we will at the end have the reverse
\textquotedblleft path\textquotedblright\ of $\boldsymbol{A}$-integrals, so
that by taking the \textit{difference} of the resulting solution and the above
solution (\ref{LambdaFull1}) (or (\ref{LambdaFull2})) will lead to the
\textit{closed} line integral of $\boldsymbol{A}$\textbf{\ }which will be
immediately related to the van Kampen's magnetic flux (at the instant $t$). By
following then this route, and by applying a similar strategy at every
intermediate step, we finally obtain the following solution (the spatially
\textquotedblleft dual\textquotedblright\ of (\ref{LambdaFull1})), which is
eq.(\ref{LambdaFull4}) of the text, namely%

\[
\Lambda(x,y,t)=\Lambda(x_{0},y_{0},t_{0})+\int_{x_{0}}^{x}A_{x}(x^{\prime
},y,t)dx^{\prime}+\int_{y_{0}}^{y}A_{y}(x_{0},y^{\prime},t)dy^{\prime}+%
%TCIMACRO{\dint \limits_{x_{0}}^{x}}%
%BeginExpansion
{\displaystyle\int\limits_{x_{0}}^{x}}
%EndExpansion
dx^{\prime}%
%TCIMACRO{\dint \limits_{y_{0}}^{y}}%
%BeginExpansion
{\displaystyle\int\limits_{y_{0}}^{y}}
%EndExpansion
dy^{\prime}B_{z}(x^{\prime},y^{\prime},t)+\hat{G}(x,t_{0})-
\]

\[
-c%
%TCIMACRO{\dint \limits_{t_{0}}^{t}}%
%BeginExpansion
{\displaystyle\int\limits_{t_{0}}^{t}}
%EndExpansion
\phi(x_{0},y_{0},t^{\prime})dt^{\prime}+c%
%TCIMACRO{\dint \limits_{t_{0}}^{t}}%
%BeginExpansion
{\displaystyle\int\limits_{t_{0}}^{t}}
%EndExpansion
dt^{\prime}%
%TCIMACRO{\dint \limits_{x_{0}}^{x}}%
%BeginExpansion
{\displaystyle\int\limits_{x_{0}}^{x}}
%EndExpansion
dx^{\prime}E_{x}(x^{\prime},y_{0},t^{\prime})+c%
%TCIMACRO{\dint \limits_{t_{0}}^{t}}%
%BeginExpansion
{\displaystyle\int\limits_{t_{0}}^{t}}
%EndExpansion
dt^{\prime}%
%TCIMACRO{\dint \limits_{y_{0}}^{y}}%
%BeginExpansion
{\displaystyle\int\limits_{y_{0}}^{y}}
%EndExpansion
dy^{\prime}E_{y}(x,y^{\prime},t^{\prime})+F(x,y)+\hat{h}(y_{0},t_{0})
\]
with the functions $\hat{G}(x,t_{0})$ \ and $\ F(x,y)$ \ to be chosen in such
a way as to satisfy the following 3 independent conditions$\boldsymbol{:}$%

\[
\left\{  \hat{G}(x,t_{0})+%
%TCIMACRO{\dint \limits_{y_{0}}^{y}}%
%BeginExpansion
{\displaystyle\int\limits_{y_{0}}^{y}}
%EndExpansion
dy^{\prime}%
%TCIMACRO{\dint \limits_{x_{0}}^{x}}%
%BeginExpansion
{\displaystyle\int\limits_{x_{0}}^{x}}
%EndExpansion
dx^{\prime}B_{z}(x^{\prime},y^{\prime},t_{0})\right\}  \boldsymbol{:}%
\ is\ \ \mathsf{independent\ of\ }\ x,
\]

\bigskip%
\[
\left\{  F(x,y)+c%
%TCIMACRO{\dint \limits_{t_{0}}^{t}}%
%BeginExpansion
{\displaystyle\int\limits_{t_{0}}^{t}}
%EndExpansion
dt^{\prime}%
%TCIMACRO{\dint \limits_{x_{0}}^{x}}%
%BeginExpansion
{\displaystyle\int\limits_{x_{0}}^{x}}
%EndExpansion
dx^{\prime}E_{x}(x^{\prime},y,t^{\prime})\right\}  \boldsymbol{:}%
\ is\ \ \mathsf{independent\ of\ }\ x,
\]

\bigskip%
\[
\left\{  F(x,y)+c%
%TCIMACRO{\dint \limits_{t_{0}}^{t}}%
%BeginExpansion
{\displaystyle\int\limits_{t_{0}}^{t}}
%EndExpansion
dt^{\prime}%
%TCIMACRO{\dint \limits_{y_{0}}^{y}}%
%BeginExpansion
{\displaystyle\int\limits_{y_{0}}^{y}}
%EndExpansion
dy^{\prime}E_{y}(x,y^{\prime},t^{\prime})\right\}  \boldsymbol{:}%
\ is\ \ \mathsf{independent\ of\ }\ y,
\]
where again for the above results the Faraday's law was crucial. The above
conditions are eqs (\ref{Gcondition3})-(\ref{F(x,y)condition4}) of the text.

Finally, the corresponding analog of the alternative form (\ref{LambdaFull2})
(i.e. with $B_{z}$ now appearing at $t_{0}$) is more important and turns out
to be%

\[
\Lambda(x,y,t)=\Lambda(x_{0},y_{0},t_{0})+\int_{x_{0}}^{x}A_{x}(x^{\prime
},y,t)dx^{\prime}+\int_{y_{0}}^{y}A_{y}(x_{0},y^{\prime},t)dy^{\prime}+%
%TCIMACRO{\dint \limits_{x_{0}}^{x}}%
%BeginExpansion
{\displaystyle\int\limits_{x_{0}}^{x}}
%EndExpansion
dx^{\prime}%
%TCIMACRO{\dint \limits_{y_{0}}^{y}}%
%BeginExpansion
{\displaystyle\int\limits_{y_{0}}^{y}}
%EndExpansion
dy^{\prime}B_{z}(x^{\prime},y^{\prime},t_{0})+\hat{G}(x,t_{0})-
\]

\[
-c%
%TCIMACRO{\dint \limits_{t_{0}}^{t}}%
%BeginExpansion
{\displaystyle\int\limits_{t_{0}}^{t}}
%EndExpansion
\phi(x_{0},y_{0},t^{\prime})dt^{\prime}+c%
%TCIMACRO{\dint \limits_{t_{0}}^{t}}%
%BeginExpansion
{\displaystyle\int\limits_{t_{0}}^{t}}
%EndExpansion
dt^{\prime}%
%TCIMACRO{\dint \limits_{x_{0}}^{x}}%
%BeginExpansion
{\displaystyle\int\limits_{x_{0}}^{x}}
%EndExpansion
dx^{\prime}E_{x}(x^{\prime},y,t^{\prime})+c%
%TCIMACRO{\dint \limits_{t_{0}}^{t}}%
%BeginExpansion
{\displaystyle\int\limits_{t_{0}}^{t}}
%EndExpansion
dt^{\prime}%
%TCIMACRO{\dint \limits_{y_{0}}^{y}}%
%BeginExpansion
{\displaystyle\int\limits_{y_{0}}^{y}}
%EndExpansion
dy^{\prime}E_{y}(x_{0},y^{\prime},t^{\prime})+F(x,y)+\hat{h}(y_{0},t_{0})
\]
which is eq.(\ref{LambdaFIN}) of the text, with $\hat{G}(x,t_{0})$ and
$F(x,y)$ following the same 3 conditions above. The constant term $\hat
{h}(y_{0},t_{0})$ again describes possible multiplicities at the instant
$t_{0}\boldsymbol{;}$ it is absent for simple-connected spacetimes, but was
crucial in the discussion of the van Kampen thought-experiment of Section XII.

In (\ref{LambdaFull4}) (and in (\ref{LambdaFIN})), note the \textquotedblleft
alternative paths\textquotedblright\ (compared to solution (\ref{LambdaFull1})
(and (\ref{LambdaFull2}))) of line integrals of $\boldsymbol{A}$'s (or of
$\boldsymbol{E}$'s). But the most crucial element for what is done in the text
is the exclusive use of forms (\ref{LambdaFull2}) and (\ref{LambdaFIN}) (where
$B_{z}$ only appears at $t_{0}$), and the fact that, within each solution, the
sense of $\boldsymbol{A}$-integrals is the \textit{same} as the sense of the
$\boldsymbol{E}$-integrals. (This is \textit{not} true in the other solutions
where $B_{z}(..,t)$ appears, as the reader can directly see). These facts were
crucial to the discussion that addresses the so called van Kampen
\textquotedblleft paradox\textquotedblright\ in Section XII. It is the
subtraction of these two forms (\ref{LambdaFull2}) and (\ref{LambdaFIN})
(where the $B_{z}$-term is evaluated at $t_{0}$) that leads to the final
\textbf{causal} result that $\Delta\Lambda(t)=\Phi(t_{0})$ in the main text
(when the spacetime point $(x,y,t)$ is such that, at the instant $t$, the
physical information (generated at $t_{0}$) has not yet reached the spatial
point $(x,y)$).

\textbf{Figure 1.} (Color online)$\boldsymbol{:}$ Examples of simple
field-configurations (in simple-connected regions), where the nonlocal terms
exist and are nontrivial, but can easily be determined$\boldsymbol{:}$ (a) a
striped case in 1+1 spacetime, where the electric flux enclosed in the
\textquotedblleft observation rectangle\textquotedblright\ is dependent on $t$
but independent of $x$; (b) a triangular distribution in 2-D space, where the
part of the magnetic flux inside the corresponding \textquotedblleft
observation rectangle\textquotedblright\ depends on \textit{both} $x$
\textit{and} $y$. The appropriate choices for the corresponding nonlocal
functions $g(x)$ and $\hat{g}(t)$ for case (a), or $g(x)$ and $h(y)$ for case
(b), are given in the text (Sections VIII and VI respectively).

\textbf{Figure 2.} (Color online)$\boldsymbol{:}$ The analog of paths of Fig.1
but now in 2+1 spacetime for the van Kampen thought-experiment, when the
instant of observation $t$ is so short that the physical information has not
yet reached the spatial point of observation $(x,y)$. The two solutions (that,
for wavepackets, have to be subtracted in order to give the phase difference
at $(x,y,t)$) are eqs (\ref{LambdaFull2}) and (\ref{LambdaFIN}) of the text,
and are here characterized through their electric field $E$-line integral
behavior$\boldsymbol{:}$ \textquotedblleft electric field path
(I)\textquotedblright\ (the red-arrow route) denotes solution (\ref{LambdaFIN}%
), and \textquotedblleft electric field path (II)\textquotedblright\ (the
green-arrow route) denotes solution (\ref{LambdaFull2}). Note that the strips
of Fig.1(a) have now given their place to a light-cone. At the point of
observation, the Aharonov-Bohm phase difference has now become
\textquotedblleft causal\textquotedblright\ due to cancellations between the
two solutions (the two \textquotedblleft electric field
paths\textquotedblright\ above).

\textbf{Figure 3.} (Color online)$\boldsymbol{:}$ The standard double-slit
apparatus with an additional strip of a perpendicular magnetic field $B$ of
width $W$ placed between the slit-region and the observation screen. In the
text we deal for simplicity with the case $W<<L,$ so that deflections (of the
semiclassical trajectories) due to the Lorentz force, shown here for a
negative charge $q$, are very small.

\textbf{Figure 4.} (Color online)$\boldsymbol{:}$ The analog of Fig.3 (again
for a negative $q$) but with an additional electric field parallel to the
observation screen that is turned on for a time interval $T$. In the text we
deal for simplicity with the case $T<<\frac{L}{v}$ (with $v=\frac{1}{m}%
\frac{h}{\lambda}$, $\lambda$ the de Broglie wavelength), so that deflections
(of the semiclassical trajectories) due to the electric force are again very
small. For both Fig.3 and 4, it is shown in the text that $\Delta
\varphi^{semiclassical}=-\Delta\varphi^{AB},$ hence we observe an extra minus
sign compared to what is usually reported in the literature.


\begin{thebibliography}{99}
\bibitem {ring}An error often made (when the 1-D path is a closed loop, such
as a nanoring) is the simultaneous imposition of spatial single-valuedness on
both $\Psi(x,t)^{(\Phi)}$ and $\Psi(x,t)^{(0)}$ leading (erroneously) to
quantization of the enclosed inaccessible flux $\Phi$. Obviously $\Phi$ can
have any arbitrary value, and spatial single-valuedness must only be imposed
on $\Psi(x,t)^{(\Phi)}$
\end{thebibliography}
\end{document}